\documentclass[11pt,showpacs]{revtex4}
\def\comment#1{}

\usepackage{graphicx}
\begin{document}
\title{Detailed discussions and calculations of quantum Regge calculus of Einstein-Cartan theory}
\author{She-Sheng Xue}
\email{xue@icra.it}
\affiliation{
ICRANeT Piazzale della Repubblica, 10-65122, Pescara, Italy\\
Department of Physics, University of Rome ``Sapienza'', Piazzale A.~Moro 5, 00185, Rome, Italy
}


\begin{abstract}
This article presents detailed discussions and calculations of the recent paper ``Quantum Regge calculus of Einstein-Cartan theory'' in Phys.~Lett.~B682 (2009) 300.  
The Euclidean space-time is discretized by a four-dimensional simplicial complex. We adopt basic tetrad and spin-connection fields to describe the simplicial complex. By introducing diffeomorphism and local Lorentz invariant holonomy fields, we construct a regularized Einstein-Cartan theory for studying the quantum dynamics of the simplicial complex and fermion fields. This regularized Einstein-Cartan action is shown to properly approach to its continuum counterpart in the continuum limit. Based on the local Lorentz invariance, we derive the dynamical equations satisfied by invariant holonomy fields. In the mean-field approximation, we show that the averaged size of 4-simplex, the element 
of the simplicial complex, is larger than the Planck length. This formulation provides a theoretical framework for analytical calculations and numerical simulations to study the quantum Einstein-Cartan theory.

\end{abstract}
\maketitle

\section{\bf Introduction.}

Since the Regge calculus \cite{regge61,wheeler1964} was proposed for the discretization of gravity theory in 1961, many progresses have been made in the approach of Quantum Regge calculus \cite{hammer_book,shamber}
and its variant dynamical triangulations \cite{loll1999}. In particular,
the renormalization-group treatment is applied to 
discuss any possible scale dependence of gravity \cite{hammer_book}. Inspired by the success of 
lattice regularization of non-Abelian gauge theories, the 
gauge-theoretic formulation \cite{smolin1979} of quantum gravity using connection variables on a flat hypercubic lattice of the space-time was studied in the Lagrangian formalism.   
The canonical quantization approaches to the Regge calculus in Hamiltonian formulation are studied in Ref.~\cite{William1986}.  
A locally finite model for gravity has been recently proposed \cite{thooft2008}. All these studies are very important steps to understand the Einstein general relativity for gravitational fields in the framework of {\it quantum field theory}.
In the brief paper \cite{xue2009} based on the scenario of quantum Regge calculus, we present a diffeomorphism and 
local Lorentz invariant (i.e., {\it local} gauge-invariant) regularization and quantization of Euclidean Einstein-Cartan (EC) theory. 
Detailed calculations and discussions are presented in this article.

The four-dimensional Euclidean space-time is discretized by a simplicial complex, analogously to the formulation of the Regge calculus. In the framework of the Einstein-Cartan theory, we adopt basic gravitational variables, i.e., a pair of tetrad and spin-connection fields to describe the simplicial complex. Introducing diffeomorphism and local Lorentz invariant (i.e., {\it local}  gauge-invariant) holonomy fields in terms of tetrad and spin-connection fields along loops, we propose an invariantly regularized EC theory for the dynamics of simplicial complex, which couples to fermion spinor fields. We show that in the continuum limit when the wavelengths of tetrad and spin-connection fields are much larger than the Planck length, this regularized EC action properly approaches to the continuum EC action. The quantum dynamics of the simplicial complex is described by the Euclidean partition function that is a Feynman path-integral overall quantum tetrad, 
spin-connection, and fermion fields with the weight of regularized EC action. Based on {\it local} gauge invariance, we derive the dynamical equations satisfied by invariant holonomy fields of tetrad, spin-connection, and fermion fields. In the mean-field approximation, we show the averaged size of 4-simplex (and its 3-simplex and 2-simplex), elements of the simplicial complex, has to be larger than the Planck length. This formulation provides a theoretical framework for analytical calculations, in particular, numerical simulations to study the Einstein-Cartan theory as a {\it quantum field theory}. 

This article is organized as follows: In Sec.~ \ref{conec}, we give a brief review of the continuum EC theory. In Sec.~ \ref{regec}, we discuss the regularized EC theory
based on (1) the description of simplicial complex by tetrad and spin-connection fields; (2) parallel transport equations in simplicial complex; (3) invariant holonomy fields and regularized EC action and their continuum limit; 
(4) the Euclidean partition function.
In Secs.~ \ref{ifermion} and \ref{dehf}, we study chiral gauge symmetric bilinear and quadralinear-fermion actions, 
and derive dynamical equations for holonomy fields. In Sec.~ \ref{meanf}, we adopt the method of the mean-field approximation to show the averaged size of the 4-simplex has to be larger than the Planck length. In the last section, we give some concluding remarks, and detailed calculations are arranged in Appendices \ref{con}, \ref{xcon}, \ref{a2}, \ref{meana} and \ref{meana1}. 

\section{\bf Continuum Einstein-Cartan theory}\label{conec}

The basic gravitational variables in the Einstein-Cartan theory constitute a pair of tetrad and spin-connection fields 
$[e_\mu^{\,\,\,a}(x), \omega^{ab}_\mu(x)]$, whose Dirac-matrix values
\begin{equation}
 e_\mu (x)= e_\mu^{\,\,\,a}(x)\gamma_a \quad {\rm  and}\quad \omega_\mu(x) = \omega^{ab}_\mu(x)\sigma_{ab}.
 \label{basic}
\end{equation}
The fields $e_\mu^{\,\,\,a}(x)$ and $\omega^{ab}_\mu(x)$ are 1-form real fields 
on the four-dimensional Euclidean space-time ${\mathcal R}^4$, taking values, respectively, in the local Lorentz vector space $V_{\mathcal L}$
and in the Lie algebra $so(4)$ of the Lorentz group $SO(4)$ of the linear transformations of
$V_{\mathcal L}$ preserving $\delta^{ab}=(+,+,+,+)$. In this local Lorentz vector space $V_{\mathcal L}$, fermions are 
spinor fields $\psi(x)$, Dirac $\gamma$ matrices obey 
\begin{equation}
\{\gamma_a,\gamma_b\}=-2\delta_{ab},
\label{gbasic}
\end{equation}
$\gamma^\dagger_a=-\gamma_a$ and  $\gamma^2_a=-1$ ($a=0,1,2,3$); 
the Hermitian $\gamma_5$ matrix 
\begin{equation}
\gamma_5=\gamma^5=\gamma^0\gamma^1\gamma^2\gamma^3=\gamma_0\gamma_1\gamma_2\gamma_3,
\label{g5basic}
\end{equation}
$\gamma_5^\dagger=\gamma_5$ and $\gamma_5^2=1$; the Hermitian spinor matrix,
\begin{equation}
\sigma^{ab}=\frac{i}{2}[\gamma^a,\gamma^b].
\label{gsbasic}
\end{equation}
Totally antisymmetric tensor 
$\epsilon_{\mu\nu\rho\sigma}=\epsilon_{abcd}e^{\,\,\,a}_\mu e^{\,\,\,b}_\nu e^{\,\,\,c}_\rho e^{\,\,\,d}_\sigma$ . The space-time metric of four-dimensional Euclidean manifold ${\mathcal R}^4$ is
\begin{equation} 
g_{\mu\nu}(x)=e^{\,\,\,a}_\mu(x)e^{\,\,\,b}_\nu(x)\delta_{ab}=-\frac{1}{2}\{e_\mu,e_\nu\}.
\label{diracg}
\end{equation}
And the Lorentz scalar components of the metric tensor are then simply  
\begin{equation} 
\delta_{ab}=g_{\mu\nu}\,\,e_{\,\,\,a}^\mu e_{\,\,\,b}^\nu,
\label{diracg1}
\end{equation}
where the inverse of the tetrad fields $e_{\,\,\,a}^\mu e^{\,\,\,a}_\nu=\delta_{\,\,\,\nu}^\mu$ and $ e^{\,\,\,b}_\mu e_{\,\,\,a}^\mu=\delta_{\,\,\,a}^b$.

Two gauge invariances due to the equivalence principle have to be respected: (1)
the diffeomorphism invariance under the general coordinate transformation $x\rightarrow x'(x)$; (2) the {\it local} 
gauge invariance under the local 
Lorentz coordinate transformation $\xi(x)\rightarrow \xi'(x)$, i.e.,
\begin{eqnarray}
\xi^{'a}(x)=[\Lambda(x)]^a_b\xi^b(x).
\label{llt}
\end{eqnarray}  
Under the local Lorentz coordinate transformation (\ref{llt}), 
the finite {\it local} gauge transformation is 
\begin{eqnarray} 
&&{\mathcal V}(\xi)=\exp i[\theta^{ab}(\xi)\sigma_{ab}]\in SO(4)\nonumber\\
&&{\mathcal V}(\xi)\gamma_a{\mathcal V}^\dagger(\xi)
=[\Lambda^{-1}(x)]_a^b\gamma_b ,
\label{lgauge}
\end{eqnarray}
where $\theta^{ab}(\xi)$ is the antisymmetric tensor and an arbitrary function of $\xi=\xi(x)$. 
The Dirac-matrix valued fields $e_\mu$, $\omega_\mu$ and fermion spinor field $\psi$ are transformed as follows
\begin{eqnarray}
e_\mu(\xi)\rightarrow e'_\mu(\xi)&=&{\mathcal V}(\xi)e_\mu(\xi){\mathcal V}^\dagger(\xi);\label{varie}\\
\omega_\mu(\xi)\rightarrow \omega'_\mu(\xi)&=&{\mathcal V}(\xi)\omega_\mu(\xi){\mathcal V}^\dagger(\xi)+
{\mathcal V}(\xi)\partial_\mu{\mathcal V}^\dagger(\xi),
\label{gtran0}\\
\psi(\xi)\rightarrow \psi'(\xi)&=&{\mathcal V}(\xi) \psi(\xi); \label{ftran}\\
{\mathcal D}'_\mu &=& {\mathcal V}(\xi){\mathcal D}_\mu
{\mathcal V}^\dagger(\xi),\label{vcovd}
\end{eqnarray}
where the derivative $\partial_\mu= e_\mu^a(\partial/\partial\xi^a)$, 
the covariant derivative 
\begin{eqnarray}
{\mathcal D}_\mu =\partial_\mu - ig\omega_\mu(\xi),
\label{cd}
\end{eqnarray}
and $g$ is the gauge coupling.
Corresponding to the finite {\it local} gauge transformations (\ref{varie}-\ref{ftran}), infinitesimal {\it local} gauge transformations for fields $e_\mu$, $\omega_\mu$ and $\psi$ are 
\begin{eqnarray} 
\delta e_\mu(\xi)&=&\theta^{ab}(\xi)d_{ab,c}e^c_\mu(\xi);\label{ivarie}\\
\delta\omega_\mu(\xi)&=&2\gamma_5\epsilon_{abcd}\omega_\mu^{ab}\theta^{cd}(\xi)-i\sigma_{ab}\partial_\mu\theta^{ab}(\xi);\label{gtran1}\\
\delta \psi(\xi)&=&i\theta^{ab}(\xi)\sigma_{ab} \psi(\xi), \label{iftran}
\end{eqnarray}
where
\begin{eqnarray}
d_{ab,c}=i[\sigma_{ab},\gamma_c]=2(\delta_{bc}\gamma_a-\delta_{ac}\gamma_b),
\label{psisi}
\end{eqnarray}
and we use the commutator relation
\begin{eqnarray}
\{\sigma^{\alpha\beta},\sigma^{\delta\gamma}\}
=-2i\gamma^5\epsilon^{\alpha\beta\delta\gamma},
\label{sisi}
\end{eqnarray}
to obtain Eq.~(\ref{gtran1}).
 
In an $SU(2)$ gauge theory, 
gauge field $A_a(\xi_E)$ can be viewed as a connection 
$\int A_a(\xi_E)d\xi_E^a$ on the global flat manifold. On a locally flat manifold, the spin connection 
$\omega_\mu dx^\mu =\omega_a(\xi)d\xi^a$, 
where $\omega_a(\xi)=\omega_\mu e^{\mu}_{\,\,\,a}$,
one can identify that the spin-connection field 
$\omega_\mu(x)$ or $\omega_a(\xi)$ is the gravity analog of gauge field
and its {\it local} curvature is given by
\begin{equation}
R^{ab}=d\omega^{ab} - g\omega^{ae}\wedge\omega^{b}{}_{e},
\label{rcurvature}
\end{equation}
and the Dirac-matrix valued curvature $R_{\mu\nu}=R^{ab}_{\mu\nu}\sigma_{ab}$.  Under the gauge transformation (\ref{varie},\ref{gtran0}),
\begin{eqnarray} 
R^{'ab}={\mathcal V}(\xi)R^{ab}(\xi){\mathcal V}^\dagger(\xi).
\label{ltr}
\end{eqnarray}
The diffeomorphism invariance under the general coordinate transformation $x\rightarrow x'(x)$ is preserved by all derivatives and $d$-form fields on ${\mathcal R}^4$ made to be coordinate scalars with the help of tetrad fields $e^{\,\,\,a}_\mu=\partial\xi^a/\partial x^\mu$ (see Ref.~\citep{wein1972}). The diffeomorphism and {\it local} gauge-invariant EC action for gravity coupling to fermions is given by 
the Palatini action $S_P$ 
and host modification $S_H$ 
for the gravitational field, 
\begin{eqnarray}
S_{EC}(e,\omega
)&=&S_P(e,\omega)+S_H(e,\omega)+S_F(e,\omega,\psi)
\label{ec0}\\
S_P(e,\omega)&=&\frac{1}{4\kappa}\int d^4x\det(e)\epsilon_{abcd}e^a\wedge e^b \wedge R^{cd},
\label{host}\\
S_H(e,\omega)&=&\frac{1}{2\kappa\tilde\gamma}\int d^4x\det(e) e_a\wedge e_b \wedge R^{ab}\label{host1},
\end{eqnarray}
and fermion action $S_F$ (see Refs.~\cite{art1989,kleinert}), 
\begin{eqnarray}
S_F(e,\omega,\psi)&=&\frac{1}{2}\int d^4x\det(e)\left[\bar\psi e^\mu {\mathcal D}_\mu\psi +{\rm h.c.}\right],
\label{art}
\end{eqnarray}
where $\kappa\equiv 8\pi G$, the Newton constant $G=1/m_{\rm Planck}^2$, $\det(e)$ is the Jacobi of mapping $x\rightarrow \xi(x)$ and the integration $\int d^4x\equiv \int_{{\mathcal R}^4} d^4x$.
\comment{
In addition, 
the diffeomorphism invariance under the general coordinate transformation $x\rightarrow x'(x)$ 
is preserved by all fields in Eqs.~(\ref{host}-\ref{art}) 
made to be coordinate scalars by using tetrad fields. The derivatives represents the propagation of fields in coordinate space is related to the connection field 
in local Lorentz frame. spin-connection is the gravity analog of gauge field, how ever it is constructed by the general coordinate derivatives of tetrad field relating to general connection. .....
}
The complex Ashtekar connection \cite{A1986} with 
reality condition and the real Barbero connection \cite{b1995} are linked by a canonical transformation of the connection with a finite complex 
Immirzi parameter $\tilde\gamma\not=0$ \cite{i1997}, which is crucial for {\it loop quantum gravity} \cite{rt1998}.
\comment{, a quantum theory of gravity in 
Hamiltonian formalism, where intrinsic discrete eigenvalues of invariant area and volume operators are obtained
in the diffeomorphism invariant Hilbert space, as a result, 
the space-time is discretized with the Planck length and
the black-hole entropy is obtained.
} 

Classical equations of motion can be obtained by the stationarity of 
the EC action (\ref{ec0})
under variations
(\ref{varie}-\ref{ftran}),
\begin{equation}
\delta S_{EC}(e,\omega,\psi)=\frac{\delta S_{EC}}{\delta e_\mu}\delta e_\mu
+\frac{\delta S_{EC}}{\delta \psi(x)}\delta \psi(x)
+\frac{\delta S_{EC}}{\delta \omega_\mu}\delta \omega_\mu
=0.
\label{inv}
\end{equation}
From Eqs.~(\ref{ivarie}-\ref{iftran}), we find that Eq.~(\ref{inv}) can be expressed in terms of independent bases $\gamma_5$, $\gamma_\mu$ and $\sigma_{ab}$ of Dirac matrices. Therefore, for arbitrary function $\theta_{ab}(\xi)$, Eq.~(\ref{inv}) leads to the following three equalities 
\begin{eqnarray}
\frac{\delta S_{EC}}{\delta \psi}=0;\quad \frac{\delta S_{EC}}{\delta e_\mu}=0;\quad  \frac{\delta S_{EC}}{\delta \omega_\mu}=0.
\label{3e}
\end{eqnarray}
The first and second equations respectively lead to the Dirac equation,
\begin{eqnarray}
e^\mu {\mathcal D}_\mu\psi(x)=0,
\label{deq}
\end{eqnarray} 
and the Einstein equation 
\begin{eqnarray}
\epsilon_{abcd}e^a\wedge e^b\wedge R^{cd}[\omega(e)]=\kappa\bar\psi(x) (e\wedge{\mathcal D})\psi(x),
\label{eineq}
\end{eqnarray} 
where the energy-momentum tensor is
\begin{eqnarray}
\bar\psi(e\wedge{\mathcal D})\psi\equiv \frac{1}{2}\bar\psi[e_\mu{\mathcal D}_\nu-{\mathcal D}_\mu e_\nu]\psi.
\label{emt}
\end{eqnarray} 
The gauge invariance of the EC action (\ref{ec0}) under 
the gauge transformation (\ref{gtran1}) leads to the third constraint equation $\delta S_{EC}/\delta \omega_\mu=0$ of Eq.~(\ref{3e}), which is the Cartan structure equation,
\begin{equation}
de^a -g\omega^{ab}\wedge e_{b}-T^a=0,
\label{werelation1}
\end{equation}
where the nonvanishing torsion field,
\begin{eqnarray}
T^a=\kappa ge_b\wedge e_cJ^{ab,c},
\label{torsion}
\end{eqnarray}
relating to the fermion spin current
\begin{eqnarray}
J^{ab,c}=i\bar\psi\{\sigma^{ab},\gamma^c\}\psi &=&\epsilon^{abcd}\bar\psi\gamma_d\gamma^5\psi,
\label{spinc}\\
\{\sigma^{ab},\gamma^c\}&=&i\epsilon^{abcd}\gamma^5\gamma_d.
\label{sgc}
\end{eqnarray} 
The fermion spin-current (\ref{spinc})
contributes only to the pseudotrace axial vector of torsion tensor, which is one of irreducible parts of torsion 
tensor \cite{torsion2001}.
The solution to Eq.~(\ref{werelation1}) is 
\begin{equation}
\omega^{ab}_\mu=\omega^{ab}_\mu(e)+\tilde\omega^{ab}_\mu,\quad
\tilde\omega^{ab}_\mu=\kappa g e_\mu^cJ^{ab}{}_c,
\label{connectiont}
\end{equation}
where the connection $\omega^{ab}_\mu(e)$ obeys Eq.~(\ref{werelation1}) for torsion-free case $T^a=0$,
\begin{equation}
de^a -g\omega^{ab}(e)\wedge e_{b}=0.
\label{werelation0}
\end{equation}
\comment{
The curvature (\ref{rcurvature}) contains quadratic term of field $\omega_\mu$ and the covariant derivative ${\mathcal D}_\mu$ contains the field $\omega_\mu$ (\ref{connectiont}), which indicate the nonlinear interaction of fermion fields in addition to the coupling of fermion fields to the 
torsion-free field $\omega_\mu(e)$. 
In fact,
}
Replacing the spin-connection field $\omega_\mu^{ab}$ in the Einstein-Cartan action (\ref{host},\ref{art}) by Eq.~(\ref{connectiont}),
\begin{eqnarray}
S_P[e,\omega]&\rightarrow & S_P[e,\omega(e)]+\kappa g^2\int d^4x\det(e)(\bar\psi\gamma^d\gamma^5\psi)(\bar\psi\gamma_d\gamma^5\psi);\label{eca0}\\
S_F[e,\omega,\psi,\bar\psi]&\rightarrow & S_F[e,\omega(e),\psi,\bar\psi]
+2\kappa g^2\int d^4x\det(e)(\bar\psi\gamma^d\gamma^5\psi)(\bar\psi\gamma_d\gamma^5\psi),
\label{eca1}
\end{eqnarray} 
one obtains the well-known Einstein-Cartan theory: 
the standard tetrad action of torsion-free gravity coupling to fermions 
with four-fermion interactions,
\begin{eqnarray}
S_{EC}[e,\omega(e),\psi,\bar\psi]&=&S_P[e,\omega(e)]+S_F[e,\omega(e),\psi,\bar\psi]\nonumber\\
&+&3\kappa g^2\int d^4x\det(e)(\bar\psi\gamma^d\gamma^5\psi)(\bar\psi\gamma_d\gamma^5\psi).
\label{eca2}
\end{eqnarray}
Note that the four-fermion interaction actually is the coupling of two fermion spin-currents (\ref{spinc}).
Taking into account the host action (\ref{host1}), one obtains
\begin{eqnarray}
S_{EC}[e,\omega(e),\psi]&=&S_P[e,\omega(e)]+S_H[e,\omega(e)]+S_F[e,\omega(e),\psi]+S_{4F}(e,\psi)\label{eca}\\
S_{4F}(e,\psi)&=&3\zeta\kappa g^2\int d^4x\det(e)(\bar\psi\gamma^d\gamma^5\psi)(\bar\psi\gamma_d\gamma^5\psi),
\label{4f}
\end{eqnarray}
where $\zeta=\tilde\gamma^2/(\tilde\gamma^2+1)$ \cite{ar05}. Using the commutator relations (\ref{sisi}) and $[\sigma_{ab},\gamma_5]=0$, 
one can show that $(\bar\psi\gamma_d\gamma^5\psi)$ is a pseudovector and (\ref{4f}) is invariant under the gauge transformation (\ref{ftran}).

As we can see from Eqs.~(\ref{art}) to (\ref{eca}), the bilinear term (\ref{art}) of massless
fermion fields coupled to the spin-connection field (\ref{cd}) is bound to yield a nonvanishing torsion field $T^a$ (\ref{werelation1}), which is local and 
static (see, for example, Refs.~\cite{kleinert,s2001}). As a result, the spin connection $\omega_\mu$ is no longer torsion-free and 
acquires a torsion-related spin-connection $\tilde\omega^{ab}_\mu$ (\ref{connectiont}), in addition to the torsion-free spin connection 
$\omega^{ab}_\mu(e)$. The torsion-related spin connection 
$\tilde\omega^{ab}_\mu$ is related to the fermion spin current (\ref{spinc}). The quadratic term 
of the spin connection field $\omega$ in the curvature (\ref{rcurvature}) and the coupling between the spin connection field $\omega$
and fermion spin current in Eqs.~(\ref{cd},\ref{art}) lead to the quadralinear terms of fermion fields 
in Eqs.~(\ref{eca0}) and (\ref{eca1}). Another way to see this is to treat the static torsion-related spin-connection $\tilde\omega^{ab}_\mu$ (\ref{connectiont}) as a static auxiliary field, which 
has its quadratic term and linear coupling to the spin current of fermion fields. 
Performing the Gaussian integral of the static auxiliary field, we exactly 
obtain the quadralinear term (\ref{4f}), in addition to the torsion-free EC action.

The action (\ref{ec0}) and classical Eqs.~ (\ref{deq}-\ref{werelation1}) can be separated into 
left- and right-handed parts  \cite{lrseparation}, 
with respect to the local $SU_L(2)$ and $SU_R(2)$ symmetries of the Lorentz group $SO(4)=SU_L(2)\otimes SU_R(2)$. 
This can be shown by writing Dirac fermions 
$\psi=\psi_L+\psi_R$, where Weyl fermions $\psi_{L,R}\equiv P_{L,R}\psi,\, P_{L,R}=(1\mp\gamma_5)/2$; and 
Dirac-matrix valued tetrad field $e^\mu= e^\mu_L+e^\mu_R\,$, $e^\mu_{L,R}\equiv P_{L,R}e^\mu$, as well as Dirac-matrix valued spin-connection fields $\omega_\mu=\omega^\mu_L+\omega^\mu_R\,$, $\omega^\mu_{L,R}\equiv P_{L,R}\omega^\mu$.

\begin{figure}[ptb]
\includegraphics[scale=1.2
]{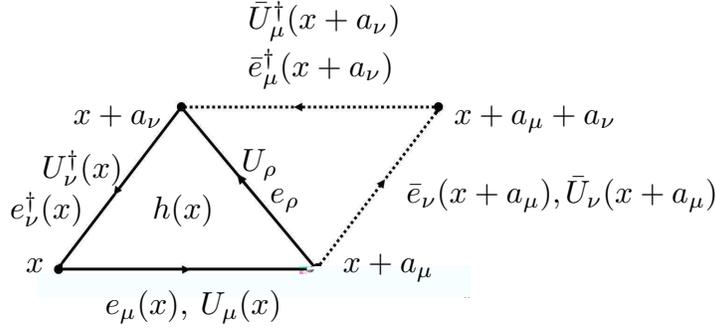}
\comment{
\put(-100,25){$h(x)$}
\put(-140,8){$x$}
\put(-40,8){$x+a_\mu$}
\put(-115,-5){$e_\mu(x)$,}
\put(-85,-5){$U_\mu(x)$}
\put(-5,55){$x+a_\mu+a_\nu$}
\put(-125,55){$x+a_\nu$}
\put(-70,70){$\bar e^\dagger_\mu(x+a_\nu)$}
\put(-70,85){$\bar U^\dagger_\mu(x+a_\nu)$}
\put(-145,25){$e^\dagger_\nu(x)$}
\put(-135,38){$U^\dagger_\nu(x)$}
\put(-63,30){$e_\rho$}
\put(-72,40){$U_\rho$}
\put(-20,30){$\bar e_\nu(x+a_\mu)$,}
\put(30,30){$\bar U_\nu(x+a_\mu)$}
}
\caption{
We sketch a 2-simplex (triangle) $h(x)$ formed by 
three edges $l_\mu(x)=ae_\mu(x)$, $l_\rho(x+a_\mu)=ae_\rho(x+a_\mu)$ and $l_\nu(x+a_\mu)=ae_\nu(x+a_\nu)$ [$a=1$] connecting three vertexes $x$, $x+a_\mu $ and $x+a_\nu$.
Assuming three edge spacings $a_{\mu}$, $a_\nu$ and $a_{\rho}$ (\ref{edged0}) are 
so small that the geometry of the interior of each 4-simplex and its subsimplex (3- and 2-simplex) is approximately flat, we assign a local Lorentz frame to each 4-simplex. On the local Lorentz manifold $\xi^a(x)$ at a space-time point 
``$x$'', we sketch a closed parallelogram ${\mathcal C}_P(x)$ lying in the 2-simplex $h(x)$. Its two edges 
$e_\mu(x)$ and $e^\dagger_\nu(x)$ are two edges of the 2-simplex $h(x)$, and other two edges (dashed lines) $\bar e^\dagger_{\mu}(x+a_\nu)$ and $\bar e_\nu(x+a_\mu)$ 
are parallel transports of $e^\dagger_{\mu}(x)$ and $e^\dagger_{\nu}(x)$ along $\nu$ and $\mu$ directions, respectively [see Eqs.~(\ref{wel1},\ref{wel2}) and (\ref{up1},\ref{up2})]. 
Each 2-simplex in the simplicial complex has a closed parallelogram lying in it.
Group-valued gauge fields $U_\mu(x)$ and $U^\dagger_\nu(x)$ are respectively associated to edges $e_\mu(x)$ and $e^\dagger_\nu(x)$ of the 2-simplex $h(x)$, as indicated. The fields $e_\rho\equiv e_\rho(x+a_\mu)$ and $U_\rho\equiv U_\rho(x+a_\mu)$ are associated to the third edge $(x+a_\mu,\rho)$ of the 2-simplex $h(x)$. The group fields $\bar U_\nu(x+a_\mu)$ and $\bar U^\dagger_\mu(x+a_\nu)$ indicate the parallel transports of $U^\dagger_\nu(x)$ and $U_\mu(x)$ [see Eqs.~(\ref{ww1p},\ref{ww2p}) and (\ref{pup11},\ref{pnup11})] for the zero curvature case. Note that
the point $(x+a_\mu +a_\nu)$ is not a vertex of the simplicial complex, points: $(x-a_{\mu})$, $(x-a_{\nu})$, $(x+a_\mu+a_\mu)$, $(x+a_\mu-a_\rho)$, and $(x+a_\nu+a_\rho)$, which are not shown in the sketch, are not vertexes of the simplicial complex as well. Parallel transports $\bar e_\nu(x+a_\mu)$ and $\bar e^\dagger_\mu(x+a_\nu)$, as well as $\bar U_\nu(x+a_\mu)$ and $\bar U^\dagger_\mu(x+a_\nu)$ are not associated to any edge of the simplicial complex. Throughout this article, the notations $\bar e$ and $\bar U$ indicates parallel transports that are not associated to any edge of the simplicial complex.   
}%
\label{pl}%
\end{figure}

\section{\bf The regularized Einstein-Cartan theory}\label{regec}
 
\subsection{Simplicial complex}\label{2-simplex}

The four-dimensional Euclidean manifold ${\mathcal R}^4$
is discretized as an ensemble of ${\mathcal N}_0$ 
space-time points (vertexes) ``${\it x}\in {\mathcal R}^4$'' and ${\mathcal N}_1$ 
links (edges) ``$l_\mu(x)$'' connecting two 
neighboring vertexes. This ensemble forms a simplicial manifold ${\mathcal M}$ embedded into the ${\mathcal R}^4$. The way to construct a simplicial manifold depends also on the assumed topology of the manifold, which gives geometric constrains on the numbers of subsimplices (${\mathcal N}_0,{\mathcal N}_1,\cdot\cdot\cdot$, see 
Ref.~\cite{loll1999}). 
In this article, analogously to the simplicial manifold adopted by the Regge calculus 
we consider the simplicial manifold ${\mathcal M}$ as a simplicial complex, whose 
elementary building block is a 4-simplex (pentachoron). 
The 4-simplex has five vertexes -- 0-simplex (a space-time point ``${\it x}$''), five ``faces'' -- 3-simplex (a tetrahedron), and each 3-simplex has four faces -- 2-simplex [a triangle $h(x)$], and each 2-simplex has three faces -- 1-simplex [an edge or a link ``$l_\mu(x)$'']. 
Different configurations of the simplicial complex correspond to variations of relative vertex-positions $\{x\}$, edges ``\{$l_\mu(x)\}$'' and 
``deficit angles'' associating to 2-simplices $h(x)$. 
These configurations will be described by the configurations of dynamical tetrad fields $e_\mu(x)$ and spin-connection fields $\omega_\mu(x)$ assigned to 1-simplexes  (edges) of the simplicial complex 
in this article. 
We are not clear now how to relate configurations of fields $e_\mu(x)$ and $\omega_\mu(x)$ to topological constrained configurations of the simplicial complex
in dynamical triangulations.

\subsubsection{Edges: 1-simplexes}

The edge (1-simplex) denoted by $(x,\mu)$, connecting two neighboring vertexes labeled by $x$ and $x+a_\mu$, can be represented as a four-vector field $l_\mu(x)$, defined at the vertex ``$x$'' by its forward direction $\mu$ pointing from $x$ to $x+a_\mu$ and its length
\begin{equation}
a_\mu(x)\equiv |l_\mu(x)|\not=0,
\label{edged0}
\end{equation}
which is the distance between two vertexes $x$ and $x+a_\mu$. The fundamental tetrad field $e_\mu(x)$ is assigned to each edge (1-simplex) of the simplicial complex to describe the edge location ``$x$,'' direction ``$\mu$'' and length $a_\mu(x)$. 
We use the tetrad field $e_\mu(x)$, defined at the vertex $x$, 
to characterize the edge (1-simplex) $l_\mu(x)$
\begin{equation}
l_\mu(x)\equiv ae_\mu(x),
\label{edged}
\end{equation}
where the Planck length $a\equiv (8\pi G)^{1/2}=\kappa^{1/2}$, and
\begin{equation}
|l_\mu(x)|\equiv \frac{a}{2}\Big\{|{\rm tr} [e_\mu(x)\cdot e_\mu(x)]|
\Big\}^{1/2}.
\label{edgel}
\end{equation}
By definition, either $l_\mu(x)$ or $e_\mu(x)$ is a Dirac-matrix valued four-vector field, defined at the vertex ``$x$''. 

\subsubsection{Triangles: 2-simplexes}

We consider an orienting 2-simplex (triangle) (see Fig.~\ref{pl}). This 2-simplex (triangle) has three edges connecting three neighboring vertexes that are labeled by 
$x$, $x+a_\mu$ and $x+a_\nu$. This triangle (2-simplex) has two orientations: (i) the anti-clocklike $h(x)$ [$\,x \stackrel{\mu}{\longmapsto} x+a_\mu \stackrel{\rho}{\longmapsto} x+a_\nu \stackrel{\nu}{\longmapsto} x\,$] and (ii) the clocklike $h^\dagger(x)$ [$\,x \stackrel{-\nu}{\longmapsto} x+a_\nu \stackrel{-\rho}{\longmapsto} x+a_\mu \stackrel{-\mu}{\longmapsto} x\,$].

Along the triangle path of the anti-clocklike  
2-simplex $h(x)$ [$\,x \stackrel{\mu}{\longmapsto} x+a_\mu \stackrel{\rho}{\longmapsto} x+a_\nu \stackrel{\nu}{\longmapsto} x\,$], 
three edges and their forward directions are represented by: (1) $l_\mu(x)$ and $\mu$ pointing from $x$ to $x+a_\mu$; (2) $l_\rho(x+a_\mu)$ and $\rho$ pointing from $x+a_\mu$ to $x+a_\nu$; (3) $l_\nu(x+a_\nu)$ and
$\nu$ pointing from $x+a_\nu$ to $x$. The lengths of three edges are respectively represented by edge spacings $a_\mu$, $a_\rho$ and $a_\nu$ [see Eqs.~(\ref{edged0},\ref{edgel})].
\comment{
\begin{eqnarray}
a_\rho(x+a_\mu) & \equiv & \frac{a}{2}\Big\{{\rm tr} [e_\rho(x+a_\mu)\cdot e_\rho(x+a_\mu)]\Big\}^{1/2}\nonumber\\
a_\nu(x+a_\nu) & \equiv & \frac{a}{2}\Big\{{\rm tr} [e_\nu(x+a_\nu)\cdot e_\nu(x+a_\nu)]\Big\}^{1/2}.
\label{edgel1}
\end{eqnarray}
}
We use the tetrad fields
\begin{eqnarray}
e_\mu(x),\quad e_\rho(x+a_\mu),\quad e_\nu(x+a_\nu),
\label{aol3e}
\end{eqnarray}
defined at $x$, $x+a_\mu$ and $x+a_\nu$, to respectively characterize locations, forward directions and lengths of three edges: (\ref{edged}) and 
\begin{eqnarray}
l_\rho(x+a_\mu) &= &ae_\rho(x+a_\mu),\nonumber\\ l_\nu(x+a_\nu)&= & ae_\nu(x+a_\nu),
\label{aol3el}
\end{eqnarray}
of the anti-clocklike  
2-simplex $h(x)$ [see Fig.~\ref{pl} and Eqs.~(\ref{edged}, \ref{edgel})]. 

\subsection{Parallel transports and curvature}\label{para}

The fundamental spin-connection fields $\{\omega_\mu(x)\}$ are assigned to
1-simplices (edges) of the simplicial complex, i.e., each edge $(x,\mu)$ we associate with it $\omega_\mu(x)$. The torsion-free Cartan Eq.~(\ref{werelation0}) is actually an equation for infinitesimal parallel transports of tetrad fields $e^a_\nu(x)$. Applying this equation to the 2-simplex $h(x)$, as shown in Fig.~{\ref{pl}},  we show that $e^a_\nu(x)$ [$e^a_\mu(x)$] undergoes its 
parallel transport to $\bar e^a_\nu(x+a_\mu)$ [$\bar e^a_\mu(x+a_\nu)$] 
along the $\mu$ ($\nu$) direction for an edge spacing $a_\mu(x)$ [$a_\nu(x)$], following 
the discretized Cartan equations
\begin{eqnarray}
\bar e^a_\nu(x+a_\mu)-e^a_\nu(x) -a_\mu g\omega_\mu^{ab}(x)\wedge e_{\nu b}(x)&=&0,
\label{wel1}\\
\bar e^a_\mu(x+a_\nu)-e^a_\mu(x) -a_\nu g\omega_\nu^{ab}(x)\wedge e_{\mu b}(x) &=&0.
\label{wel2}
\end{eqnarray}
The parallel transports $\bar e^a_\nu(x+a_\mu)$ and $\bar e^a_\mu(x+a_\nu)$ are neither independent fields, nor assigned to any edges of the simplicial complex. 
They are related 
to $e^\dagger_\nu(x)[e_\mu(x)]$ and $\omega_\mu(x)[\omega_\nu(x)]$ fields assigned to the edges $(x,-\nu)$ and $(x,\mu)$ of the 2-simplex $h(x)$ by the Cartan Eq.~(\ref{wel1},\ref{wel2}). Because of 
torsion-free, $e_\mu(x),e^\dagger_\nu(x)$ and their parallel transports $\bar e^\dagger_\mu(x+a_\nu), \bar e_\nu(x+a_\mu)$ form a {\it closed} parallelogram ${\mathcal C}_P(x)$ (Fig.~\ref{pl}). Otherwise this would means the curved space-time could not be approximated locally by a flat space-time \cite{hrbook}. Note that the point $(x+a_\mu+a_\nu)$ at the {\it closed} parallelogram ${\mathcal C}_P(x)$ (Fig.~\ref{pl}) is not any vertex of the simplicial complex. 
\comment{
In addition, we consider the following parallel transports
\begin{eqnarray}
\bar e^a_\nu(x+a_\nu)-e^a_\nu(x) -a_\nu g\omega_\nu^{ab}(x)\wedge e_{\nu b}(x)&=&0,
\label{wel1a}\\
\bar e^a_\mu(x+a_\mu)-e^a_\mu(x) -a_\mu g\omega_\mu^{ab}(x)\wedge e_{\mu b}(x) &=&0.
\label{wel2a}
\end{eqnarray}
}
\comment{
Thus, for each 2-simplex, there is a closed parallelogram, whose two edges lying in the  2-simplex and other two edges of parallel transports not lying in any 2-simplex. }

For the zero curvature case $R^{ab}_{\nu\mu}(x)=0$, the curvature Eq.~(\ref{rcurvature}) can be discretized as,
\begin{eqnarray}
\bar\omega^{ab}_\nu(x+a_\mu) 
- \omega^{ab}_\nu(x) - a_\mu g\omega^{ae}_\mu(x)\wedge \omega^b_{e\nu}(x) &=&0,
\label{ww1p}\\
\bar\omega^{ab}_\mu(x+a_\nu) - \omega^{ab}_\mu(x) - a_\nu g\omega^{ae}_\nu(x)\wedge \omega^b_{e\mu}(x) &=& 0, 
\label{ww2p}
\end{eqnarray}
where $\bar\omega^{ab}_\nu(x+a_\mu)$ and  $\bar\omega^{ab}_\mu(x+a_\nu)$ are 
respectively parallel transports of $\omega^{ab}_\nu(x)$ and  $\omega^{ab}_\mu(x)$ in the $\mu$- and $\nu$-directions.  
Analogously to the parallel transports $\bar e^a_\nu(x+a_\mu)$ and $\bar e^a_\mu(x+a_\nu)$ given by Eqs.~(\ref{wel1}) and (\ref{wel2}), parallel transports $\bar\omega^{ab}_\nu(x+a_\mu)$ and  $\bar\omega^{ab}_\mu(x+a_\nu)$ are neither independent fields, nor assigned to any edge of the simplicial complex. 
They are related to $\omega_\mu(x)$ and $\omega_\nu(x)$ fields assigned to  the edges $(x,\mu)$ and $(x+a_\nu,\nu)$ of the 2-simplex $h(x)$ by the parallel transport Eqs.~ (\ref{ww1p}) and (\ref{ww2p}). The fields $\omega_\mu(x), \omega_\nu(x)$ and their parallel transports $\bar \omega_\mu(x+a_\nu), \bar \omega_\nu(x+a_\mu)$ also form a {\it closed} parallelogram, analogously to the one ${\mathcal C}_P(x)$ formed by the tetrad fields $e_\mu(x),e_\nu(x)$ and their parallel transports $\bar e_\mu(x+a_\nu), \bar e_\nu(x+a_\mu)$ (see Fig.~\ref{pl}).

Whereas, for the nonzero curvature case $R^{ab}_{\nu\mu}(x)\not=0$, the curvature Eq.~(\ref{rcurvature}) can be discretized as
\begin{eqnarray}
\omega^{ab}_\nu(x\!+\!a_\mu) 
\!-\! \omega^{ab}_\nu(x) \!-\! a_\mu g\omega^{ae}_\mu(x)\wedge \omega^b_{e\nu}(x)\!\! &=& \!\! a_\mu R^{ab}_{\mu\nu}(x),\label{ww1}\\
\omega^{ab}_\mu(x\!+\!a_\nu) \!-\! \omega^{ab}_\mu(x) \!-\! a_\nu g\omega^{ae}_\nu(x)\wedge \omega^b_{e\mu}(x) \!\! &=&\!\! a_\nu R^{ab}_{\nu\mu}(x), 
\label{ww2}
\end{eqnarray}
which define fields $\omega^{ab}_\nu(x+a_\mu)$ and $\omega^{ab}_\mu(x+a_\nu)$ in terms of fields $\omega^{ab}_\nu(x)$, $\omega^{ab}_\mu(x)$ and curvature $R^{ab}_{\nu\mu}(x)$.
These fields $\omega^{ab}_\nu(x\!+\!a_\mu)$ and $\omega^{ab}_\mu(x\!+\!a_\nu)$ are neither independent fields, nor assigned to any edge of the simplicial complex. They are related not only
to $\omega^{ab}_\mu(x)$ and $\omega^{ab}_\nu(x)$ fields assigned to the edges $(x,\mu)$ and $(x+a_\nu,\nu)$ of the 2-simplex $h(x)$, but also to the curvature $R_{\mu\nu}^{ab}$ (\ref{ww1}) and $R_{\nu\mu}^{ab}$ (\ref{ww2}).

These fields $\omega^{ab}_\nu(x+a_\mu)$ and $\omega^{ab}_\mu(x+a_\nu)$ are no longer parallel transports $\bar\omega^{ab}_\nu(x+a_\mu)$ and  $\bar\omega^{ab}_\mu(x+a_\nu)$ defined by 
Eqs.~(\ref{ww1p}) and(\ref{ww2p}). The difference 
between $\omega^{ab}_\nu(x+a_\mu)$ and $\bar\omega^{ab}_\nu(x+a_\mu)$ [or between $\omega^{ab}_\mu(x+a_\nu)$ and $\bar\omega^{ab}_\mu(x+a_\nu)$] is the curvature $a_\mu R^{ab}_{\mu\nu}(x)$ [$a_\nu R^{ab}_{\nu\mu}(x)$],
\begin{eqnarray}
\omega^{ab}_\nu(x\!+\!a_\mu)-\bar \omega^{ab}_\nu(x\!+\!a_\mu) 
&=& \!\! a_\mu R^{ab}_{\mu\nu}(x),\label{dww1}\\
\omega^{ab}_\mu(x\!+\!a_\nu) - \bar \omega^{ab}_\mu(x\!+\!a_\nu) 
&=&\!\! a_\nu R^{ab}_{\nu\mu}(x). 
\label{dww2}
\end{eqnarray}
The fields $\omega_\mu(x), \omega_\nu(x)$ and fields $\omega_\mu(x+a_\nu), \omega_\nu(x+a_\mu)$ do not form a {\it closed} parallelogram, due to the nonzero curvature $R^{ab}_{\nu\mu}(x)\not=0$.

\subsection{Group-valued fields}\label{gvf}

Instead of a $\omega_\mu(x)$ field, we assign a group-valued field $U_\mu(x)$ to each edge (1-simplex) of the simplicial complex. On the edge $(x,\mu)$ connecting two vertexes $x$ and $x+a_\mu$ in the forward direction $\mu$, we place an 
$SO(4)$ group-valued spin-connection fields,
\begin{eqnarray}
U_\mu(x)= e^{iga\omega_\mu(x)},
\label{0link0}
\end{eqnarray}
whereas the same edge $(x+a_\mu,-\mu)$ in the backward direction $-\mu$, we associate with it 
\begin{eqnarray}
U_{-\mu}(x+a_\mu) \equiv U_\mu^\dagger(x) = U_\mu^{-1}(x),
\label{1link0}
\end{eqnarray}
analogously to the definition of link fields in lattice gauge theories.
On the three edges in forward directions $(x,\mu)$, $(x+a_\mu,\rho)$ and $(x+a_\nu,\nu)$ of the anti-clocklike 2-simplex $h(x)$ ($\mu\not=\nu\not=\rho$ see Fig.~\ref{pl}), 
we define $SO(4)$ group-valued spin-connection fields,
\begin{eqnarray}
U_\mu(x)&=& e^{iga\omega_\mu(x)},\label{link0}\\ 
U_\rho(x+a_\mu) &=& e^{iga\omega_\rho(x+a_\mu)},\label{link0_0}\\ 
U_\nu(x+a_\nu) &=& e^{iga\omega_\nu(x+a_\nu)},
\label{link000}
\end{eqnarray}
which 
take values of the fundamental representation of the compact group $SO(4)$. 
On the three edges in backward directions $(x,-\nu)$, $(x+a_\nu,-\rho)$ and $(x+a_\mu,-\mu)$ of the clocklike 2-simplex $h^\dagger(x)$ (see Fig.~\ref{pl}), 
we define $SO(4)$ group-valued spin-connection fields,
\begin{eqnarray}
U_{-\nu}(x)&=& U^\dagger_{\nu}(x+a_\nu)= e^{-iga\omega_\nu(x+a_\nu)},\label{orinu1}\\ 
U_{-\rho}(x+a_\nu) &=& U^\dagger_{\rho}(x+a_\mu)= e^{-iga\omega_\rho(x+a_\mu)},\label{orirho1}\\
U_{-\mu}(x+a_\mu)&=& U_\mu^\dagger(x)=e^{-iga\omega_\mu(x)}.
\label{orimu1}
\end{eqnarray}
These uniquely define group-valued spin-connection fields on the anti-clocklike and clocklike 2-simplex. 
\comment{
Analogously to the three tetrad $e$-fields Eqs.~(\ref{orimu}-\ref{orirho}), these $U$-fields are  in directions $\mu,\nu$ and $\rho$, we have 
\begin{eqnarray}
U_\mu(x)&=& U^\dagger_\mu(x+a_\mu)=U_{-\mu}(x+a_\mu),
\label{orimu1}
\\
U_\nu(x+a_\nu)&=& U^\dagger_\nu(x)=U_{-\nu}(x),
\label{orinu1}\\
U_\rho(x+a_\mu)&=& U^\dagger_\rho(x+a_\nu)=U_{-\rho}(x+a_\nu).
\label{orirho1}
\end{eqnarray}
}

\subsubsection{Unitary operators for parallel transports of $e_\mu(x)$ fields}

Actually, these group-valued fields (\ref{link0}-\ref{link000}) and (\ref{orinu1}-\ref{orimu1}) can be viewed as unitary operators for finite parallel transportations. The parallel transportation 
(Cartan) Eqs.~(\ref{wel1}) and (\ref{wel2}) can be 
generalized to $(\mu\not=\nu)$
\begin{eqnarray}
\bar e_\nu(x+a_\mu)&=& U^\dagger_\mu(x)e_\nu(x)U_\mu(x),
\label{up1}\\
\bar  e_\mu(x+a_\nu)&=& U^\dagger_\nu(x)e_\mu(x)U_\nu(x),
\label{up2}
\end{eqnarray}
and using Eq.~(\ref{1link0}) these equations can be equivalently rewritten as
\begin{eqnarray}
e_\nu(x)&=& U^\dagger_{-\mu}(x+a_\mu)\bar e_\nu(x+a_\mu)U_{-\mu}(x+a_\mu),
\label{up1'}\\
e_\mu(x)&=& U^\dagger_{-\nu}(x+a_\nu)\bar e_\mu(x+a_\nu)U_{-\nu}(x+a_\nu).
\label{up2'}
\end{eqnarray}
While for $(\mu=\nu)$, we similarly have the following parallel transportation 
equations
\begin{eqnarray}
\bar e_\mu(x+a_\mu)&=& U^\dagger_\mu(x)e_\mu(x)U_\mu(x),
\nonumber\\
e_\mu(x)&=& U^\dagger_{-\mu}(x+a_\mu)\bar e_\mu(x+a_\mu)U_{-\mu}(x+a_\mu),
\label{up2''}
\end{eqnarray}
indicating that $e_\mu(x)$ is parallel transported to $\bar e_\mu(x+a_\mu)$ in the $\mu$ forward direction, and $\bar e_\mu(x+a_\mu)$ is parallel transported to $e_\mu(x)$ in the $-\mu$ backward direction. Similar discussions can be made for parallel transports with the unitary operator $U_\rho(x+a_\mu)$. 

\subsubsection{Unitary operators for parallel transports of $e^\dagger_\mu(x)$ fields}

In the simplicial complex, each edge (1-simplex) connecting two vertexes has only one direction. One can identify each edge by its starting vertex and direction pointing to its ending vertex. 
On the basis of the  tetrad field $e_\mu(x)$ (\ref{edged})
defined at the vertex ``$x$'' for the edge ($x,\mu$) starting from the vertex ``$x$'' in the forward direction ($\mu$) to the vertex ``$x+a_\mu$,'' below, using the unitary operator $U_\mu(x)$ for parallel transports, 
we will uniquely introduce the ``conjugated'' field $e^\dagger_\mu(x)$ defined at the vertex ``$x$'' to describe the same edge ($x+a_\mu,-\mu$) but in the backward direction $-\mu$ starting from the vertex ``$x+a_\mu$'' to the vertex ``$x$.'' 
Analogously to Eq.~(\ref{edged}), this edge starting from the vertex ``$x+a_\mu$'' in the backward direction ($-\mu$) can be formally represented by
\begin{eqnarray}
l_{-\mu}(x+a_\mu)\equiv ae_{-\mu}(x+a_\mu).
\label{lre} 
\end{eqnarray}
By the parallel transport, we define the field $e_{-\mu}(x+a_\mu)$ as
\begin{eqnarray}
e_{-\mu}(x+a_\mu)&\equiv & U^\dagger_{\mu}(x) e^\dagger_\mu(x)U_{\mu}(x)= e^\dagger_{\mu}(x+a_\mu),
\label{up2'''}
\end{eqnarray}
in terms of the unitary operator $U_\mu(x)$ and conjugated tetrad fields $e^\dagger_\mu(x)$ defined at the vertex ``$x$''. From the definition in Eq.~(\ref{up2'''}), we rewrite 
\begin{eqnarray}
e^\dagger_\mu(x) &\equiv & U_\mu(x)e_{-\mu}(x+a_\mu)U^\dagger_\mu(x) =\bar e_{-\mu}(x).
\label{up1'''}
\end{eqnarray}
The second equalities in Eq.~(\ref{up2'''}) and Eq.~(\ref{up1'''}) are given by the definition of parallel transports by unitary operators [see Eq.~(\ref{up1})].
Equation (\ref{up2'''}) means that we can associate the conjugated field 
\begin{eqnarray}
e^\dagger_\mu(x) &= & U_{\mu}(x) e^\dagger_{\mu}(x+a_\mu)U^\dagger_{\mu}(x),
\label{up2''''}
\end{eqnarray}
with the same edge $(x+a_\mu,-\mu)$ but in backward direction $-\mu$ and write 
\begin{eqnarray}
l^{\dagger}_\mu(x) \equiv a e^\dagger_\mu(x). 
\label{invl}
\end{eqnarray}
As a result, the edge ($x,\mu$) [($x+a_\mu,-\mu$)] in the forward (backward) direction is uniquely described by the field $e_\mu(x)$ [$e^\dagger_\mu(x)$] defined at the vertex $x$. Note that the conjugated field $e^\dagger_\mu(x)$ is given by the parallel transport (\ref{up2''''}) from $x+a_\mu$ to $x$ in the direction $(-\mu)$. In addition,
Eqs.~(\ref{up2'''}) and (\ref{up1'''}) indicate that conjugated fields mean the inverse of field's direction $(\mu\rightarrow -\mu)$. 

This prescription shows that the edge ($x,\mu$) is completely described by the fields $e_\mu(x)$ and $e^\dagger_\mu(x)$, latter is a function of fields $e_\mu(x)$ and $U_\mu(x)$, as required by the principle of local gauge symmetries and the gauge field $U_\mu(x)$ corresponds a parallel transport between $x$ and $x+a_\mu$. In consequence, any edge (1-simplex) of the simplicial complex is uniquely identified by its location and direction ($z,\sigma$), and described by the fields $e_\sigma(z)$ and $U_\sigma(z)$.  
\comment{ 
\begin{eqnarray}
l_{-\mu}(x+a_\mu)\equiv l^{\dagger}_{\mu}(x),\quad e_{-\mu}(x+a_\mu)\equiv e^{\dagger}_{\mu}(x).
\label{lre1} 
\end{eqnarray}
}

Using the properties 
$(\gamma_a)^{\dagger}=-\gamma_a$ [see Eq.~(\ref{gbasic})] and the definition of tetrad field 
$e_\mu(x)=e^{\,\,\, a}_\mu(x)\gamma_a$, where the index $\mu$ is fixed, 
we have 
\begin{eqnarray}
e^{\dagger}_{\mu}(x)&=& [e_\mu^{\,\,\,a}(x)\gamma_a]^{\dagger} =(\gamma_a)^\dagger[e^{\,\,\,a}_\mu(x)]^\dagger,\nonumber\\
&=&-e_\mu(x),
\label{einv} 
\end{eqnarray}
\comment{
its inverse  
\begin{eqnarray}
e^{-1}_{\mu}(x)&=& [e_\mu^{\,\,\,a}(x)\gamma_a]^{-1}=(\gamma_a)^{-1}[e_\mu^{\,\,\,a}(x)]^{-1}\nonumber\\
&=&(\gamma_a)^\dagger e^\mu_{\,\,\,a}(x) =(\gamma_a)^\dagger[e^{\,\,\,a}_\mu(x)]^\dagger,\nonumber\\
&=&e_\mu^\dagger(x)=-e_\mu(x),
\label{einv} 
\end{eqnarray}
}
where because of the index $\mu$ being fixed, the real tetrad-field component 
$e_\mu^{\,\,\,a}(x)\equiv \partial\xi^a/\partial x^\mu $ 
can be viewed as a one-row matrix 
$(e_\mu^{\,\,\,0},e_\mu^{\,\,\,1},e_\mu^{\,\,\,2},e_\mu^{\,\,\,3})$ and $[e^{\,\,\,a}_\mu(x)]^\dagger$ a one-column matrix 
$ (e_\mu^{\,\,\,0},e_\mu^{\,\,\,1},e_\mu^{\,\,\,2},e_\mu^{\,\,\,3})^\dagger$.
\comment{ 
so that we define $[e_\mu^{\,\,\,a}(x)]^\dagger \equiv e^\mu_{\,\,\,a}(x)$ in Eq.~(\ref{einv}). One can show 
\begin{eqnarray}
e^{-1}_{\mu}(x)e_{\mu}(x) = e^\dagger_{\mu}(x)e_{\mu}(x) =1,\quad \mu -{\rm fixed} 
\label{erel0}
\end{eqnarray}
by using $e_\mu^{\,\,\,a}(x) e^\mu_{\,\,\,b}(x)=\partial \xi^a/\partial\xi^b=\delta ^a_{\,\, b} $, $e_\mu^{\,\,\,b}(x) e^\mu_{\,\,\,a}(x)=\partial \xi^b/\partial\xi^a=\delta ^b_{\,\, a}$ and Eq.~(\ref{gbasic}).
}
Analogously to Eq.~(\ref{edgel}), the length of the edge (\ref{invl}) in backward direction $-\mu$,
\begin{eqnarray}
|l^{\dagger}_\mu(x)| &= & 
\frac{a}{2}\Big\{|{\rm tr} [e^\dagger_\mu(x)\cdot e^\dagger_\mu(x)]|\Big\}^{1/2}\nonumber\\
&=&|l_\mu(x)|.
\label{edgel3}
\end{eqnarray}
which is the same as the length of the edge in the forward direction $\mu$.
\comment{
As a result, the edge in the backward direction $-\mu$ is defined as
\begin{eqnarray}
e_{-\mu}(x+a_\mu)\equiv e^\dagger_{\mu}(x).
\label{edre1}
\end{eqnarray}
We will give the gauge covariant definitions of Eq.~(\ref{lre1},\ref{edre1}) by using unitary operators for parallel transport in Sec.~\ref{gvf}.
}

We turn to the discussion of other two backward-direction edges $(x+a_\nu,-\nu)$ and $(x+a_\mu,-\rho)$ of the clocklike 2-simplex $h^\dagger(x)$ (see Fig.~\ref{pl}).
Analogously to Eqs.~(\ref{up2'''},\ref{up1'''}), we have in the $(-\nu)$ direction,
\begin{eqnarray}
e_{-\nu}(x)&\equiv & U_{\nu}(x) e^\dagger_\nu(x+a_\nu)U^\dagger_{\nu}(x)= e^\dagger_{\nu}(x),
\label{nup2'''}\\
e^\dagger_\nu(x+a_\nu)& \equiv & U^\dagger_\nu(x)e_{-\nu}(x)U_\nu(x)=\bar e_{-\nu}(x+a_\nu),\nonumber
\end{eqnarray}
and in the $(-\rho)$ direction 
\begin{eqnarray}
e_{-\rho}(x+a_\nu)&\equiv & U^\dagger_{\rho}(x+a_\mu) e^\dagger_\rho(x+a_\mu) U_{\rho}(x+a_\mu)
=e^\dagger_{\rho}(x+a_\nu).
\label{rup2'''}\\
e^\dagger_\rho(x+a_\mu)& \equiv  & U_\rho(x+a_\mu)e_{-\rho}(x+a_\nu)U^\dagger_\rho(x+a_\mu) =\bar e_{-\rho}(x+a_\mu),
\nonumber
\end{eqnarray}
As a result, the edge ($x+a_\nu,\nu$) [($x+a_\nu,-\nu$)] in the forward (backward) direction is uniquely described by the field $e_\nu(x+a_\nu)$ [$e^\dagger_\nu(x+a_\nu)$] 
defined at the vertex $x+a_\nu$ 
\begin{eqnarray}
e^\dagger_\nu(x+a_\nu)=U^\dagger_{\nu}(x) e^\dagger_{\nu}(x) U_{\nu}(x),
\label{nup2''''}
\end{eqnarray}
see Eq.~(\ref{nup2'''}). Note that the conjugated field $e^\dagger_\nu(x+a_\nu)$ is given by the parallel transport (\ref{nup2''''}) from $x$ to $x+a_\nu$ in the direction $(\nu)$. We can write 
\begin{eqnarray}
l^{\dagger}_\nu(x+a_\nu) \equiv a e^\dagger_\nu(x+a_\nu).
\label{invl1}
\end{eqnarray}
Similarly, the edge ($x+a_\mu,\rho$) [($x+a_\mu,-\rho$)] in the forward (backward) direction is uniquely described by the field $e_\rho(x+a_\mu)$ [$e^\dagger_\rho(x+a_\mu)$] defined at the vertex $x+a_\mu$
\begin{eqnarray}
e^\dagger_\rho(x+a_\mu) = U_{\rho}(x+a_\mu) e^\dagger_{\rho}(x+a_\nu) U^\dagger_{\rho}(x+a_\mu),
\label{rup2''''}
\end{eqnarray}
see Eq.~(\ref{rup2'''}). Note that the conjugated field $e^\dagger_\rho(x+a_\mu)$ is given by the parallel transport (\ref{rup2''''}) from $x+a_\nu$ to $x+a_\mu$ in the direction $(-\rho)$.
We can write 
\begin{eqnarray}
l^{\dagger}_\rho(x+a_\mu) \equiv ae^\dagger_{\rho}(x+a_\mu).
\label{invl2}
\end{eqnarray}
This prescription shows that the edge ($x+a_\nu,\nu$) is completely described by the fields $e_\nu(x+a_\nu)$ and $U_\nu(x+a_\nu)$, and the edge ($x+a_\mu,\rho$) by the fields $e_\rho(x+a_\mu)$ and $U_\rho(x+a_\mu)$. The field $U_\nu(x+a_\nu)$ [$U_\rho(x+a_\mu)$] corresponds a parallel transport between $x$ and $x+a_\mu$ ($x+a_\mu$ and $x+a_\nu$). 

Along the triangle path of the clocklike  
2-simplex $h^\dagger(x)$ [$\,x \stackrel{-\nu}{\longmapsto} x+a_\nu \stackrel{-\rho}{\longmapsto} x+a_\mu \stackrel{-\mu}{\longmapsto} x\,$] (see Fig.~\ref{pl}), these three edges and their backward directions are formally represented by (1) $l_{-\mu}(x+a_\mu)$ and
$-\mu$ pointing from $x+a_\mu$ to $x$; (2) $l_{-\nu}(x)$ and $-\nu$ pointing from $x$ to $x+a_\nu$; (3) $l_{-\rho}(x+a_\nu)$ and $-\rho$ pointing from $x+a_\nu$ to $x+a_\mu$. 
Based on Eqs.~(\ref{up2'''}), (\ref{nup2'''}), (\ref{rup2'''}), (\ref{up2''''}), (\ref{nup2''''}) and (\ref{rup2''''}),
we use the conjugated tetrad fields
\begin{eqnarray}
e^\dagger_\mu(x),\quad
e^\dagger_\nu(x+a_\nu),\quad
e^\dagger_\rho(x+a_\mu),
\label{ol3e}
\end{eqnarray}
which are respectively defined at vertexes $x$, $x+a_\nu$, $x+a_\mu$,
\comment{
\begin{eqnarray}
e_{-\mu}(x+a_\mu)&\equiv & 
e^\dagger_\mu(x),
\label{orimu}\\
e_{-\nu}(x)&\equiv & 
e^\dagger_\nu(x+a_\nu),
\label{orinu}\\
e_{-\rho}(x+a_\nu) &\equiv & 
e^\dagger_\rho(x+a_\mu),
\label{orirho}
\end{eqnarray}
whose corresponding gauge covariant definitions by using unitary operators for parallel transports will be given in Sec.~\ref{gvf}.  
Thus we use the tetrad fields 
\begin{eqnarray}
e^\dagger_\mu(x+a_\mu)&=&U^\dagger_{\mu}(x) e^\dagger_\mu(x)U_{\mu}(x),\label{orimu}\\
e^\dagger_\nu(x)&=&U_{\nu}(x) e^\dagger_\nu(x+a_\nu)U^\dagger_{\nu}(x), \label{orinu}\\
e^\dagger_\rho(x+a_\nu)&=& U^\dagger_{\rho}(x+a_\mu) e^\dagger_\rho(x+a_\mu) U_{\rho}(x+a_\mu),
\label{orirho}
\end{eqnarray}
}
to characterize both backward directions and lengths of three edges (\ref{invl}), (\ref{invl1}), and (\ref{invl2}) of the clocklike  
2-simplex $h^\dagger(x)$. 

In the simplicial complex, each edge (1-simplex), described by tetrad field $e_\mu(x)$, is uniquely identified by its location and direction ($x,\mu$), and each triangle (2-simplex) $h(x)$ has a definite orientation, as indicated in Fig.~\ref{pl}, either anti-clocklike or clocklike. Thus each triangle, for example, the one presented in Fig.~\ref{pl} is completely described by the tetrad fields $e_\mu(x)$, $e_\nu(x+a_\nu)$, $e_\rho(x+a_\mu)$, and unitary operators $U_\mu(x)$, $U_\nu(x+a_\nu)$, $U_\rho(x+a_\mu)$. 

\comment{The three tetrad fields (\ref{ol3e}) are related to the three tetrad fields (\ref{aol3e}) by relationships (\ref{einv}) and (\ref{edgel3}).
In Sec.~\ref{gaugetran}, we will see that the conjugated tetrad fields $e^\dagger_\mu(x)$, $e^\dagger_{\mu}(x+a_\mu)$, $e^\dagger_{\nu}(x)$, $e^\dagger_\nu(x+a_\nu)$, $e^\dagger_\rho(x+a_\mu)$, and $e^\dagger_{\rho}(x+a_\nu)$ are gauge covariant, and properly transformed under local gauge transformations.
}

\subsubsection{Unitary operators and curvature}

In the zero curvature case, the group-valued fields for parallel transports
$\bar\omega_\mu(x+a_\nu)$ and $\bar \omega_\nu(x+a_\mu)$, defined by parallel transport Eqs.~(\ref{ww1p}) and (\ref{ww2p}), are given by
\begin{eqnarray}
\bar U_\mu(x+a_\nu) = e^{iga\bar\omega_\mu(x+a_\nu)},\quad \bar U_\nu(x+a_\mu) = e^{iga\bar\omega_\nu(x+a_\mu)}.
\label{plink00}
\end{eqnarray} 
Similarly to Eqs.~(\ref{up1},\ref{up2}), the parallel transport Eqs.~(\ref{ww1p}) and (\ref{ww2p}) can be generalized to 
\begin{eqnarray}
\bar U_\nu(x+a_\mu)& = &U^\dagger_\mu(x)U_\nu(x)U_\mu(x),\label{pup11}\\
\bar  U_\mu(x+a_\nu)& = &U^\dagger_\nu(x)U_\mu(x)U_\nu(x).\label{pnup11}
\end{eqnarray}
The parallel transport fields $\bar  U_\nu(x+a_\mu)$ and $\bar  U_\mu(x+a_\nu)$ together with $U_\mu(x)$ and $U_\nu(x)$ form a {\it closed} parallelogram, see Fig.~\ref{pl}. This closed parallelogram is not the same as 
the parallelogram ${\mathcal C}_P(x)$ formed by $e$ and $\bar e$ fields.

In the nonzero curvature case, corresponding to the fields $\omega_\mu(x+a_\nu)$ and $\omega_\nu(x+a_\mu)$ defined by Eqs.~(\ref{ww1}) and (\ref{ww2}), the group-valued fields can be similarly given by
\begin{eqnarray}
U_\mu(x+a_\nu) = e^{iga\omega_\mu(x+a_\nu)},\quad U_\nu(x+a_\mu) = e^{iga\omega_\nu(x+a_\mu)},
\label{link00}
\end{eqnarray}
whose values obviously depend on the curvature $R_{\mu\nu}(x)$.
The same as the fields $\omega_\mu(x+a_\nu)$ and $\omega_\nu(x+a_\mu)$, these group-valued fields $U_\nu(x+a_\mu)$ and $ U_\mu(x+a_\nu)$ are neither independent fields, nor assigned to any edge of the simplicial complex. They are related 
to $U_\mu(x)$ and $U_\nu(x)$ fields assigned to the edges $(x,\mu)$ and $(x,\nu)$ of the 2-simplex $h(x)$ 
by 
\begin{eqnarray}
U_\nu(x+a_\mu)& \equiv &U^\dagger_\mu(x)U_\nu(x)U_\mu(x),\label{up11}\\
U_\mu(x+a_\nu)& \equiv &U^\dagger_\nu(x)U_\mu(x)U_\nu(x),\label{nup11}
\end{eqnarray}
which are generalized from Eqs.~(\ref{ww1}) and (\ref{ww2}). 
The fields $U_\nu(x+a_\mu)$ and $U_\mu(x+a_\nu)$ defined in Eqs.~(\ref{up11},\ref{nup11}) encode the information of a nontrivial curvature. They
do not form a {\it closed} parallelogram together with $U_\mu(x)$ and $U_\nu(x)$, at the point $(x+a_\mu + a_\nu)$ (see Fig.~\ref{pl}). 

\comment{
Corresponding to the curvature equation (\ref{ww1}) for the field $\omega_\nu(x+a_\mu)$, we define
the group-valued field
\begin{eqnarray}
U_\nu(x+a_\mu)&\equiv &U_\mu(x)U_\nu(x)U^\dagger_\mu(x),\label{up11}
\end{eqnarray}
Analogously, corresponding to the curvature equation (\ref{ww2}) for the field $\omega_\mu(x+a_\nu)$, we define the group-valued field
\begin{eqnarray}
 U_\mu(x+a_\nu)&\equiv &U_\nu(x)U_\mu(x)U^\dagger_\nu(x).\label{nup11}
\end{eqnarray}
For the sake of simplicity, we introduce notations
\begin{eqnarray}
U_{\mu\nu}(x)&\equiv &U_\mu(x) U_\nu(x) = U_\nu(x+a_\mu)U_\mu(x),
\label{p00}\\
U_{\nu\mu}(x)&\equiv &U_\nu(x) U_\mu(x) = U_\mu(x+a_\nu)U_\nu(x),
\label{p001}
\end{eqnarray}
which, as will be shown bellow, characterize the ``deficit angle'': $\omega_\mu(x+a_\nu)-\bar \omega_\mu(x+a_\nu)$, see Eqs.~(\ref{ww1}-\ref{ww2p}).
}


In order to see the nontrivial curvature information encoded in the fields $U_\nu(x+a_\mu)$ and $U_\mu(x+a_\nu)$ defined by Eqs.~(\ref{link00})-(\ref{nup11}), based on Eqs.~(\ref{up11}) and (\ref{nup11}), we introduce quantities
\begin{eqnarray}
U_{\mu\nu}(x)&\equiv & U_\nu(x) U_\mu(x) = U_\mu(x)U_\nu(x+a_\mu),
\label{p00}\\
U_{\nu\mu}(x)&\equiv & U_\mu(x) U_\nu(x) = U_\nu(x)U_\mu(x+a_\nu),
\label{p001}
\end{eqnarray} 
and calculate their expressions in the naive continuum limit.
In the {\it naive continuum limit}: 
$ag\omega_\mu \ll 1$ (small coupling $g$ or weak $\omega_\mu$ field), indicating that the wavelengths of 
weak and slow-varying fields $\omega_\mu(x)$  
are much larger than the edge spacing $a_{\mu}$, 
we obtain (see Appendix \ref{con})
\begin{eqnarray}
U_{\mu\nu}(x)
&=&\exp\Big\{iga[\omega_\mu(x)+\omega_\nu(x)]
+iga^2\partial_\mu \omega_\nu(x)\nonumber\\
&-&\frac{1}{2}(ga)^2\left[\omega_\mu(x),\omega_\nu(x)\right]+{\mathcal O}(a^3)\Big\},
\label{uu0}
\end{eqnarray}
where ${\mathcal O}(a^3)$ indicates 
high-order powers of $ag\omega_\mu$. It is shown that the quantity $U_{\mu\nu}(x)$
[Eq.~(\ref{uu0})] is related to the curvature $R_{\mu\nu}(x)$ in Appendix \ref{con}.
For the sake of simplicity in the following calculations to show the naive continuum limit, the quantities introduced by (\ref{p00}) and (\ref{p001}) and their expressions in the naive continuum limit (\ref{uu0})
are useful.

\subsection{Triangle constrain and area}\label{2-simplex-area}

Three tetrad fields $e_\mu(x)$, $e_\rho(x+a_\mu)$ and $e_\nu(x+a_\nu)$ [see Eq.~(\ref{aol3e})] are three edges of the anti-clocklike  
2-simplex $h(x)$, satisfying the triangle 
constraint
\begin{eqnarray}
e_\rho(x+a_\mu)= e_{-\nu}(x)-e_\mu(x) = e^\dagger_\nu(x)-e_\mu(x).
\label{threec}
\end{eqnarray}
Equivalently, three tetrad fields $e^\dagger_\mu(x)$, $e^\dagger_\nu(x+a_\nu)$ and $e^\dagger_\rho(x+a_\mu)$ [see Eqs.~(\ref{up2''''}), (\ref{nup2''''}), and (\ref{rup2''''}) or (\ref{ol3e})] of the clocklike  
2-simplex $h^\dagger(x)$, satisfying the triangle constraint
\begin{eqnarray}
e_{-\rho}(x+a_\nu)= e_{\mu}(x)-e_{-\nu}(x) = e_\mu(x) - e^\dagger_\nu(x),
\label{threec1}
\end{eqnarray}
where $e_{-\rho}(x+a_\nu)=e^\dagger_{\rho}(x+a_\nu)$ [see Eq.~(\ref{rup2'''})]. Also, Eq.~(\ref{nup2'''}) is used for $e_{-\nu}(x)=e^\dagger_{\nu}(x)$ in the second equality of Eqs.~(\ref{threec},\ref{threec1}).
Two of three edges are independent for a given anti-clocklike (clocklike) 
2-simplex $h(x)$ [$h^\dagger(x)$]. 

However, in Eqs.~(\ref{threec},\ref{threec1}), vector fields defined at different vertexes are related without being parallel  transported to the same vertex, thus these relationships are not proper and does not
properly transform under local gauge transformations. This is an exactly essential point of local gauge symmetries, that gauge fields $U$ for parallel transports are needed to relate variations of gauge freedom at different coordinate points.
Using the parallel transport by the unitary operator $U_\mu(x)$, we rewrite the triangle constraint (\ref{threec}) 
for the anti-clocklike 2-simplex $h(x)$ as 
\begin{eqnarray}
U_\mu(x)e_\rho(x+a_\mu)U^\dagger_\mu(x)
= e^\dagger_{\nu}(x) -e_\mu(x),
\label{gthreec}
\end{eqnarray}
where in the left-handed side, $e_\rho(x+a_\mu)$ is parallel transported from the vertex $x+a_\mu$ to the vertex $x$ to be related to $e^\dagger_{\nu}(x)$ and $e_\mu(x)$ at the same vertex $x$ in the right-handed side. Using $\bar e_\rho(x)=U_\mu(x)e_\rho(x+a_\mu)U^\dagger_\mu(x)$, we rewrite Eq.~(\ref{gthreec}) as
\begin{eqnarray}
e_{\nu}(x) + e_\mu(x)+\bar e_\rho(x)=0.
\label{gthreec'}
\end{eqnarray}
Using the parallel transport by the unitary operator $U_\nu(x)$, we rewrite the triangle constraint (\ref{threec1}) 
for the clocklike  
2-simplex $h^\dagger(x)$ as  
\begin{eqnarray}
U_\nu(x)e^\dagger_{\rho}(x+a_\nu)U^\dagger_\nu(x)
= e_\mu(x) - e^\dagger_{\nu}(x).
\label{gthreec1}
\end{eqnarray}
where in the left-handed side $e^\dagger_\rho(x+a_\nu)$ is 
parallel transported from the vertex $x+a_\nu$ to the vertex $x$ to be related to $e^\dagger_{\nu}(x)$ and $e_\mu(x)$ at the same vertex $x$ in the right-handed side. Equation (\ref{gthreec1}) is identical to Eq.~(\ref{gthreec}) or Eq.~(\ref{gthreec'}), if we consider $\bar e^\dagger_{\rho}(x)=U_\nu(x)e^\dagger_{\rho}(x+a_\nu)U^\dagger_\nu(x)$ and $\bar e^\dagger_{\rho}(x)= -\bar e_{\rho}(x)$. The proper parallel transports by unitary operators can shift the triangle constrain to other vertexes 
for example, $x+a_\mu$ and $x+a_\nu$.  

We are now in the position of discussing the area of 
the 2-simplex $h(x)$.
We define the fundamental area operator of the anti-clocklike 2-simplex $h(x)$ (see Fig.~\ref{pl}) 
\begin{eqnarray}
S^{\rm h}_{\mu\nu}(x)&\equiv & \,\,
a^2 \,\, e_\mu(x)\wedge e_{-\nu}(x)\label{pareaod}
\end{eqnarray}
at the vertex $x$. In addition,
we can also define the following area operators
\begin{eqnarray}
S^{\rm h}_{\rho\mu}(x+a_\mu)&\equiv &
a^2 \,\, e_\rho(x+a_\mu)\wedge e_{-\mu}(x+a_\mu)
\label{areaod1}
\end{eqnarray}
at the vertex $x+a_\mu$, and
\begin{eqnarray}
S^{\rm h}_{\nu\rho}(x+a_\nu)&\equiv & 
a^2 \,\, e_\nu(x+a_\nu)\wedge e_{-\rho}(x+a_\nu)
\label{areaod2}
\end{eqnarray}
at the vertex $x+a_\nu$. 
Using Eqs.~(\ref{up2'''}), (\ref{nup2'''}), and (\ref{rup2'''}), we rewrite the area operators (\ref{pareaod})-
(\ref{areaod2}) of the anti-clocklike 2-simplex $h(x)$ as
\begin{eqnarray}
S^{\rm h}_{\mu\nu}(x)&\equiv & \,\,
a^2 \,\, e_\mu(x)\wedge e^\dagger_{\nu}(x),\label{gpareaod}\\
S^{\rm h}_{\rho\mu}(x+a_\mu)&\equiv &
a^2 \,\, e_\rho(x+a_\mu)\wedge e^\dagger_{\mu}(x+a_\mu),
\label{gareaod1}\\
S^{\rm h}_{\nu\rho}(x+a_\nu)&\equiv & 
a^2 \,\, e_\nu(x+a_\nu)\wedge e^\dagger_{\rho}(x+a_\nu).
\label{gareaod2}
\end{eqnarray}
In the following, we show that area operators (\ref{gpareaod})-
(\ref{gareaod2}) defined at three vertexes $x$, $x+a_\mu$, and $x+a_\nu$ are universal up to parallel transports by unitary operators.
Using Eqs.~(\ref{up2'''}) and (\ref{gthreec}), we obtain
\begin{eqnarray}
S^{\rm h}_{\rho\mu}(x+a_\mu)&=&
a^2 \,\, U^\dagger_\mu(x)[e^\dagger_{\nu}(x) -e_\mu(x)]U_\mu(x)\nonumber\\
&\wedge& U^\dagger_\mu(x)e^\dagger_\mu(x)U_\mu(x),\nonumber\\
&=&a^2 \,\, U^\dagger_\mu(x)[e^\dagger_{\nu}(x)
\wedge e^\dagger_\mu(x)]U_\mu(x),\nonumber\\
&=& U^\dagger_\mu(x)S^{\rm h}_{\mu\nu}(x)U_\mu(x).
\label{gareaod1s}
\end{eqnarray}
Analogously, using Eqs.~(\ref{nup2'''}) and (\ref{gthreec1}), we obtain
\begin{eqnarray}
S^{\rm h}_{\nu\rho}(x+a_\nu)
&=&a^2 \,\, U^\dagger_\nu(x)e_\nu(x)U_\nu(x)\nonumber\\
&\wedge & U^\dagger_\nu(x)[e_\mu(x)-e^\dagger_{\nu}(x)]U_\nu(x)\nonumber\\
&=&a^2 \,\, U^\dagger_\nu(x)e_\nu(x)\wedge e_\mu(x)U_\nu(x)\nonumber\\
&=&  U^\dagger_\nu(x) S^{\rm h}_{\mu\nu}(x) U_\nu(x). 
\label{gareaod2s}
\end{eqnarray}
In Eqs.~(\ref{gareaod1s}) and (\ref{gareaod2s}), we use $e^\dagger_\mu(x)=-e_\mu(x)$, $e_\mu(x)\wedge e_\mu(x)=e_\mu^\dagger(x)\wedge e^\dagger_\mu(x)=e_\mu^\dagger(x)\wedge e_\mu(x)=0$ and the same for $(\mu\rightarrow\nu)$.
This shows that the area operators (\ref{gpareaod}), (\ref{gareaod1}) and (\ref{gareaod2}) defined at three vertexes of the 2-simplex $h(x)$ are universal up to parallel transports. 

Therefore, Eq.~(\ref{pareaod}) or (\ref{gpareaod}) defines the area operator of the 2-simplex $h(x)$
\begin{eqnarray}
S^{\rm h}_{\mu\nu}(x)
&\equiv& \frac{a^2}{2} \,\,\left[ e_\mu(x) e^\dagger_{\nu}(x)-e^\dagger_{\nu}(x) e_\mu(x)\right]\nonumber\\
&=& a^2\frac{i}{2} \,\,\sigma_{ab}\left[ e^a_\mu(x) e^b_{\nu}(x)-e^a_{\nu}(x) e^b_\mu(x)\right],
\label{areaod}
\end{eqnarray}
up to parallel transports. As consequence, the area of the 2-simplex $h(x)$ is uniquely determined by 
\begin{eqnarray}
S_{\rm h}(x)\equiv |S^{\rm h}_{\mu\nu}(x)|,\quad
S^2_{\rm h}(x)\equiv
\frac{1}{8}\,{\rm tr}\left[S^{\rm h}_{\mu\nu}(x)\cdot S^{{\rm h}\dagger}_{\mu\nu}(x)\right]\,.\label{aread}
\end{eqnarray}
Its uniqueness [independence of the vertexes $x$, $x+a_\mu$ and $x+a_\nu$ 
of the 2-simplex $h(x)$], i.e.,
\begin{eqnarray}
S_{\rm h}(x)&\equiv& |S^{\rm h}_{\mu\nu}(x)|=|S^{\rm h}_{\rho\mu}(x+a_\mu)|
=|S^{\rm h}_{\nu\rho}(x+a_\nu)|\,,\label{aread=}
\end{eqnarray}
can be shown by using Eqs.~(\ref{gareaod1s},\ref{gareaod2s}).

In the same way as Eqs.~(\ref{pareaod})-(\ref{areaod2}), we define the area operators of the clocklike 2-simplex $h^\dagger(x)$: 
\begin{eqnarray}
S^{\rm h}_{\nu\mu}(x)&\equiv &a^2 \,\, e_{-\nu}(x)\wedge e_\mu(x)\label{areaod0}\\
& =& -S^{\rm h}_{\mu\nu}(x)=S^{{\rm h}\dagger}_{\mu\nu}(x),\nonumber\\
S^{\rm h}_{\mu\rho}(x+a_\mu)&\equiv & a^2 \,\, e_{-\mu}(x+a_\mu) \wedge  e_\rho(x+a_\mu) \nonumber\\
&=&-S^{\rm h}_{\rho\mu}(x+a_\mu)=S^{{\rm h}\dagger}_{\rho\mu}(x+a_\mu),\nonumber\\
S^{\rm h}_{\rho\nu}(x+a_\nu)&\equiv & a^2 \,\, e_{-\rho}(x+a_\nu)\wedge  e_\nu(x+a_\nu)\nonumber\\
&=&-S^{\rm h}_{\nu\rho}(x+a_\nu)= S^{{\rm h}\dagger}_{\nu\rho}(x+a_\nu),\nonumber
\end{eqnarray}   
whose directions are opposite to the counterparts of  
anti-clocklike 2-simplex $h(x)$. However,
the area of the clocklike 2-simplex $h^\dagger(x)$ is equal to the area (\ref{aread}). 

Based on the definition of 2-simplex $h(x)$ area (\ref{aread}), we can define a 
volume element around the vertex ``$x$''
\begin{equation}
dV(x)=\sum_{h(x)}dV_h(x),\quad dV_h(x)\equiv S_{\rm h}^2(x)
\label{vold}
\end{equation}
where $dV_h(x)$ indicates the volume element contributed from a 2-simplex $h(x)$, and $\sum_{h(x)}$ indicates the sum over all 2-simplices $h(x)$ that share the same vertex $x$. This definition of volume element (\ref{vold})  indicates that a 2-simplex $h(x)$ contributes the volume element $S_{\rm h}^2$ at its three vertexes $x$, $x+a_\mu$ and $x+a_\nu$. 

Before ending this section, we note that
using the parallel transports (\ref{up2'''}), (\ref{nup2'''}), and (\ref{rup2'''}), 
one can obtain parallel transports  of area operators (\ref{pareaod},\ref{areaod1},\ref{areaod2}) 
of triangles (2-simplexes),
\begin{eqnarray}
\bar S_{\mu\nu}(x+a_\mu)&=& U^\dagger_\mu(x)
S^{\rm h}_{\mu\nu}(x)U_\mu(x),\nonumber\\
\bar S_{\mu\nu}(x+a_\nu)&=& U^\dagger_\nu(x)S^{\rm h}_{\mu\nu}(x)U_\nu(x),
\label{apt}\\&\cdot\cdot\cdot& ,\nonumber
\end{eqnarray}
which are consistent with the definitions of unitary operators $U_\mu(x)$ and $U_\nu(x)$ for parallel transports (\ref{up1},\ref{up2}) of edges (1-simplexes).
The notation ``$\bar S_{\mu\nu}$'' instead of $S^{\rm h}_{\mu\nu}$ in the left-handed side 
of Eqs.~(\ref{apt}) indicates that the parallel transport ``$\bar S_{\mu\nu}$'' is not associated to any triangle of the simplicial complex.

\subsection{Local gauge transformations}\label{gaugetran}

In accordance with Eq.~(\ref{gtran0}), the bilocal gauge transformations of three $U$ fields (\ref{link0})-
(\ref{link000}) of the anti-clocklike 2-simplex $h(x)$ 
are,
\begin{eqnarray}
U_\mu(x)&\rightarrow &{\mathcal V}(x)U_\mu(x){\mathcal V}^\dagger(x+a_\mu),
\nonumber\\
U_\nu(x+a_\nu)&\rightarrow &{\mathcal V}(x+a_\nu)U_\nu(x+a_\nu){\mathcal V}^\dagger(x),
\nonumber\\
U_\rho(x+a_\mu)&\rightarrow &{\mathcal V}(x+a_\mu)U_\rho(x+a_\mu){\mathcal V}^\dagger(x+a_\nu),
\label{gtranu}
\end{eqnarray}
and their inverses
(\ref{orinu1})-
(\ref{orimu1}) of the clocklike 2-simplex $h^\dagger(x)$ transform as
\begin{eqnarray}
U^\dagger_\mu(x)&\rightarrow &{\mathcal V}(x+a_\mu)U^\dagger_\mu(x){\mathcal V}^\dagger(x),
\nonumber\\
U^\dagger_\nu(x+a_\nu)&\rightarrow &{\mathcal V}(x)U^\dagger_\nu(x+a_\nu){\mathcal V}^\dagger(x+a_\nu),
\nonumber\\
U^\dagger_\rho(x+a_\mu)&\rightarrow &{\mathcal V}(x+a_\nu)U^\dagger_\rho(x+a_\mu){\mathcal V}^\dagger(x+a_\mu).
\label{agtranu}
\end{eqnarray}

In accordance with Eq.~(\ref{varie}), the tetrad fields $e_\mu(x), e_\nu(x+a_\nu)$ and $e_\rho(x+a_\mu)$ 
for the anti-clocklike 2-simplex $h(x)$ transform under local gauge transformations 
\begin{eqnarray}
e_\mu(x)&\rightarrow & e'_\mu(x)= {\mathcal V}(x)e_\mu(x){\mathcal V}^\dagger(x),\nonumber\\
e_\nu(x+a_\nu)&\rightarrow & e'_\nu(x+a_\nu)= {\mathcal V}(x+a_\nu)e_\nu(x+a_\nu){\mathcal V}^\dagger(x+a_\nu),
\nonumber\\
e_\rho(x+a_\mu)&\rightarrow& e'_\rho(x+a_\mu)= {\mathcal V}(x+a_\mu)e_\rho(x+a_\mu){\mathcal V}^\dagger(x+a_\mu),
\label{varie1}
\end{eqnarray}
respectively at the vertexes $x$, $x+a_\nu$ and $x+a_\mu$ where they are defined. Using above local gauge transformations (\ref{gtranu})-
(\ref{varie1}), we obtain the following 
local gauge transformations of the conjugated fields  $e^\dagger_\mu(x)$, $e^\dagger_\nu(x+a_\nu)$ and $e^\dagger_\rho(x+a_\mu)$ defined by Eqs.~(\ref{up2'''}), (\ref{nup2'''}), and (\ref{rup2'''}) 
for the clocklike 2-simplex $h^\dagger(x)$,
\begin{eqnarray}
e^\dagger_\mu(x)&\rightarrow & e^{\dagger\, '}_\mu(x)= {\mathcal V}(x)e^\dagger_\mu(x){\mathcal V}^\dagger(x),\nonumber\\
e^\dagger_\nu(x+a_\nu)&\rightarrow & e^{\dagger\, '}_\nu(x+a_\nu)= {\mathcal V}(x+a_\nu)e^\dagger_\nu(x+a_\nu){\mathcal V}^\dagger(x+a_\nu),
\nonumber\\
e^\dagger_\rho(x+a_\mu)&\rightarrow& e^{\dagger\, '}_\rho(x+a_\mu)= {\mathcal V}(x+a_\mu)e^\dagger_\rho(x+a_\mu){\mathcal V}^\dagger(x+a_\mu).
\label{bvarie1}
\end{eqnarray}
These local gauge transformations (\ref{bvarie1}) of the conjugated fields at the vertexes $x$, $x+a_\nu$ and $x+a_\mu$   
are in the same manner as that given by Eqs.~(\ref{varie1}). This means that each edge (1-simplex) $l_\mu(x)$ of the simplicial complex is uniquely described by tetrad fields $e_\mu(x)$ and $e^\dagger_\mu(x)$, that are defined at the vertex $x$, and covariantly transformed under local gauge transformation. 

It is worthwhile to mention that the transformations (\ref{bvarie1}) are just conjugated transformations (\ref{varie1}), and consistent with 
\comment{
(i) the definitions
(\ref{up2'''},\ref{nup2'''},\ref{rup2'''}) of $e^\dagger_\mu(x)$, $e^\dagger_\nu(x+a_\nu)$ and $e^\dagger_\rho(x+a_\mu)$, (ii) the bilocal gauge transformation (\ref{gtranu},\ref{agtranu})
for $U$-fields and (iii)
} 
the following local gauge transformations: 
\begin{eqnarray}
e_{-\mu}(x+a_\mu)&\rightarrow & e_{-\mu}'(x+a_\mu)= {\mathcal V}(x+a_\mu)e_{-\mu}(x+a_\mu){\mathcal V}^\dagger(x+a_\mu),\nonumber\\
e_{-\nu}(x)&\rightarrow & e_{-\nu}'(x)= {\mathcal V}(x)e_{-\nu}(x){\mathcal V}^\dagger(x),
\nonumber\\
e_{-\rho}(x+a_\nu)&\rightarrow& e_{-\rho}'(x+a_\nu)= {\mathcal V}(x+a_\nu)e_{-\rho}(x+a_\nu){\mathcal V}^\dagger(x+a_\nu),
\label{bvarie2}
\end{eqnarray}
which follow the transformation rules of Eq.~(\ref{varie1}).

It is shown that the tetrad fields (\ref{aol3e}) and their conjugated fields (\ref{ol3e}) given by Eqs.~(\ref{up2''''}), (\ref{nup2''''}), and (\ref{rup2''''}), as well as the triangle constraints (\ref{gthreec},\ref{gthreec1}), are gauge covariant, and properly transformed
under local gauge transformations (\ref{gtranu})-(\ref{bvarie1}).
The length (\ref{edgel}) or (\ref{edgel3}) of edges (1-simplexes) is unique and invariant under local gauge transformations (\ref{gtranu})-(\ref{bvarie1}). 

Under local gauge transformations (\ref{gtranu})-(\ref{bvarie1}), the fundamental area operators (\ref{gpareaod})-
(\ref{gareaod2}) of the anti-clocklike 2-simplex $h(x)$ are gauge covariant and transform 
\begin{eqnarray}
S^{\rm h}_{\mu\nu}(x)&\rightarrow & S^{\rm h'}_{\mu\nu}(x)= {\mathcal V}(x)S^{\rm h}_{\mu\nu}(x){\mathcal V}^\dagger(x),\nonumber\\
S^{\rm h}_{\nu\rho}(x+a_\nu)&\rightarrow & S^{\rm h'}_{\nu\rho}(x+a_\nu)= {\mathcal V}(x+a_\nu)S^{\rm h}_{\nu\rho}(x+a_\nu){\mathcal V}^\dagger(x+a_\nu),
\nonumber\\
S^{\rm h}_{\rho\mu}(x+a_\mu)&\rightarrow& S^{\rm h'}_{\rho\mu}(x+a_\mu)= {\mathcal V}(x+a_\mu)S^{\rm h}_{\rho\mu}(x+a_\mu){\mathcal V}^\dagger(x+a_\mu),
\label{varis1}
\end{eqnarray}
which are consistent with Eqs.~(\ref{gareaod1s}), (\ref{gareaod2s}), (\ref{gtranu}), and (\ref{agtranu}),
and their counterparts [see
Eq.~(\ref{areaod0})] of the clocklike 2-simplex $h^\dagger(x)$ transform in the same manner. The parallel transports (\ref{apt}) of area operators transform consistently with Eqs.~(\ref{gtranu}), (\ref{agtranu}) and (\ref{varis1}). However, the area (\ref{aread}) of the 2-simplex $h(x)$ is unique and invariant under local gauge transformations. 

It is worthwhile to mention that
under local gauge transformation (\ref{gtranu})-
(\ref{varie1}), parallel transport fields (\ref{up1}) and (\ref{up2}) transform locally 
\begin{eqnarray}
\bar e_\mu(x+a_\nu)&\rightarrow & \bar e'_\mu(x+a_\nu)={\mathcal V}(x+a_\nu)\bar e_\mu(x+a_\nu) {\mathcal V}^\dagger(x+a_\nu),\nonumber\\
\bar e_\nu(x+a_\mu)&\rightarrow & \bar e'_\nu(x+a_\mu)={\mathcal V}(x+a_\mu)\bar e_\nu(x+a_\mu) {\mathcal V}^\dagger(x+a_\mu),
\label{varie2}
\end{eqnarray}
in accordance with local gauge transformations (\ref{varie1}) for tetrad fields.
Therefore, the {\it closed} parallelogram ${\mathcal C}_P(x)$ (see Fig.~\ref{pl}), formed by $e_\mu(x),e_\nu(x)$ and their parallel transports $\bar e_\mu(x+a_\nu), \bar e_\nu(x+a_\mu)$, is invariant under local gauge transformation. This is consistent with 
the torsion-free condition for the existence of local Lorentz frames at each points of a curved space-time. 

The prescription of using tetrad fields $e_\sigma(z)$ and gauge fields $U_\sigma(z)$ for parallel transports to describe edges (1-simplexes) and triangles (2-simplexes) of
the simplicial complex fully respects the principle of local gauge symmetries. Therefore, this prescription is independent of a particular vertex $z$, oriented edge $l_\sigma(z)$ and triangle $h(z)$, because of the gauge invariance. The formulation of defining tetrad fields $e_\sigma(z)$ at one of edge endpoints ``$z$'' and direction ``$\sigma$,'' and each triangle has a definite orientation is gauge invariant.  

However, the gauge transformation properties of fields $U_\nu(x+a_\mu)$ and $U_\mu(x+a_\nu)$ defined by Eqs.~(\ref{up11}) and (\ref{nup11}), as well as $U_{\mu\nu}(x)$ and $U_{\nu\mu}(x)$ introduced by Eqs.~(\ref{p00}) and (\ref{p001}),  are very complicate under the bilocal gauge transformations (\ref{gtranu}) and (\ref{agtranu}). This implies that we could not use these fields 
to construct a gauge-invariant object. We need to study the object of three $U$ fields, $U_\mu(x)$, $U_\rho(x+a_\mu)$ and $U_\nu(x+a_\nu)$ along a closed triangle path of each 2-simplex $h(x)$ (see Fig.~\ref{pl}), which will be discussed in the next section.

\subsection{Regularized EC action}\label{naivec}

\begin{figure}[ptb]
\includegraphics[scale=1.2 
]{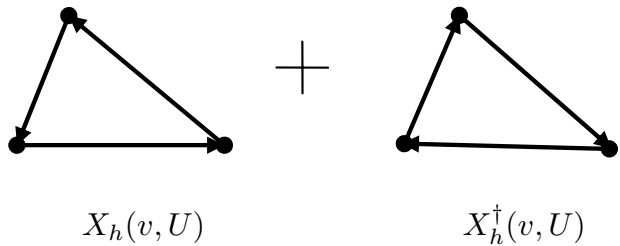}
\comment{
\put(-215,10){$X_h(v,U)$}
\put(-95,10){$X^\dagger_h(v,U)$}
}
\caption{The smallest holonomy field along a closed triangle path of the 2-simplex $h(x)$: the anti-clocklike orientation $X_h(v,U)$ [left]; the clocklike orientation $X^\dagger_h(v,U)$ [right]. 
}%
\label{xh}%
\end{figure}

To illustrate how to construct a gauge-invariantly regularized EC theory describing dynamical configurations of the simplicial complex, we consider anti-clocklike 2-simplex (triangle) $h(x)$ and clocklike 2-simplex (triangle) $h^\dagger(x)$ (see Fig.~\ref{pl} and Fig.~\ref{xh}).

For simplifying notations, we henceforth do not explicitly write negative signs $-\mu,-\nu,-\rho$ to indicate the backward directions of edges. In terms of the tetrad fields $e_\mu(x)$ and $e_\nu(x)$ of the 2-simplex $h(x)$ (see Fig.~\ref{pl}), we introduce the following vertex fields $v_{\mu\nu}(x)$:
\begin{eqnarray}
v_{\mu\nu}(x)&\equiv &\gamma_5e_{\mu\nu}(x),\label{v2}\\
e_{\mu\nu}(x)&\equiv & \sigma_{ab}\left[e^a(x)\wedge e^b(x)\right]_{\mu\nu}\nonumber\\
&\equiv&\frac{1}{2}\sigma_{ab}\left[e^a_\mu(x) e^b_\nu(x)
-e^a_\nu(x) e^b_\mu(x)\right]\nonumber\\
&=&\frac{i}{2}\left[e_\mu(x) e_\nu(x)-e_\nu(x) e_\mu(x)\right],
\label{dirace}
\end{eqnarray}
which have properties: $v_{\mu\nu}(x)=-v_{\nu\mu}(x)$, ${\rm tr}[v_{\mu\nu}(x)]=0$ and $v^\dagger_{\mu\nu} (x)= v_{\nu\mu} (x)$ (see Appendix \ref{xcon}). Under the local gauge transformation (\ref{varie},\ref{varie1}), the vertex fields (\ref{v2}) and (\ref{dirace}) transform locally at a vertex $x$, 
\begin{eqnarray}
v_{\mu\nu}(x)&\rightarrow &{\mathcal V}(x)v_{\mu\nu}(x){\mathcal V}^\dagger(x),
\label{gtranl}
\end{eqnarray}
which is transformed in the same manner as area operators (\ref{varis1}). 
In addition to the vertex field $e_{\mu\nu}(x)$ (\ref{dirace}) at the vertex $(x)$, we can define in the same way the vertex fields $e_{\rho\mu}(x+a_\mu)$ at the vertex $(x+a_\mu)$, and $e_{\nu\rho}(x+a_\nu)$ at the vertex $(x+a_\nu)$ of the anti-clocklike 2-simplex $h(x)$ (see Fig.~\ref{pl}). Actually, the vertex fields $e_{\mu\nu}(x)$ (\ref{dirace}), $e_{\rho\mu}(x+a_\mu)$ and $e_{\nu\rho}(x+a_\nu)$ are
related to the fundamental area operators $S_{\mu\nu}^{\rm h}(x)$  (\ref{gpareaod}), $S_{\rho\mu}^{\rm h}(x+a_\mu)$ (\ref{gareaod1}) 
and $S_{\nu\rho}^{\rm h}(x+a_\nu)$ (\ref{gareaod2}), e.g.,
\begin{equation}
S_{\mu\nu}^{\rm h}(x)=i a^2 e_{\mu\nu}(x).
\label{seeq}
\end{equation}
As discussions for three area operators in Eqs.~(\ref{pareaod})-(\ref{areaod}), 
only one of three vertex-fields $e_{\mu\nu}(x)$, $e_{\rho\mu}(x+a_\mu)$ and $e_{\nu\rho}(x+a_\nu)$ is independent for the anti-clocklike 2-simplex $h(x)$. As for an clocklike 2-simplex $h^\dagger(x)$, vertex fields can be obtained by using the relations $e^\dagger_{\mu\nu} (x)= e_{\nu\mu} (x)$ and $e_{\mu\nu}(x)=-e_{\nu\mu}(x)$. 

Using the tetrad fields $e_\mu(x)$ and vertex fields $v_{\mu\nu}(x)$ to construct coordinate and Lorentz scalars 
to preserve the diffeomorphism 
and {\it local} gauge invariance, we define 
a smallest holonomy field along the closed triangle path of the 2-simplex $h(x)$ (see Fig.~\ref{pl}):
\begin{eqnarray}
X_{h} (v,U)&=& 
{\rm tr}\left[v_{\nu\mu}(x)U_{\mu}(x)v_{\mu\rho}(x+a_\mu)U_{\rho}(x+a_\mu)
v_{\rho\nu}(x+a_\nu)U_{\nu}(x+a_\nu)\right],
\label{xs}
\end{eqnarray}
whose orientation is anti-clocklike, as shown the left graphic in Fig.~\ref{xh}.
Considering the clocklike orientation, as shown the right graphic in Fig.~\ref{xh}, 
we have 
\begin{eqnarray}
X^{\rm clocklike}_{h} (v,U)&=& 
{\rm tr}\left[v_{\mu\nu}(x)U_{\nu}(x)v_{\nu\rho}(x+a_\nu)U_{\rho}(x+a_\nu)
v_{\rho\mu}(x+a_\mu)U_{\mu}(x+a_\mu)\right]\nonumber\\
&=&X_{h} (v,U)|_{\mu\leftrightarrow\nu}.
\label{xsc}
\end{eqnarray}
On the other hand,
\begin{eqnarray}
X^\dagger_{h} (v,U)&=& 
{\rm tr}\left[U^\dagger_{\nu}(x+a_\nu)v^\dagger_{\rho\nu}(x+a_\nu)U^\dagger_{\rho}(x+a_\mu)v^\dagger_{\mu\rho}(x+a_\mu)U^\dagger_{\mu}(x)
v^\dagger_{\nu\mu}(x)\right]\nonumber\\
&=&{\rm tr}\left[U_{\nu}(x)v_{\nu\rho}(x+a_\nu)U_{\rho}(x+a_\nu)v_{\rho\mu}(x+a_\mu)
U_{\mu}(x+a_\mu)v_{\mu\nu}(x)\right] \nonumber\\
&=&{\rm tr}\left[v_{\mu\nu}(x)U_{\nu}(x)v_{\nu\rho}(x+a_\nu)U_{\rho}(x+a_\nu)v_{\rho\mu}(x+a_\mu)
U_{\mu}(x+a_\mu)\right] \nonumber\\
&=& X^{\rm clocklike}_{h} (v,U)
\label{xsh1}
\end{eqnarray}
where in the second line of equation, we use the properties $U^\dagger_{\nu}(x+a_\nu)=U_{\nu}(x)$, $U^\dagger_{\rho}(x+a_\mu)=
U_{\rho}(x+a_\nu)$, $U^\dagger_{\mu}(x)=U_{\mu}(x+a_\mu)$ and $v^\dagger_{\mu\nu} (x)= v_{\nu\mu} (x)$. Therefore we have 
\begin{eqnarray}
X_{h} (v,U)+{\rm h.c.} = X_{h} (v,U)+ X^{\rm clocklike}_{h} (v,U).
\label{xsc+}
\end{eqnarray}
Equations (\ref{xsc})-
(\ref{xsc+}) are invariant under gauge transformations  (\ref{gtranu}), (\ref{agtranu}), and (\ref{gtranl}).

Using Eqs.~(\ref{xs})-(\ref{xsc+}), we are ready to construct the diffeomorphism 
and {\it local} gauge-invariant regularized EC action. First we consider the case
$v_{\mu\nu}(x)  =  e_{\mu\nu}(x)\gamma_5$:   
\begin{eqnarray}
{\mathcal A}_P(e,U)&=&\frac{1}{8g^2}\sum_{h\in {\mathcal M}}
\left\{X_{h} (v,U)+{\rm h.c.}\right\},
\label{pact}
\end{eqnarray}
where 
$\sum_{h\in {\mathcal M}}$ is the sum over all 2-simplices $h$ of the simplicial complex. 
In the naive continuum limit: $ag\omega_\mu \ll 1$, Eq.~(\ref{pact}) becomes (see Appendix \ref{xcon})
\begin{eqnarray}
{\mathcal A}_P(e,U_\mu)
&= &\frac{1}{a^2} \sum_{h\in {\mathcal M}} S_{\rm h}^2(x) \epsilon_{cdab}\, e^c\wedge e^d \wedge R^{ab}+{\mathcal O}(a^4),
\label{pa2}
\end{eqnarray}
where the 2-simplex $h(x)$ contributed volume element $S_{\rm h}^2(x)$ is given in Eq.~(\ref{aread}) or Eq.~(\ref{xsv1}).
\comment{
the fundamental area  of the 2-simplex (triangle) $h(x)$ (see Fig.~\ref{pl}), 
is expressed in terms of the  vertex-field $v_{\mu\nu}(x)$ (\ref{v2},\ref{dirace}),
\begin{eqnarray}
S^2_{\rm h}(x)\equiv a^4{\rm tr} [e_{\mu\nu}(x)e_{\mu\nu}(x)], 
\quad  |S_{\rm h}(x)|= a^2 \sqrt{{\rm tr} [e_{\mu\nu}(x)e_{\mu\nu}(x)]},
\label{ao}
\end{eqnarray} 
consistently with Eq.~(\ref{aread}).  
} 
Based the volume element $dV(x)$ (\ref{vold}) around the vertex ``$x$''  
\begin{eqnarray}
\sum_{h\in {\mathcal M}} S_{\rm h}^2(x)=\frac{1}{3}\sum_x dV(x)
\label{vo}
\end{eqnarray}
where $\sum_x$ stands for a sum overall vertexes (0-simplices) of the simplicial complex, and the factor $1/3$ is due to each 2-simplex contributing its area to its three vertexes. 
The interior of the 4-simplex is approximately flat, leading to
\begin{eqnarray}
\sum_{x}dV(x) \Rightarrow \int d^4\xi(x)
=\int d^4x{\rm det}[e(x)].
\label{vint}
\end{eqnarray}
As a result, Eq.~(\ref{pa2}) approaches to $S_P(e,\omega)$ (\ref{host}) with an effective Newton constant 
\begin{eqnarray}
G_{\rm eff}=\frac{3}{4}\,g\, G,
\label{effg}
\end{eqnarray}
and $\kappa_{\rm eff}\equiv 8\pi G_{\rm eff}$.
The second we consider the case $v_{\mu\nu}(x)  =  e_{\mu\nu}(x)$:
\begin{eqnarray}
{\mathcal A}_H(e,U_\mu)&=&\frac{1}{8g^2\gamma}\sum_{h\in {\mathcal M}}
\left[X_h(v,U)+{\rm h.c.}\right],\label{hact}
\label{diract}
\end{eqnarray}
where the real parameter $\gamma=i\tilde\gamma$ [see Eq.~(\ref{host1})].
Analogously, in the naive continuum limit: $ag\omega_\mu \ll 1$, Eq.~(\ref{hact}) approaches to $S_H(e,\omega)$ (\ref{host1}) [see Appendix \ref{xcon}],
\begin{eqnarray}
{\mathcal A}_H(e,U_\mu)
&= & \frac{1}{2\kappa_{\rm eff} \tilde\gamma}\int d^4x{\rm det}[e(x)] 
 e_a\wedge e_b \wedge R^{ab}+{\mathcal O}(a^4),
\label{hpa2}
\end{eqnarray}
with the effective Newton constant $\kappa_{\rm eff}\equiv 8\pi G_{\rm eff}$ (\ref{effg}).
The diffeomorphism 
and {\it local} gauge-invariant regularized EC action is then given by
\begin{eqnarray}
{\mathcal A}_{EC}={\mathcal A}_{P}+{\mathcal A}_{H}.
\label{ecp}
\end{eqnarray}

In addition,
we can generalize the link field $U_\mu(x)$
to be all irreducible representations $j$ of the gauge group $SO(4)$. 
The regularized EC action (\ref{ecp}) should be a sum over all irreducible representations $j$, 
\begin{equation}
{\mathcal A}_{EC}=\sum_j\frac{4}{d_j}\left[
{\mathcal A}^j_P(e_\mu,U_\mu)+{\mathcal A}^j_H(e_\mu,U_\mu)
\right],
\label{allj}
\end{equation}
where $d_j$ is the dimensions of the irreducible representations $j$ and $d_j=4$ for the fundamental representation, which is the dimension of the Dirac spinor space.

\subsection{Invariant holonomy fields along a large loop}\label{lloop}

We consider the following diffeomorphism and {\it local} gauge-invariant holonomy fields along a loop ${\mathcal C}$ on the Euclidean manifold ${\mathcal R}^4$ 
\begin{eqnarray}
X_{\mathcal C}(v,\omega)&=&
{\mathcal P}_C{\rm tr}\exp\left\{ ig\oint_{\mathcal C}v_{\mu\nu}(x)
\omega^\mu(x) dx^\nu\right\},
\label{pa0s}
\end{eqnarray} 
where ${\mathcal P}_C$ is the path-ordering and ``${\rm tr}$'' denotes 
the trace over spinor space. We attempt to regularize these holonomy fields
(\ref{pa0s}) on the simplicial complex ${\mathcal M}$. 
Suppose that an orientating loop ${\mathcal C}$ 
passes space-time points (vertexes) $x_1,x_2,x_3,\cdot\cdot\cdot, x_N=x_1$ and edges connecting between neighboring points in the simplicial complex ${\mathcal M}$ (see the diagram in the left-hand side of graphic equation, Fig.~\ref{sdf1}). 
At each point $x_i$
two tetrad fields $e_\mu(x_i)$ and $e_{\mu'}(x_i)$ $(\mu\not=\mu')$ respectively 
orientating path incoming to ($i-1\rightarrow i$) and outgoing from 
($i\rightarrow i+1$) the point $x_i$, we have the 
vertex field $v_{\mu\mu'}(x_i)$ defined by Eqs.~(\ref{v2}) and (\ref{dirace}).  
Link fields $U_\mu(x_i)$ are defined on edges lying in the loop ${\mathcal C}$. Recalling the relationship 
$U_{-\mu}(x_{i+1})=U^\dagger_{\mu}(x_{i})$ [see Eqs.~(\ref{orinu1}-\ref{orimu1})], 
we can write the regularization of the holonomy fields (\ref{pa0s}) as
\begin{eqnarray}
X_{\mathcal C}(v,U)&\!\!=\!\!&{\mathcal P}_C{\rm tr} \Big[v_{\mu\mu'}(x_1)U_{\mu'}(x_1)v_{\mu'\nu}(x_2)U_{\nu}(x_2)\nonumber\\
&\cdot\cdot\cdot &
v_{\rho\rho'}(x_i)U_{\rho'}(x_i)v_{\rho'\sigma}(x_{i+1})\nonumber\\
&\cdot\cdot\cdot &v_{\lambda\mu}(x_{N-1})U^\dagger_{\mu}(x_{N-1})\Big],
\label{rpa0s}
\end{eqnarray} 
which preserve diffeomorphism and {\it local} gauge invariances.  
The holonomy fields $X_{\mathcal C}(e,U)$ are functionals of fields ($v,U$) and  loop ${\mathcal C}$. Consistently with the holonomy fields $X_{\mathcal C}(e,U)$ [Eq.~(\ref{rpa0s})], the holonomy field $X_h(e,U)$ [Eq.~(\ref{xs})] is the one with the smallest loop, i.e., the closed path of the 2-simplex (triangle) $h(x)$, see Fig.~\ref{pl}.  


\subsection{\bf Euclidean partition function}\label{epf}

The partition function $Z_{EC}$ and effective action 
${\mathcal A}^{\rm eff}_{EC}$ are given by
\begin{eqnarray}
Z_{EC}=\exp-{\mathcal A}^{\rm eff}_{EC}=\int {\mathcal D}e{\mathcal D}U
\exp -{\mathcal A}_{EC},
\label{par}
\end{eqnarray}
with the diffeomorphism and {\it local} gauge-invariant measure  
\begin{eqnarray}
\int \!\!{\mathcal D}e{\mathcal D}U 
\!\equiv\!\! \prod_{l_\mu(x)\,\in \, {\mathcal M}}\,\,\,\int_{l_\mu(x)}\,\,\, de_\mu(x)dU_\mu(x)\,\, \delta(\Delta)
\label{mea1}
\end{eqnarray}
where $\prod_{l_\mu(x)\,\in \, {\mathcal M}}$ indicates the product of overall edges (1-simplices) of the four-dimensional simplicial complex ${\mathcal M}$. As already mentioned, the configuration $\{l_\mu(x)\,\in \, {\mathcal M}\}$ is formulated such that each edge $l_\mu(x)=ae_\mu(x)$ is defined by giving its coordinate (vertex) $x\,\in \, {\mathcal M}$ in one of the endpoint coordinates $x$ and $x+a_\mu$, and giving its forward direction $\mu$ pointing from $x$ to $x+a_\mu$. This endpoint coordinate $x$ and forward direction $\mu$ have to be uniquely chosen for each edge $l_\mu(x)\,\in \, {\mathcal M}$. Beside, on such defined edge $l_\mu(x)$, we place an independent gauge field $U_\mu(x)$ corresponding a parallel transport between $x$ and $x+a_\mu$. The gauge invariant properties, discussed in Sec.~ \ref{gaugetran}, guarantee that the change of a formulation does not lead to the change in the measure of the configuration $\{l_\mu(x)\,\in \, {\mathcal M}\}$. In addition, the triangle constraint (\ref{gthreec}) and (\ref{gthreec'}) must be imposed in the measure (\ref{mea1}), symbolically indicated as $\delta(\Delta)$, a $\delta$ functional of Eq.~(\ref{gthreec}) or Eq.~(\ref{gthreec'}).    

In the single edge measure [see Eq.~(\ref{mea1})]
\begin{eqnarray}
\int_{l_\mu(x)}\,\,\, de_\mu(x)dU_\mu(x),
\label{smea1}
\end{eqnarray}
$dU_\mu(x)$ is the invariant Haar measure of the compact gauge group $SO(4)$ or $SU_L(2)\otimes SU_R(2)$, and $de_\mu(x)$ is the measure of the Dirac-matrix valued field $e_\mu(x)=\sum_a e_\mu^a(x)\gamma_a$, determined by the functional measure $de_\mu^a(x)$ of the bosonic field $e_\mu^a(x)$. The single edge measure has to be the measure over fields only $e_\mu(x)$ and $U_\mu(x)$ of the edge in the forward direction $\mu$, because $e^\dagger_{\mu}(x)$ and $U^\dagger_{\mu}(x)$ of the edge in the backward direction $-\mu$ are related to the fields $e_\mu(x)$ and $U_\mu(x)$ by Eqs.~(\ref{1link0}), (\ref{up2'''}), (\ref{up2''''}), and (\ref{einv}) so that the single edge measure (\ref{smea1}) is actually over all degrees of fields assigned on the edge.

It should be mentioned that the measure (\ref{mea1}) is just a lattice form of the standard DeWitt functional measure \cite{dewitt67} over the continuum degrees, with the integral of the spin-connection field $\omega_\mu(x)$ replaced by the Haar integral over the $U_\mu(x)$'s, analytical integration or numerical simulations runs overall configuration space of continuum degrees and no gauge fixing is needed. In addition, it should be noted that the measure (\ref{mea1}) does not contain parallel transport fields $\bar e$ and $\bar U$, for examples $\bar e_\nu(x+a_\mu)$ and $\bar e_\mu(x+a_\nu)$ (see Fig.~\ref{pl}) given by the Cartan Eqs.~ (\ref{wel1}) and (\ref{wel2}), since parallel transport fields are not associated to any edges of the four-dimensional simplicial complex. This means that the torsion-free Cartan equation has been taken into account.
\comment{   
Note that the measure ${\mathcal D}U_\mu(x)$ 
includes all link fields lying in both edges ($e_\mu,e_\nu$) of 2-simplices and their parallel transports ($e_\mu,e_\nu$), as shown in Fig.~\ref{pl}.
\begin{eqnarray} 
[e^a_\mu(x),e^b_\nu(x')]=\delta_{\mu\nu}(x)\delta^{ab}\delta(x-x'),
\label{qe0}
\end{eqnarray}
and equivalently
\begin{eqnarray} 
\{e_\mu(x),e^\dagger_\nu(x')\}=\delta_{\mu\nu}(x)\delta(x-x').
\label{qe}
\end{eqnarray}
}

In this path-integral quantization formalism, the partition function (\ref{par}) presents 
all dynamical configurations of the simplicial complex, described by the configurations of dynamical fields $e_\mu(x)$ and ${U_\mu(x)}$ in the weight of $\exp -{\mathcal A}_{EC}$.
The effective action ${\mathcal A}_{EC}^{\rm eff}$ (\ref{par}) contains all one-particle irreducible (1PI) functions(operators), i.e., all truncated $n$-point Green-functions. The vacuum expectational 
values (v.e.v.) of diffeomorphism and {\it local} gauge-invariant quantities, for instance holonomy fields (\ref{rpa0s}), are given by
\begin{eqnarray}
\langle X_{\mathcal C}(v,U)\rangle=\frac{1}{Z_{EC}}\int {\mathcal D}e{\mathcal D}U
\Big[X_{\mathcal C}(v,U)\Big]\exp -{\mathcal A}_{EC}\,.
\label{eve}
\end{eqnarray}    
In the action (\ref{pact},\ref{hact}), $X_h(v,U)$ [Eq.~(\ref{xs})] contains the quadratic term of $e_\mu(x)$-field associated to each edge of 2-simplex $h(x)$,
the partition function $Z_{EC}$ (\ref{par}) and v.e.v.~(\ref{eve}) are not divergent for large fluctuating $e_\mu$ fields, provided the action ${\mathcal A}_{EC}$ is positive definite, see discussions below. On the other hand, all edge lengths do not vanish [$|e_\mu(x)|\not=0$, see Eqs.~(\ref{edged0},\ref{edged})], and all simplicial triangle inequalities and their higher dimensional analogs should be imposed \cite{wheeler1964, hammer_book}. Integrating spin-connection fields $U_\mu$ 
over the Haar measure of compact gauge groups is similar to that in the Wilson-lattice QCD, the difference is that the $X_h(v,U)$ (\ref{xs}) contains 
three $U$ fields in a 2-simplex $h$, while the Wilson action contains four $U$ fields in a plaquette.     
Equation (\ref{eve}) can be calculated by 
numerical Monte-Carlo simulations. We are trying do some numerical Monte-Carlo simulations, it will
take time so that the results will be published in a separate paper.
 
Before ending this section, we make some discussions on the convergences of the partition function (\ref{par}) and v.e.v.~ (\ref{eve}). 
Suppose that we first integrate Eqs.~(\ref{par}) and (\ref{eve}) over the compact Haar measure of the $SO(4)$ gauge group, roughly speaking, the result gives, in addition to a polynomial of tetrad fields $e$, a combination of both decreasing exponents $\exp \,[-{\mathcal A}^{(+)}(e)]$ and increasing exponents 
$\exp \,[-{\mathcal A}^{(-)}(e)]$ as functions of increasing tetrad fields $e$. 
From the regularized action (\ref{xs}), one can find that 
${\mathcal A}^{(\pm)}(e)$ depend on 2-simplex area operators $S_{\rm h}$ (\ref{aread}) and are the sum over all 2-simplexes.
${\mathcal A}^{(\pm)}(e)$ are either some extremal values of the action ${\mathcal A}_{EC}$ (\ref{ecp}) with respect to group-valued $U$ fields, or those values taken at the boundary points of the compact $SO(4)$ gauge group. 
Clearly, for the case of decreasing exponents $\exp \,[-{\mathcal A}^{(+)}(e)]$, integrations Eqs.~(\ref{par}) and (\ref{eve}) over tetrad fields $e$ are convergent. This is certainly the case for perturbative weak $U$ fields, i.e., $U\sim 1$. While  
for the case of increasing exponents $\exp\,[-{\mathcal A}^{(-)}(e)]$, integrations Eqs.~(\ref{par}) and (\ref{eve}) over tetrad fields $e$ are divergent. 
 
To avoid these possible divergences,  it is necessary to add into the regularized action ${\mathcal A}_{EC}$ (\ref{ecp}) an additional term of another dimensionality: either a curvature squared $R^2$ term: $X_h^2(v,U)+{\rm h.c.}$ with a new coupling parameter; or a bare cosmological term: ${\mathcal A}_\Lambda(e)$. We consider here an additional bare cosmological term ${\mathcal A}_\Lambda$ to the regularized action ${\mathcal A}_{EC}$ (\ref{ecp}):  ${\mathcal A}_{EC}\rightarrow {\mathcal A}_{EC}+{\mathcal A}_\Lambda$, 
\begin{eqnarray}
{\mathcal A}_\Lambda (e) &=& \frac{\lambda}{4\cdot (4!)^2}\epsilon^{\mu\nu\rho\sigma}\sum _x{\rm tr}\Big[\gamma_5e_\mu(x) e_\nu(x) e_\rho(x) e_\sigma(x)\Big]+{\rm h.c.}\nonumber\\
&=&\lambda \sum_x{\rm det} [e_\mu^a(x)] + {\rm h.c.}
\label{cosmo}
\end{eqnarray}
where the cosmological parameter $\lambda \equiv \Lambda a^2$ and $\Lambda$ is the bare cosmological constant. The bare cosmological term ${\mathcal A}_\Lambda(e)$ is a four-dimensional volume term (sum over all vertexes $x$), which is independent of configurations of group-valued $U$ fields. The exponent $\exp\,[ -{\mathcal A}_\Lambda(e)]$ decreases with strong tetrad fields $e$, large volume configurations. Bare parameters $g$, $\gamma$ and $\lambda$  play an important role for convergences of the partition function (\ref{par}) and vacuum expectational values (\ref{eve}). It needs further studies to find the region of bare parameters $g$, $\gamma$ and $\lambda$ for the convergences, and the scaling invariant region ($g_c,\,\gamma_c,\,\lambda_c$) for the physically sensible continuum limit, see the discussions in the last Sec.~ \ref{remarks}.

\subsection{\bf {\it Local} gauge symmetry}\label{local}

Analogously to Eq.~(\ref{inv}), 
the {\it local} gauge invariance of the partition function (\ref{par}), i.e., $\delta Z_{EC}=0$ under the gauge transformation (\ref{gtranu}) and (\ref{gtranl}), leads to (no summation over index $\mu$)
\begin{equation}
\langle \frac{\delta{\mathcal A}_{EC}}{\delta e_\mu}\delta e_\mu
+\delta\omega_\mu\frac{\delta {\mathcal A}_{EC}}{\delta \omega_\mu}
+{\rm h.c.}\rangle =0.
\label{inv1}
\end{equation}
Based on $\delta e_\mu$ and $\delta \omega_\mu$ (\ref{ivarie}) and (\ref{gtran1}) for an arbitrary function 
$\theta^{ab}(x)$ and the independent bases of Dirac matrices $\gamma_5$, $\gamma_\mu$ and $\sigma_{ab}$,
we obtain   
 the ``averaged'' Cartan Eq.~(\ref{werelation0}) for the torsion-free case,
\begin{equation}
\langle U_\mu\frac{\delta {\mathcal A}_{EC}}{\delta U_\mu}
- U^{\dagger }_\mu\frac{\delta {\mathcal A}_{EC}}{\delta U^{\dagger}_\mu}\rangle =0,
\label{inv2}
\end{equation}
where we use
\begin{equation}
\frac{\delta {\mathcal A}_{EC}}{\delta \omega_\mu}=iag\Big\{ U_\mu\frac{\delta {\mathcal A}_{EC}}{\delta U_\mu}
- U^{\dagger }_\mu\frac{\delta {\mathcal A}_{EC}}{\delta U^{\dagger}_\mu}\Big\},
\label{inv3}
\end{equation}
for the group-valued field $U_\mu(x)=\exp [ iga\omega_\mu(x)]$ (\ref{link0}).
The averaged torsion-free Cartan Eq.~(\ref{inv2}) actually
shows the impossibility of spontaneous breaking of the {\it local} gauge symmetry. This should not be surprised, since the torsion-free (\ref{werelation1}) is a necessary condition to 
have a {\it local} Lorentz frame, therefore a {\it local} gauge-invariance, as required by the equivalence principle.
\comment{
the ``averaged'' Einstein equation 
\begin{equation}
\langle\frac{\delta {\mathcal A}_{EC}}{\delta e_\mu}\delta e_\mu\rangle +{\rm h.c.}=0,
\label{ae0}
\end{equation}
``averaged'' Dirac equation 
\begin{equation}
\langle\frac{\delta {\mathcal A}_{EC}}{\delta \psi}\delta \psi\rangle +{\rm h.c.}=0,
\label{ad0}
\end{equation}
}

\section{Including fermion fields}\label{ifermion}

\subsection{Bilinear and quadralinear-fermion actions}

Introducing dimensionless fermion field 
$\psi'(x)\equiv a^{3/2}\psi(x)$ (drop ``prime'' henceforth) and using the relations
$\gamma^0(\gamma_a)^\dagger\gamma^0=\gamma_a$, $\gamma^0(\sigma_{ab})^\dagger\gamma^0=\sigma_{ab}$ and 
\comment{
\begin{eqnarray} 
\gamma^0(\gamma_a)^\dagger\gamma^0=\gamma_a;\quad  \gamma^0(\sigma_{ab})^\dagger\gamma^0=\sigma_{ab}
\label{drf1}
\end{eqnarray}
} 
\begin{eqnarray} 
\gamma^0e_\mu^\dagger\gamma^0=e_\mu;\quad  \gamma^0U_\mu^\dagger\gamma^0=U^\dagger_\mu,
\label{drf2}
\end{eqnarray}  
we consider the following regularized kinetic action of fermion fields, 
\begin{eqnarray}
{\mathcal A}_F(e,U,\psi)
&\!\!=\!\!&\frac{1}{2}\sum_{x,\,\mu}\Big[\bar\psi(x) 
e^\mu(x) U_\mu(x)\psi(x+a_\mu)\nonumber\\
&\! \!-\!\!&\bar\psi(x+a_\mu)  U^\dagger_\mu(x)e^\mu(x)\psi(x)\Big],
\label{plart}
\end{eqnarray}
where fermion fields $\psi(x)$ and $\psi(x+a_\mu)$ are defined at two neighboring
points (vertexes) of the edge $(x,x+a_\mu)$, (see Fig.~\ref{pl}), where fields $U_\mu(x)$ and $e_{\mu}(x)$ are added to preserve {\it local} gauge and diffeomorphism invariances, and $\sum_{x,\,\mu}$ is the sum over all edges (1-simplexes) of the  simplicial complex. 

Using Eq.~(\ref{inv3}) and performing a variation of the regularized fermion action (\ref{plart}) with respect to the spin-connection field $\omega_\mu(x)$, i.e., $\delta {\mathcal A}_F(e,U,\psi)/\delta \omega_\mu $, we obtain the nonvanishing torsion field $T^a=\kappa ge_b\wedge e_c{\mathcal J}^{ab,c}$, where the regularized fermion spin current is
\begin{eqnarray}
{\mathcal J}^{ab,c}&=&\epsilon^{abcd}\bar\psi(x)\gamma_d\gamma^5U_\mu(x)\psi(x+a_\mu),\quad \mu \hskip0.2cm{\rm fixed},
\label{spinc1}
\end{eqnarray}
[see Eq.~(\ref{spinc})]. Instead of solving regularized Cartan equation and finding an effective theory, as what is done in the continuum case (\ref{inv})-(\ref{spinc}), 
we assume that the $U_\mu(x)$ in Eqs.~(\ref{plart}) and (\ref{spinc1}) is the group-valued spin-connection field $\omega_\mu(e)$ for the torsion-free case (\ref{werelation0}), i.e., $U_\mu(x)=\exp [iag \omega_\mu(e)]$. Thus, the regularization of the effective EC theory (\ref{eca}) and (\ref{4f}) is given by Eqs.~(\ref{ecp}) and (\ref{plart}) and  the regularized four-fermion interaction
\begin{eqnarray}
{\mathcal A}_{4F}(U,\psi) &=&
3\zeta g^2\sum_{x,\,\mu}[\bar\psi(x)\gamma^d\gamma^5U^\mu(x)\psi(x+a_\mu)]
[\bar\psi(x+a_\mu)U^\dagger_\mu(x)\gamma_d\gamma^5\psi(x)],
\label{4fl1}
\end{eqnarray}
where $\zeta =\tilde\gamma^2/(\tilde\gamma^2+1) = \gamma^2/(\gamma^2+1)$ [see Eq.~(\ref{4f})].
In the naive continuum limit $ag\omega_\mu\ll 1$, the regularized fermion action ${\mathcal A}_F(e,U,\psi)$ (\ref{plart}) approaches to the continuum fermion action $S_F(e,\omega_\mu,\psi)$ (\ref{art}), 
and Eqs.~(\ref{spinc1}) and (\ref{4fl1}), respectively 
approach to their continuum counterparts $J^{ab,c}$ (\ref{spinc}) and $S_{4F}$ (\ref{4f}). 
\comment{
This bilinear fermion action (\ref{plart}) introduces a nonvanishing torsion field \cite{kleinert,s2001}. 
We need to study whether the regularized 
EC action (\ref{ecp}) with fermion fields (\ref{plart}) can be written in form 
of a torsion-free part and four fermion 
interactions, analogously to its continuum counterparts \cite{ar05}.  
In addition, the bilinear fermion action (\ref{plart}) 
has the problem of either fermion doubling or chiral (parity) 
gauge symmetry breaking, due to the No-Go theorem \cite{nn1981}. Resultant four fermion interactions can possibly be resolution to this 
problem \cite{ep1986,xue1997}.
We adopt the weak-field expansion $U_\mu(x)\approx 1+iga\omega_\mu(x)$, 
apply it to ${\mathcal A}_F$ (\ref{plart}) and 
follow the same procedure from Eqs.~(\ref{inv}) to (\ref{spinc}), 
leading to 
}
The diffeomorphism 
and {\it local} gauge-invariant regularized EC action is then given by
\begin{eqnarray}
{\mathcal A}_{EC}={\mathcal A}_{P}+{\mathcal A}_{H}+{\mathcal A}_{F}+
{\mathcal A}_{4F}.
\label{ecpf}
\end{eqnarray}
The partition function $Z_{EC}$ and effective action 
${\mathcal A}^{\rm eff}_{EC}$ are
\begin{eqnarray}
Z_{EC}=\exp-{\mathcal A}^{\rm eff}_{EC}=\int {\mathcal D}e{\mathcal D}U {\mathcal D}\psi
\exp -{\mathcal A}_{EC},
\label{parf}
\end{eqnarray}
with the diffeomorphism and {\it local} gauge-invariant measure  
\begin{eqnarray}
\int \!\!{\mathcal D}e{\mathcal D}U {\mathcal D}\psi
\!\equiv\,\, \prod_{l_\mu(x)\in \,{\mathcal M}}\,\,\int_{l_\mu(x)}\,\, de_\mu(x)dU_\mu(x)\delta(\Delta) \cdot\prod_{x\in \,{\mathcal M}}\,\,\int
d\psi(x)d\bar\psi(x),
\label{mea1f}
\end{eqnarray}
where $d\psi(x)d\bar\psi(x)$ is the measure of Grassmann anticommuting fields.
Analogously to Eq.~(\ref{allj}), Eqs.~(\ref{ecpf})-(\ref{mea1f}) can be straightforwardly generalized to include all irreducible representations $j$ of the gauge group $SO(4)$ that couple to corresponding spinor states of fermion fields. 

\subsection{Holonomy fields with fermions}

We consider the following diffeomorphism and {\it local} gauge-invariant quantities 
\begin{eqnarray}
X_{\mathcal L}(e,\omega,\psi)&=&\bar\psi(x_1){\mathcal P}
\exp\left\{ ig\int_{\mathcal L}v_{\mu\nu}(x)
\omega^\mu(x) dx^\nu\right\}\psi(x_N),
\label{fa0s}
\end{eqnarray} 
where ${\mathcal L}$ stands for an orientating (${\mathcal P}$) path connecting two vertexes $x_1$ and $x_N$ ($x_1\not=x_N$) on the simplicial complex ${\mathcal M}$. In Eq.~(\ref{fa0s}),  $X_{\mathcal L}(e,\omega,\psi)$ represents the evolution of the spin of fermion fields from the vertex $x_N$ to the vertex $x_1$ under the gravitational field influence. Analogously to discussions in Sec.~(\ref{lloop}) for the holonomy fields (\ref{pa0s}), we regularize these quantities (\ref{fa0s}) on the simplicial complex as follows:
\begin{eqnarray}
X_{\mathcal L}(e,U,\psi)&\!\!=\!\!&\bar\psi(x_1){\mathcal P} \Big[U_{\mu'}(x_1)v_{\mu'\nu}(x_2)U_{\nu}(x_2)\nonumber\\
&\cdot\cdot\cdot &
v_{\rho\rho'}(x_i)U_{\rho'}(x_i)v_{\rho'\sigma}(x_{i+1})\nonumber\\
&\cdot\cdot\cdot &v_{\lambda\mu}(x_{N})U^\dagger_{\mu}(x_{N})\Big]\psi(x_N),
\label{rfa0s}
\end{eqnarray}
which preserves diffeomorphism and {\it local} gauge invariances. The graphic representation of $X_{\mathcal L}(e,U,\psi)$ can be found in Fig.~\ref{sdf2} (see the diagram in the left-hand side of graphic equation).

\comment{
Since $e_\nu(x)$ and $e^\dagger_\mu(y)$ are treated as 
Grassmann anticommuting fields, the partition function $Z_{EC}$ (\ref{par}) is bound and we have the following formula:
\begin{eqnarray}
\int {\mathcal D}e 
e^{ -e_l\Delta^{lk}(U)e^\dagger_k}&=& \det[\Delta(U)],
\label{eint1}\\
\int {\mathcal D}e (e_i e^\dagger_j)e^{ -e_l\Delta^{lk}(U)e^\dagger_k}
&=& \Delta_{ij}(U),
\label{eint2}\\
\int {\mathcal D}e [e_i \Lambda^{ij}(U) e^\dagger_j]e^{ -e_l\Delta^{lk}(U)e^\dagger_k}
&=& \text{Tr}[\Lambda(U)\Delta(U)],
\label{eint3}
\end{eqnarray}
where $\Delta(U)$ and $\Lambda(U)$ are operators in terms of links fields $\{U_\mu(x)\}$.
}

\subsection{Chiral gauge symmetries}

\comment{
We separately consider the left and right part of 
fields: $e^\mu\rightarrow P_{L,R}e^\mu$,  
$\omega_\mu\rightarrow P_{L,R}\omega_\mu$ and 
$\psi_{L,R}\rightarrow P_{L,R}\psi$, where the chiral projector
$P_{L,R}=(1\mp\gamma_5)/2$, then $SO(4)$ is 
split into two commuting and independent groups $SU_L(2)\otimes SU_R(2)$, and the 
link field $U_\mu(x)=U^R_\mu(x)\oplus U^L_\mu(x)$, where $U^R_\mu(x)\in SU_R(2)$ and $U^L_\mu(x)\in SU_L(2)$ respectively. As consequence, 
the parquet field 
$U_p^{\mu\nu}(x)=U_{p_R}^{\mu\nu}(x)\oplus U_{p_L}^{\mu\nu}(x)$.
where $U_{p_R}^{\mu\nu}(x)$ [$U_{p_L}^{\mu\nu}(x)$] is the product (\ref{pa}) of
$U^R_\mu(x)[ U^L_\mu(x)]$ along the Planck path ${\mathcal C}_P$.
This can be proved by the commutators $[\sigma^{ab}, P_{L(R)}]=0$,
$[\gamma^a\gamma^b, P_{L(R)}]=0$ and 
the chiral representation $\sigma^{ab}$
splitting into two independent Lie algebra of sub-groups $SU_L(2)\otimes SU_R(2)$.
Analogous to the explicit chirality of the EC action (\ref{ec0}-\ref{art}) \cite{lrseparation}, the left and right parts of regularized EC action (\ref{ecp}) can be completely separated. 
}


Analogously to the discussions in the continuum EC theory (see the end of Sec.~\ref{conec}), the regularized EC action (\ref{ecpf}) can be separated into left- and right-handed parts. Fermion fields $\psi$ are decomposed into their left- and right-handed Weyl fields: $\psi=\psi_{L}+\psi_{R}$ and $\psi_{L,R}\equiv P_{L,R}\psi$,
where the chiral projector $P_{L,R}=(1\mp\gamma_5)/2$ and the commutators $[\sigma^{ab}, P_{L,R}]=0$ and $[\gamma^a\gamma^b, P_{L,R}]=0$. The $4\times 4$ Dirac spinor space is split into two independent left- and right-handed $2\times 2$ Weyl spinor spaces.  
In the chiral representation of matrices $\gamma^a$ and $\sigma^{ab}$
\begin{eqnarray}
\gamma^0&=&i\left(\matrix{0&-I\cr -I&0}\right),\quad 
\gamma^i=\left(\matrix{0&\sigma^i\cr -\sigma^i&0}\right),\quad
\gamma_5=\left(\matrix{I&0\cr 0&-I}\right),\label{chiralg}\\
\sigma^{ij}&=&\left(\matrix{\Sigma^{ij}&0\cr 0&\Sigma^{ij}}\right), \quad
\sigma^{0i}=i\left(\matrix{\sigma^i&0\cr 0&-\sigma^i}\right);
\label{pauli}
\end{eqnarray}
where $\Sigma^{ij}=\epsilon^{ij}_{~~k}\sigma^k$ and 
$\sigma^i ~~ (i=1,2,3)$ are the Pauli matrices, we define $\gamma^a_{L,R}\equiv P_{L,R}\gamma^a$:
\begin{eqnarray}
P_L\gamma^{0}&=&i\left(\matrix{0&0\cr -I&0}\right), \quad
P_L\gamma^{i}=\left(\matrix{0&0\cr -\sigma^i&0}\right),\nonumber\\
P_R\gamma^{0}&=&i\left(\matrix{0&-I\cr 0&0}\right), \quad
P_R\gamma^{i}=\left(\matrix{0&\sigma^i\cr 0&0}\right);
\label{prog}
\end{eqnarray} 
and $\sigma^{ab}_{L,R}\equiv P_{L,R}\sigma^{ab}$:
\begin{eqnarray}
P_L\sigma^{ij}&=&\left(\matrix{0&0\cr 0&\Sigma^{ij}}\right), \quad
P_L\sigma^{0i}=i\left(\matrix{0&0\cr 0&-\sigma^i}\right);\nonumber\\
P_R\sigma^{ij}&=&\left(\matrix{\Sigma^{ij}&0\cr 0&0}\right), \quad
P_R\sigma^{0i}=i\left(\matrix{\sigma^i&0\cr 0&0}\right).
\label{proj}
\end{eqnarray}
Using Eq.~(\ref{prog}), we separate tetrad fields $e^\mu$ into their left- and right-handed fields: $e^\mu=e^\mu_{L}+e^\mu_{R}$, $e^\mu_{L,R}\equiv P_{L,R}e^\mu$. Using Eq.~(\ref{proj}), we separate spin-connection fields 
$\omega^\mu$ and vertex-fields $v_{\mu\nu}$ into their left- and right-handed fields:
$\omega^\mu=\omega^\mu_{L}+\omega^\mu_{R}$, $\omega^\mu_{L,R}\equiv P_{L,R}\omega^\mu$; and  $v_{\mu\nu}=v_{\mu\nu}^{L}+v_{\mu\nu}^{R}$, $v_{\mu\nu}^{L,R}\equiv P_{L,R}v_{\mu\nu}$. This splits the Lie algebra of the group $SO(4)$ into two independent Lie algebra of sub groups $SU_L(2)\otimes SU_R(2)$. Therefore,
the four-dimensional rotational group 
$SO(4)$ is split into two commuting and independent groups $SU_L(2)\otimes SU_R(2)$. The link fields $U_\mu(x)=U^R_\mu(x)\oplus U^L_\mu(x)$, where $U^R_\mu(x)\in SU_R(2)$ and $U^L_\mu(x)\in SU_L(2)$ respectively. 

The regularized EC theory (\ref{ecpf})-(\ref{mea1f}) possesses exact chiral gauge symmetries, as consequences, the holonomy fields (\ref{xs}), (\ref{rpa0s}), and (\ref{rfa0s}) can be split into the left- and right-handed parts: 
\begin{eqnarray}
X_h(e,U) &=& X^L_h(e^L,U^L)+ X^R_h(e^R,U^R);
\label{twop0}\\
X_{\mathcal C}(e,U) &=& X^L_{\mathcal C}(e^L,U^L)+ X^R_{\mathcal C}(e^R,U^R);
\label{twop1}\\
X_{\mathcal L}(e,U,\psi) &=& X^L_{\mathcal L}(e^L,U^L,\psi_L)+ X^R_{\mathcal L}(e^R,U^R,\psi_R),
\label{twop2}
\end{eqnarray}
where notations in the right-handed side of equations, for instance, $X^L_{\mathcal L}(e^L,U^L,\psi_L)$ indicates the same function 
$X_{\mathcal L}(e,U,\psi)$ (\ref{rfa0s}) with replacements $e\rightarrow e^L$, $U\rightarrow U^L$ and $\psi\rightarrow \psi_L$. 
The fermion action (\ref{plart}) and four-fermion interaction (\ref{4fl1}) are also separated into the left- and right-handed parts:
\begin{eqnarray}
{\mathcal A}_F(e,U,\psi)&=&{\mathcal A}_F^L(e^L,U^L,\psi_L)+ {\mathcal A}_F^R(e^R,U^R,\psi_R);\label{flr}\\
{\mathcal A}_{4F}(U,\psi)&=&{\mathcal A}_{4F}^L(U^L,\psi_L)+ {\mathcal A}_{4F}^R(U^R,\psi_R).
\label{twop3}
\end{eqnarray}
\comment{
In the Planck-lattice regularized actions (\ref{pact}) and (\ref{hact}),
the left and right parts can be completely separated , as in continuum actions
(\ref{host}) and (\ref{host1}) \cite{lrseparation}. Chiral gauge symmetries and flavor symmetries of massless fermions are easily implemented.
} 
The chiral gauge symmetries of the regularized EC theory (\ref{ecpf})-(\ref{mea1f}) are crucial for formulating the parity-violating (chiral) gauge symmetries $SU_L(2)\otimes U_Y(1)$, e.g., the standard model for particle physics, onto the simplicial complex described by the dynamical tetrad fields $e_\mu(x)$ and group-valued spin-connection fields $U_\mu(x)$. We only discuss the case of Weyl fermions (massless Dirac fermions), and the discussions on the case of Majorana fermions are the same, thus not presented in this article.

\section{Dynamical equations for holonomy fields}\label{dehf}

\comment{
The {\it local} gauge-invariance of (\ref{eve}) ($\delta \langle X\rangle =0 $) leads to dynamical equations for holonomies (\ref{rpa0s}), which can be formally written as
\begin{equation}
\langle \frac{\delta X}{\delta e_\mu}\delta e_\mu +X\frac{\delta{\mathcal A}_{EC}}{\delta e_\mu}\delta e_\mu
+X+XU_\mu\frac{\delta {\mathcal A}_{EC}}{\delta U_\mu}
+{\rm h.c.}\rangle =0,
\label{sinv1}
\end{equation}
leading to $\langle \delta X/\delta e_\mu+X{\delta \mathcal A}_{EC}/\delta e_\mu\rangle +{\rm h.c.}=0$, and
\begin{equation}
\langle X\rangle + \langle X\Big( 
U_\mu\frac{\delta {\mathcal A}_{EC}}{\delta U_\mu}
- U^{\dagger }_\mu\frac{\delta {\mathcal A}_{EC}}{\delta U^{\dagger}_\mu}\Big)\rangle =0.
\label{sinv2}
\end{equation}
Eq.~(\ref{sinv2}) has the same form as the Schwinger-Dyson equation for Wilson loops in lattice gauge theories.
}

\begin{figure}[ptb]
\includegraphics[scale=1.2
]{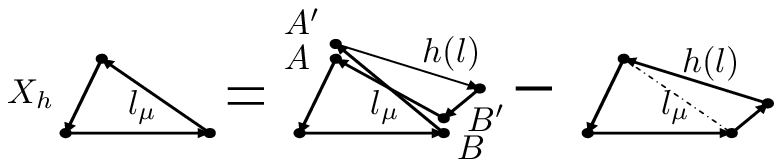}
\comment{
\put(-245,33){$X_h$}
\put(-210,30){$l_\mu$}
\put(-140,30){$l_\mu$}
\put(-56,30){$l_\mu$}
\put(-125,45){$h(l)$}
\put(-50,42){$h(l)$}
\put(-165,43){$A$}
\put(-165,53){$A'$}
\put(-115,17){$B$}
\put(-112,25){$B'$}
}
\caption{We sketch a graphic representation of the dynamical Eq.~(\ref{sinv2}) for the smallest holonomy field $X_h(v,U)$ (\ref{xs}). 
The diagram in the left-hand side of the graphic equation indicates the first term in Eq.~(\ref{sinv2}). The first and second diagrams
in the right-hand side of the graphic equation, respectively indicate the third and second terms in Eq.~(\ref{sinv2}). Note that $A$ and $A'$ are the same vertex, so are $B$ and $B'$. We indicate the edge $l_\mu$, where the {\it local} gauge 
transformation is made. 
In the right-hand side of the graphic equation, the summation over all 2-simplices
$h(l)$ associated to this edge $l_\mu$ is made.
}%
\label{sdf0}%
\end{figure}

\begin{figure}[ptb]
\includegraphics[scale=1.2 
]{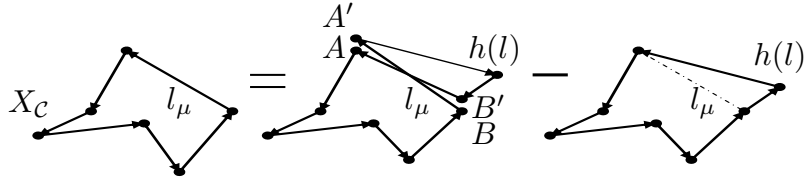}
\comment{
\put(-270,33){$X_{\mathcal C}$}
\put(-220,34){$l_\mu$}
\put(-145,34){$l_\mu$}
\put(-55,34){$l_\mu$}
\put(-125,50){$h(l)$}
\put(-35,48){$h(l)$}
\put(-171,50){$A$}
\put(-171,60){$A'$}
\put(-124,23){$B$}
\put(-124,31){$B'$}
}
\caption{We sketch a graphic representation of the dynamical Eq.~(\ref{sinv2}) for the general holonomy field $X_{\mathcal C}$ (\ref{rpa0s}). 
The diagram in the left-hand side of the graphic equation indicates the first term in Eq.~(\ref{sinv2}). The first and second diagrams in the right-hand side of the graphic equation, respectively, indicate the third and second terms in Eq.~(\ref{sinv2}). We indicate the edge $l_\mu$, where the {\it local} gauge 
transformation is made. 
In the right-hand side of graphic equation, the summation over all 2-simplices $h(l)$ associated to this edge $l_\mu$ is made.
}%
\label{sdf1}%
\end{figure}

\begin{figure}[ptb]
\includegraphics[scale=1.2
]{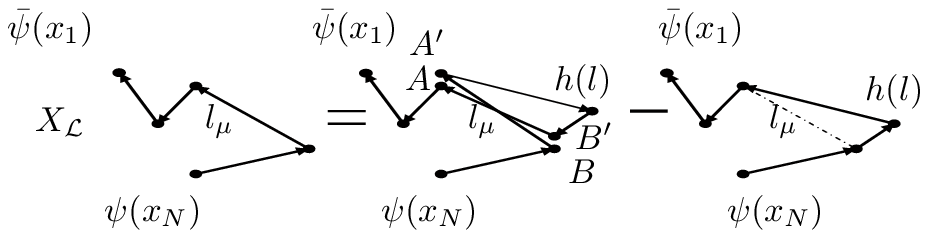}
\comment{
\put(-280,33){$X_{\mathcal L}$}
\put(-231,34){$l_\mu$}
\put(-155,34){$l_\mu$}
\put(-68,34){$l_\mu$}
\put(-130,45){$h(l)$}
\put(-40,42){$h(l)$}
\put(-173,45){$A$}
\put(-172,55){$A'$}
\put(-126,18){$B$}
\put(-124,28){$B'$}
\put(-288,60){$\bar\psi(x_1)$}
\put(-200,60){$\bar\psi(x_1)$}
\put(-100,60){$\bar\psi(x_1)$}
\put(-260,7){$\psi(x_N)$}
\put(-180,7){$\psi(x_N)$}
\put(-80,7){$\psi(x_N)$}
}
\caption{We sketch a graphic representation of the dynamical Eq.~(\ref{sinv2}) for the field $X_{\mathcal L}$ (\ref{rfa0s}). 
The diagram in the left-hand side of the graphic equation indicates the first term in Eq.~(\ref{sinv2}). The first and second diagrams in the right-hand side of the graphic equation respectively indicate the third and second terms in Eq.~(\ref{sinv2}). Note that $A$ and $A'$ are the same vertex, so are $B$ and $B'$. We indicate the edge $l_\mu$, where the {\it local} gauge 
transformation is made. 
We also indicate the fermion field $\psi(x_N)$ at staring point $x_N$ and the fermion field $\bar\psi(x_1)$ at ending point $x_1$ of the path ${\mathcal L}$. 
In the right-hand side of the graphic equation, the summation over all 2-simplices $h(l)$ associated to this edge $l_\mu$ is made.
}%
\label{sdf2}%
\end{figure}

Under a {\it local} gauge transformation (\ref{varie})-(\ref{ftran}), equivalently (\ref{varie}), (\ref{gtranu}), and (\ref{ftran}), the {\it local} gauge invariance of holonomy fields $\langle X\rangle$ [Eq.~(\ref{eve})], i.e., $\delta \langle X\rangle =0 $, leads to the dynamical equations for the holonomy fields $X_h$ (\ref{xs}), $X_{\mathcal C}$ (\ref{rpa0s}) and $X_{\mathcal L}$ (\ref{rfa0s}),
\begin{eqnarray}
&&\langle \frac{\delta X}{\delta e_\mu}\delta e_\mu -X\frac{\delta{\mathcal A}_{EC}}{\delta e_\mu}\delta e_\mu\rangle
+\langle\frac{\delta X}{\delta \psi}\delta \psi -X\frac{\delta{\mathcal A}_{EC}}{\delta \psi}\delta \psi\rangle\nonumber\\
&&+iag\langle X\delta\omega_\mu \rangle - \langle X\frac{\delta {\mathcal A}_{EC}}{\delta \omega_\mu}\delta \omega_\mu
\rangle +{\rm h.c.}=0,
\label{sinv1}
\end{eqnarray}
where the index $\mu$ is fixed, and for the variation $\delta X/\delta\omega_\mu$ we use Eq.~(\ref{inv3}) and the relationship
\begin{equation}
\sum_{ab}U^{ab}_\mu\frac{\delta X}{\delta U^{ab}_\mu}= X;\quad {\rm or}\quad
\sum_{ab} U^{ab\dagger}_\mu\frac{\delta X}{\delta U^{ab\dagger}_\mu} = X.
\label{inv3d}
\end{equation}
Analogously to the analysis in Sec.~(\ref{local}), we obtain the dynamical equations for the holonomy fields $X=X_h,X_{\mathcal C}$ and $X_{\mathcal L}$
\begin{eqnarray}
\langle \frac{\delta X}{\delta e_\mu}\delta e_\mu-X\frac{\delta { \mathcal A}_{EC}}{\delta e_\mu}\delta e_\mu\rangle +{\rm h.c.}&=&0,\label{ae1}\\
\langle \frac{\delta X}{\delta \psi}\delta\psi-X\frac{\delta { \mathcal A}_{EC}}{\delta \psi}\delta\psi\rangle +{\rm h.c.}&=&0,\label{ad1}
\end{eqnarray}
and
\begin{eqnarray}
\langle X\rangle + \langle X \Big(U^{\dagger }_\mu\frac{\delta {\mathcal A}_{EC}}{\delta U^{\dagger}_\mu}\Big)\rangle - \langle X\Big( 
U_\mu\frac{\delta {\mathcal A}_{EC}}{\delta U_\mu}\Big)\rangle
 &=&0.
\label{sinv2}
\end{eqnarray}
Equation (\ref{sinv2}) has the same form as the Dyson-Schwinger equation for the Wilson loops in lattice gauge theories. In Figs.~\ref{sdf0},\ref{sdf1},\ref{sdf2}, we show the graphic representations of the dynamical Eqs.~ (\ref{sinv2}) for the holonomy fields
$X_{h}$ (\ref{xs}) and $X_{\mathcal C}$ (\ref{rpa0s}), as well as $X_{\mathcal L}$ (\ref{rfa0s}).

\section{Mean-field approximation}\label{meanf}

\subsection{Mean-field approach}\label{meanapp}

\comment{
We consider a simplicial complex, i.e., random simplicial surface, 
whose elementary building block is a triangle $h(x)$ (see Fig.~\ref{pl}).  
We have the area operator of triangle $h(x)$,
\begin{eqnarray}
S^{\rm h}_{\mu\nu}(x)&\equiv & i\frac{1}{2} l_\mu(x)\wedge l_\nu(x) = ia^2 \frac{1}{2} e_\mu(x)\wedge e_\nu(x) = a^2e_{\mu\nu}\nonumber\\
S^{\rm h}_{\mu\nu}(x)&\equiv& i\gamma_5\frac{1}{2} l_\mu(x)\wedge l_\nu(x) = ia^2 \frac{1}{2} e_\mu(x)\wedge e_\nu(x) = a^2\gamma_5 e_{\mu\nu}
\label{ao}
\end{eqnarray}
where $\mu\not=\nu$ indicates edges of the 2-simplex $h$. 
}


In this section, we try to approximately calculate the partition function (\ref{par}), the vacuum expectational values of the 2-simplex area (\ref{aread}) and the volume element (\ref{vold}) by using the approach of the mean-field approximation. In the regularized action $X_{h}(v,U)$ (\ref{xs}) associating to the 2-simplex $h(x)$ (Fig.~\ref{pl}), we replace the vertex fields  $v_{\mu\rho}(x+a_{\mu})$ and $v_{\rho\nu}(x+a_\nu)$ by assuming a nonvanishing mean-field value $M_{\rm h}>0$, 
\begin{eqnarray}
(M^2_{\rm h})\delta^{\alpha\beta} &\equiv &  \Big[\langle v_{\mu\rho}v_{\rho\nu}\rangle\Big]^{\alpha\beta},
\label{aoa}
\end{eqnarray}
where $\alpha, \beta$ are Dirac spinor indexes. The definition of mean-field value (\ref{aoa}) does not depend on whether $v_{\mu\rho}$ and $v_{\rho\nu}$ contain the matrix $\gamma_5$ or not, due to $\gamma_5^2=1$ and $[\gamma_5,\, \sigma_{ab}]=0$. The mean-field value $M_{\rm h}$ is 
independent of any specific vertex, edge and 2-simplex of the simplicial complex.  
Based on the definitions of the 2-simplex area (\ref{aread}) and the volume element (\ref{vold}), the mean-field values for the 2-simplex area and 
the volume element are given by
\begin{eqnarray}
\langle S_{\rm h}(x) \rangle &=& a^2 M_{\rm h},\nonumber\\
\langle dV(x) \rangle &=& a^4 N_{\rm h}M^2_{\rm h},
\label{aov}
\end{eqnarray} 
where $N_{\rm h}$ is the mean value of the number of 2-simplices $h(x)$ that share the same vertex. Note that in this preliminary calculations in the mean-field approximation, we do not take into account the cosmological term (\ref{cosmo}), since the path integrals are convergent (see below) for positive mean-field value $M_{\rm h} >0$. 

Based on the mean-field value (\ref{aoa}), the smallest holonomy field $X_{h}(v,U)$ (\ref{xs}) is approximated by its mean-field counterpart 
\begin{eqnarray}
\bar X_{h} (v,U)&=& 
{\rm tr}\left[v_{\nu\mu}(x)U_{\mu}(x)U_{\rho}(x+a_\mu)U_{\nu}(x+a_\nu)\right]M^2_{\rm h}\label{xsgm}\\
\bar X^\dagger_{h} (v,U) &=& 
{\rm tr}\left[v_{\mu\nu}(x)U_{\nu}(x)U_{\rho}(x+a_\nu)U_{\mu}(x+a_\mu)\right]M^2_{\rm h}\label{xsgmh}
\end{eqnarray}
where using Eqs.~(\ref{xsc}) and (\ref{xsh1}) for $\mu\not=\nu$ we obtain $\bar X^\dagger_{h} (v,U)$. Note that two of three vertex fields $v(x)$ in the $X_{h}(v,U)$ (\ref{xs}), i.e., $v_{\mu\rho}(x+a_{\mu})$ and $v_{\rho\nu}(x+a_\nu)$  are replaced by their mean-field values $M_{\rm h}$, and the 2-simplex $h(x)$ shown in Fig.~\ref{pl} can also be identified by three different indexes $\mu\not=\nu\not=\rho$ (no summation over these indexes).  Equations (\ref{xsgm}) and (\ref{xsgmh}) depend on $U_\rho$, and the fields $(e_\mu,U_\mu)$ and $(e_\nu,U_\nu)$ associated to two edges $(x,\mu)$ and $(x,\nu)$ of the 2-simplex (triangle) $h(x)$ (see Fig.~\ref{pl}). Using Eqs.~(\ref{xsgm},\ref{xsgmh}), we define the {\it local} mean-field action $\bar {\mathcal A}_h$
for the 2-simplex $h(x)$
\begin{eqnarray}
\bar {\mathcal A}_h& = & \frac{1}{8g^2} \Big[\bar X_{h} (v,U)+\bar X_{h}^\dagger(v,U)\Big]_{v_{\mu\nu}=\gamma_5e_{\mu\nu}}\nonumber\\
&+& \frac{1}{8g^2\gamma} \Big[\bar X_{h} (v,U)+\bar X_{h}^\dagger(v,U)\Big]_{v_{\mu\nu}=e_{\mu\nu}}\nonumber\\
&=& {\rm tr}\left[e_\nu(x)\Gamma^h_{\nu\mu}(x)e_\mu(x)-e_\mu(x)\Gamma^h_{\nu\mu}(x)e_\nu(x)\right],
\label{action2d}
\end{eqnarray}
where 
\begin{eqnarray}
\Gamma^h_{\nu\mu}(x) &=& \frac{1}{8g^2}
\left(\gamma_5-\frac{1}{\gamma}\right)H_{\nu\mu}(x)\nonumber\\
&=& \frac{1}{8g^2}
\left(\frac{i}{2}\right)M^2_{\rm h}\left(\gamma_5-\frac{1}{\gamma}\right)\left[U_{\nu}(x)U_{\rho}(x+a_\nu)U^\dagger_{\mu}(x)\right]+{\rm h.c.}
\label{chg2t}
\end{eqnarray}
The detailed derivation is given 
in Appendix \ref{meana}.  
In this mean-field approximation, all 2-simplices $\{h(x)\}$ in the simplicial  complex ${\mathcal M}$ have the same {\it local} action (\ref{action2d}), namely, the single 2-simplex mean-field action $\bar {\mathcal A}_h$ (\ref{action2d}) and operator $\Gamma^h_{\nu\mu}$ (\ref{chg2t}) are independent of the vertex ``$x$''.  With the {\it local} mean-field action (\ref{action2d}), we define the {\it local} mean-field partition function
\begin{eqnarray}
\bar Z_h=\int_h{\mathcal D}U{\mathcal D}e \exp -\bar{\mathcal A}_h,
\label{meanpar0}
\end{eqnarray}
where the {\it local} mean-field measure is defined by 
\begin{eqnarray}
\int_h{\mathcal D}U{\mathcal D}e \equiv \int_h dU_\mu dU_\nu dU_\rho de_\mu de_\nu,
\label{meanm}
\end{eqnarray} 
for each 2-simplex $h$.
Thus, the regularized EC action ${\mathcal A}_{EC}$ (
\ref{ecp}) is approximated by its mean-field counterpart,
\begin{eqnarray}
\bar{\mathcal A}_{EC}=\sum_{h\in {\mathcal M}}\bar{\mathcal A}_h,
\label{meanact}
\end{eqnarray}
which is the sum of the mean-field actions $\bar{\mathcal A}_h$ over all 2-simplices $h$.
With the mean-field approximated action (\ref{meanact}), we define the mean-field approximated partition function
\begin{eqnarray}
\bar Z_{EC}=\prod_{h\in{\mathcal M}}\int_h{\mathcal D}U{\mathcal D}e \exp -\bar{\mathcal A}_{EC}=\prod_{h\in{\mathcal M}}\bar Z_h,
\label{meanpar}
\end{eqnarray}
which is the mean-field counterpart of the partition function (\ref{par}). 

Using the mean-field EC action $\bar {\mathcal A}_{EC}$ (\ref{action2d}) 
and partition function $\bar Z_{EC}$ (\ref{meanpar}), we have the following identity
\begin{eqnarray}
Z_{EC}\equiv \bar Z_{EC} \langle e^{-({\mathcal A}_{EC}-\bar {\mathcal A}_{EC})}\rangle_\circ,
\label{meanid}
\end{eqnarray}
where $\langle \cdot\cdot\cdot\rangle_\circ$ is the vacuum expectational value with respect to the mean-field partition function $\bar Z_{EC}$ (\ref{meanpar}).
Using the convexity inequality \cite{feynman1972}
\begin{eqnarray}
\langle e^{-({\mathcal A}_{EC}-\bar {\mathcal A}_{EC})}\rangle_\circ \ge e^{-\langle {\mathcal A}_{EC}-\bar {\mathcal A}_{EC}\rangle_\circ},
\label{convin}
\end{eqnarray}
one can derive the following inequality
\begin{eqnarray}
-\ln Z_{EC}\le -\ln\bar Z_{EC} + \langle {\mathcal A}_{EC}- \bar{\mathcal A}_{EC}\rangle_\circ,
\label{ineq}
\end{eqnarray}
where $-\ln Z_{EC}$  and $-\ln \bar Z_{EC}$ are proportional to the free energies. We define the right-handed side of the inequality (\ref{ineq}) as an approximate free energy (or approximate effective action)
\begin{eqnarray}
{\mathcal F}_{EC}^{\rm app}(M_{\rm h},g,\gamma) \equiv -\ln\bar Z_{EC} + \langle {\mathcal A}_{EC}- \bar{\mathcal A}_{EC}\rangle_\circ.
\label{appf}
\end{eqnarray} 
The validity of the mean-field approximation approach bases on the inequality (\ref{ineq}) that gives a low bound of the approximate free energy ${\mathcal F}_{EC}^{\rm app}(M_{\rm h},g,\gamma)$. We determine the mean-field 
value $M^*_h(g,\gamma)$ of the {\it local} mean-field action (\ref{action2d}), 
which minimizes the approximate free energy (\ref{appf}) and thus optimizes the low bound in Eq.~(\ref{ineq}), by satisfying the condition
\begin{eqnarray}
\left[\frac{\delta}{\delta M_{\rm h}}{\mathcal F}_{EC}^{\rm app}(M_{\rm h},g,\gamma)\right]_{M_{\rm h}=M^*_{\rm h}}=0.
\label{meanmin}
\end{eqnarray}
Using the mean-field value $M^*_{\rm h}(g,\gamma)$ and corresponding minimum of the
approximate free energy ${\mathcal F}_{EC}^{\rm app}[M^*_{\rm h}(g,\gamma),g,\gamma]$ (\ref{appf}), we can gain some insights into the value of the 2-simplex area (\ref{aoa},\ref{aov}),
and the critical points of the second-order phase transition, in terms of the gauge coupling $g$ and Immirzi parameter $\gamma$. In addition, we can use the mean-field action (\ref{action2d}) with the value $M^*_{\rm h}$ to calculate mean-field vacuum expectational values $\langle\cdot\cdot\cdot\rangle_\circ$ to 
approximate true vacuum expectational values $\langle\cdot\cdot\cdot\rangle$ that we discussed in Secs.~\ref{epf}, \ref{local} and \ref{dehf}.

\subsection{Analytical calculations}\label{analcal}

We can analytically calculate the mean-field 
partition function (\ref{meanpar}).
First we integrate over quantized tetrad $e_\mu(x)$ and $e_\nu(x)$ fields, which is quadratic in Eq.~(\ref{action2d}) (see Appendix \ref{meana1}).  
Using the formula (\ref{det1}), we have 
\begin{eqnarray}
\prod_{h\in{\mathcal M}}\int de_\mu de_\nu \exp -\bar{\mathcal A}_{EC}&=&\prod_{h\in{\mathcal M}}{\rm det}^{-1}[I-\Gamma^h] 
\label{det0}
\end{eqnarray}
and the Cayley-Hamilton formula for a determinant \cite{iz4-86} 
\begin{eqnarray}
{\rm det}^{-1}[I-\Gamma^h]&=&\exp[-{\rm tr}\ln(I-\Gamma^h)]\nonumber\\
&=&1+\sum_a\Gamma^h_{aa}+\frac{1}{2}\sum_{a,b}(\Gamma^h_{aa}\Gamma^h_{bb}+\Gamma^h_{ab}\Gamma^h_{ba})+\cdot\cdot\cdot\nonumber\\
&+&\frac{1}{n!}\sum_{a_1\cdot\cdot\cdot a_n}\sum_{P}\Gamma^h_{a_1a_{P_1}}\cdot\cdot\cdot\Gamma^h_{a_na_{P_n}} 
\label{det}
\end{eqnarray}
where 
$P$ indicates permutations of $(1,\cdot\cdot\cdot,n)$ and  Eq.~(\ref{det}) is a sum of traces of symmetrized tensor products. The expression (\ref{det}) stops at the $n$-th order for a finite $n$-dimensional matrix $\Gamma^h$ in the space of the gauge group.

Second we integrate over
group-valued spin-connection $U_\rho(x+a_\mu)$, $U_\mu(x)$ and $U_\nu(x)$ fields defined at edges $(x+a_\mu,\rho)$, $(x,\mu)$ and $(x,\nu)$ of the 2-simplex $h(x)$ 
by using 
the properties of the invariant Haar measure: 
\begin{eqnarray}
\int dU_\mu(x)&=&1
\label{mea20}\\
\int dU_\mu(x)U_\mu(x)&=&0\label{mea21}\\
\int dU_\mu(x) U^{ab}_\mu(x)U^{\dagger cd}_\sigma(x')&=&\frac{1}{d_j}\delta_{\mu\sigma}\delta^{ac}\delta^{bd}\delta(x-x'),
\label{mea2}
\end{eqnarray}
where $d_j=n_{j_L}n_{j_R}$ ($n_{j_L,j_R}=2j_{L,R}+1; j_{L,R}=1/2,3/2,\cdot\cdot\cdot$) is the dimensions of irreducible representations $j=(j_L,j_R)$ of the gauge group $SU_L(2)\otimes SU_R(2)$, $j_R=j_L=1/2$ and $d_j=4$
for the fundamental representation. In Appendix \ref{meana1}, we give more detailed calculations to obtain
the mean-field partition function (\ref{meanpar}),
\begin{eqnarray}
\bar Z_{EC} &=& \prod_{h\in {\mathcal M}}\left[1+\frac{\gamma^2+1}{64g^4\gamma^2d_j^3} M^4_{\rm h}\right],
\label{par2}
\end{eqnarray}
where 
$\prod_{h\in {\mathcal M}}$ is the product of all 2-simplices $h$ of the simplicial complex ${\mathcal M}$. The mean-field entropy is given by 
\begin{eqnarray}
\bar {\mathcal S}=\ln \bar Z_{EC} &=&\sum_{h\in {\mathcal M}}\ln\left[1+\frac{\gamma^2+1}{64g^4\gamma^2d_j^3} M^4_{\rm h}\right]\nonumber\\
&=&{\mathcal N}\ln\left[1+\frac{\gamma^2+1}{64g^4\gamma^2d_j^3} M^4_{\rm h}\right],
\label{entropy}
\end{eqnarray}
where ${\mathcal N}=\sum_{h\in {\mathcal M}}$ is the total number of 2-simplexes, and the mean-field free energy
\begin{eqnarray}
\bar {\mathcal F}=-\frac{1}{\beta}\ln \bar Z_{EC}=-\frac{1}{\beta}{\mathcal N}\ln\left[1+\frac{\gamma^2+1}{64g^4\gamma^2d_j^3} M^4_{\rm h}\right],
\label{free}
\end{eqnarray}
where the inverse ``temperature'' $\beta=1/g^2$, see Eqs.~(\ref{pact}) and (\ref{hact}).

We turn to calculate $ \langle \bar {\mathcal A}_{EC}\rangle_\circ $ in Eq.~(\ref{ineq}).
The mean-field value of $\bar {\mathcal A}_{EC}$ (\ref{meanact}) is calculated in Appendix \ref{meana1} [see Eq.~(\ref{va1})],
\begin{eqnarray}
\langle\bar {\mathcal A}_{EC}\rangle_\circ &=&\sum_{h\in {\mathcal M}} \langle\bar {\mathcal A}_h\rangle^h_\circ \nonumber\\
&=&{\mathcal N}\frac{\gamma^2+1}{32g^4\gamma^2d_j^3} M^4_{\rm h}\left[1+\frac{\gamma^2+1}{64g^4\gamma^2d_j^3} M^4_{\rm h}\right]^{-1},
\label{aqj}
\end{eqnarray}
where 
the vacuum expectational value with respect to the {\it local} mean-field partition function $\bar Z_h$ (\ref{meanpar0}) is defined by
\begin{eqnarray}
\langle \cdot\cdot\cdot \rangle^h_\circ &= & \frac{1}{\bar Z_h}\int_h{\mathcal D}U{\mathcal D}e \,(\cdot\cdot\cdot)\, e^{-\bar{\mathcal A}_h}.
\label{aves}
\end{eqnarray}   
The mean-field value $\langle\bar {\mathcal A}_{EC}\rangle_\circ$ (\ref{aqj}) has discrete values depending on the discrete values $d_j=4,\cdot\cdot\cdot$
of the fundamental state $j_{L,R}=1/2$ and excitation states $j_{L,R}=3/2,\cdot\cdot\cdot$, coupling to different fermion spinor states $\psi^j_{L,R}$.
 
\begin{figure}[ptb]
\includegraphics[scale=1.2]{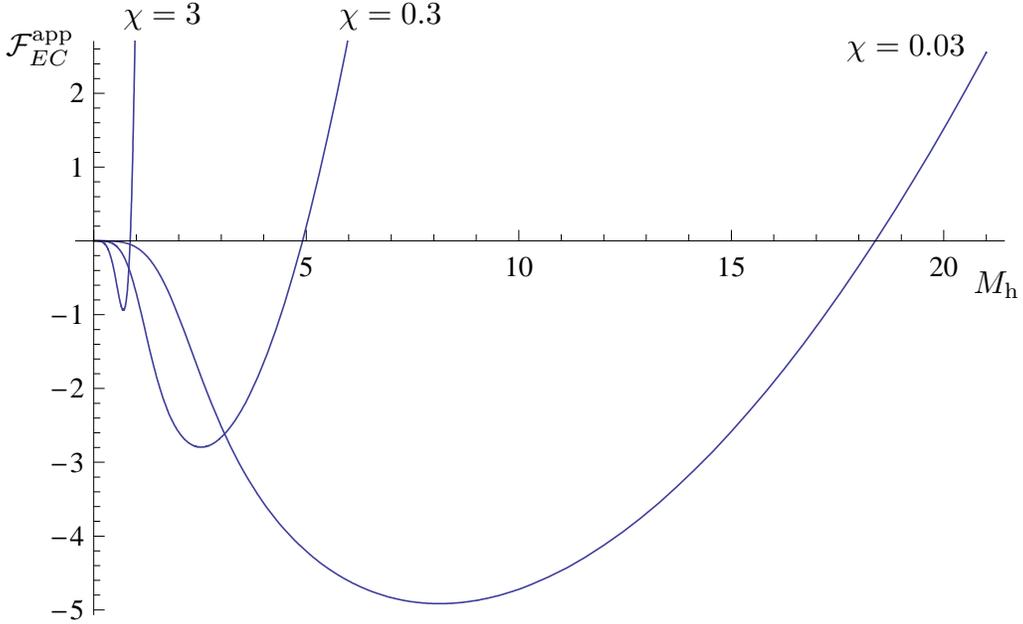}
\comment{
\put(-10,105){$M_{\rm h}$}
\put(-315,180){${\mathcal F}^{\rm app}_{EC}$}
\put(-278,190){$\chi=3$}
\put(-210,190){$\chi=0.3$}
\put(-50,180){$\chi=0.03$}
}
\caption{In the Planck unit $a=1$, the approximate free energy (\ref{appf}) as a function of the mean-field value $M_{\rm h}$ (\ref{aoa}) is plotted for selected values 
$\chi=0.03,0.3,3$. The minimal values of the approximate free energy ${\mathcal F}^{\rm app}_{EC}$ locate at the nonvanishing mean-field value $M^*_{\rm h}$. The minimal locations are $M^*_{\rm h}(\chi=0.03)\approx 7.9,\, M^*_{\rm h}(\chi=0.3)\approx 2.1,\, M^*_{\rm h}(\chi=3)\approx 0.8$.
}%
\label{pl_size}%
\end{figure}

We are in the position to calculate $ \langle {\mathcal A}_{EC}\rangle_\circ $ in Eq.~(\ref{ineq}). Since there are three vertex fields in the smallest holonomy field $X_h(v,U)$ (\ref{xs}) that constitutes the regularized EC action ${\mathcal A}_{EC}$ (\ref{pact}), (\ref{diract}), and (\ref{ecp}), while there is only one vertex field $v_{\nu\mu}$ in the mean-field action (\ref{xsgm})-(\ref{action2d}), we assign the vertex field $v_{\mu\nu}$ to the {\it local} mean-field action (\ref{xsgm}-\ref{chg2t}) of the 2-simplex $h$,
the vertex-fields $v_{\mu\rho}, v_{\rho\nu}$ to the {\it local} mean-field actions of neighboring 2-simplices, and approximate
\begin{eqnarray}
&& \langle {\rm tr}\left[v_{\nu\mu}(x)U_{\mu}(x)v_{\mu\rho}(x+a_\mu)U_{\rho}(x+a_\mu)
v_{\rho\nu}(x+a_\nu)U_{\nu}(x+a_\nu)\right]\rangle_\circ +{\rm h.c.}\nonumber\\
&&={\rm tr}\Big[ \langle v_{\nu\mu}U_{\mu}U_{\rho}
U_{\nu} v_{\mu\rho} v_{\rho\nu}\rangle^h_\circ\Big]+{\rm h.c.}\nonumber\\
&&\approx (Z_h)^2{\rm tr}\Big[ \langle v_{\nu\mu}U_{\mu}U_{\rho}
U_{\nu}\rangle^h_\circ\,\, 
\langle v_{\mu\rho}\rangle^h_\circ\,\, \langle v_{\rho\nu}\rangle^h_\circ\Big]+{\rm h.c.}\nonumber\\
&&\approx
(Z_h)^2{\rm tr}\Big[\Big( \langle v_{\nu\mu}U_{\mu}U_{\rho}
U_{\nu}\rangle^h_\circ+{\rm h.c.}\Big)\,\, 
\langle v_{\mu\rho}\rangle^h_\circ\,\, \langle v_{\rho\nu}\rangle^h_\circ\Big].
\label{ao300}
\end{eqnarray}
where $(v_{\mu\rho}v_{\rho\nu})^\dagger=(v_{\rho\nu}v_{\mu\rho})$. Using Eqs.~(\ref{action2d}) and (\ref{chg2t}), we have
\begin{eqnarray}
\langle {\mathcal A}_{EC}\rangle_\circ &\approx & \sum_{h\in {\mathcal M}} \frac{(Z_h)^2}{4M^2_{\rm h}}\left\{\langle{\rm tr}\left[e_\nu\Gamma^h_{\nu\mu}e_\mu-e_\mu\Gamma^h_{\nu\mu}e_\nu\right]\rangle_\circ {\rm tr}\Big[\langle [ e_{\mu\rho}]\rangle^h_\circ\,\, \langle [e_{\rho\nu}]\rangle^h_\circ\Big]\right\}.
\label{ao30}
\end{eqnarray}
In the last part of Appendix \ref{meana1}, we obtain
\begin{eqnarray}
\langle {\mathcal A}_{EC}\rangle_\circ &\approx & 
{\mathcal N} \frac{1}{M^2_{\rm h}}
\left(\frac{1}{\bar Z_h}\right)\left(\frac{1}{8g^2}\right)^6(M_{\rm h}^4)^3 \left(\frac{1}{4}\right)\left(    
\frac{2}{d_j^3}\right)^3\Big(\frac{\gamma^2+1}{\gamma^2}\Big)^3.
\label{fao}
\end{eqnarray}
Putting Eqs.~(\ref{entropy}), (\ref{aqj}) and (\ref{fao}) into the approximate free energy (\ref{appf}), we obtain
\begin{eqnarray}
{\mathcal F}_{EC}^{\rm app}(M_{\rm h},g,\gamma) = -\ln(1+y) - \frac{2y}{1+y}
+\chi\frac{y^{5/2}}{(1+y)},
\label{appff}
\end{eqnarray}
where 
\begin{eqnarray}
y =\frac{\gamma^2+1}{64g^4\gamma^2d_j^3} M^4_{\rm h},\quad
\chi=2\sqrt{\frac{\gamma^2+1}{64g^4\gamma^2d_j^3}}. 
\label{min0}
\end{eqnarray}
In Fig.~\ref{pl_size}, we plot the approximate free energy (\ref{appf}) as a function of the mean-field value $M_{\rm h}$ (\ref{aoa}) for selected values of the parameter $\chi$ (\ref{min0}). The minimal values of the approximate free energy ${\mathcal F}^{\rm app}_{EC}$ (\ref{appf}) locate at the nonvanishing mean-field value $M^*_{\rm h}\not=0$, which increases as the parameter $\chi$ decreases, namely, the gauge coupling increases.   The gauge coupling $g$ and Immirzi parameter $\gamma$ remain to be determined. These two parameters $(g,\gamma)$ 
should be determined at critical points of the second-order phase transition, as discussed in the last section. The mean-field approximation approach adopted here needs to be improved to see whether we can have a critical value $\chi_c$, and for $\chi>\chi_c$ the minimal value of the approximate free energy ${\mathcal F}^{\rm app}_{EC}$ locates at the vanishing mean-field value $M^*_{\rm h}\not=0$. It is usually difficult to study the vicinity of critical points of the second-order phase transition by the mean-field approximation approach.   

Considering  the case that $\gamma\gg 1$, $d_j=4$,
$g \rightarrow 4/3$ for $G_{\rm eff}\rightarrow G$ [see Eq.~(\ref{effg}) in Sec.~\ref{naivec}], and $\chi\approx 0.02$,  we have 
\begin{eqnarray}
M^*_{\rm h} &>& 1,
\label{ieq2}
\end{eqnarray}
see the curve for $\chi=0.03$ in Fig.~\ref{pl_size}, since $M^*_{\rm h}$ becomes larger as $\chi$ decreases. For larger gauge coupling $g$ and higher dimensions $d_j$ of irreducible representations, the values of $\chi$ (\ref{min0}) become smaller, and $M^*_{\rm h}$ becomes larger. 

Therefore, the mean-field value of the 2-simplex area (\ref{aoa}) 
\begin{eqnarray}
\langle S_{\rm h} \rangle =a^2M^*_{\rm h}> a^2=\frac{8\pi}{m_{\rm Planck}^2}, 
\label{ieq3}
\end{eqnarray}
and the mean-field value of the volume element (\ref{aov}) 
\begin{eqnarray}
\langle dV(x)\rangle =a^4N_{\rm h}(M^*_{\rm h})^2> N_{\rm h}\frac{(8\pi)^2}{m_{\rm Planck}^4}. 
\label{ieq4}
\end{eqnarray}
Equations (\ref{ieq3}) and(\ref{ieq4}) indicate that the averaged sizes of 2-simplex, 3-simplex, and 4-simplex, i.e., elements of the simplicial complex, are larger than the Planck length, which is
probed by short wavelengths of quantum fields $e_\mu,U_\mu,\psi$ in strong gauge couplings $g$.
This 
implies that  due to the quantum gravity, the
Planck length sets the scale for the minimal separation between two space-time points \cite{preparata91}. We end this section by noting that the mean-field approximation is not only a poor approximation, but also breaks diffeomorphism and {\it local} gauge symmetries. 

\section{\bf Some remarks}\label{remarks}
  
In addition to the Planck length $a$, the regularized EC action (\ref{ecpf}) proposed in this article contains three dimensionless parameters: the gauge coupling $g$; the Immirzi parameter $\gamma$ and the cosmological parameter $\lambda$. In the view of the naive continuum limit, the regularized EC action (\ref{ecpf}) proposed in this article is not unique. In principle, permitted by the diffeomorphism and {\it local} gauge invariances, the regularized action (\ref{ecpf}) is allowed to contain nonlocal high-dimensional ($d>6$) operators of fields $e_\mu$, $U_\mu$ and $\psi$ with extra free parameters. On the other hand,  
although the regularized EC action (\ref{ecpf}) approaches to the continuum EC action (\ref{ec0}) in the naive continuous limit, it has not been 
clear yet whether the regularized EC theory is physically sensible. The regularized EC theory is physically sensible, only if only it has a nontrivial continuum limit, where we could possibly explore the relationship to the Minkowski counterpart. Therefore
it is crucial, on the basis of nonperturbative methods and 
renormalization-group invariance, to find: 
\begin{enumerate}

\item the scaling invariant region (nontrivial ultraviolet fix points) ($g_c,\,\gamma_c,\,\lambda_c$), where the singularity in the free energy appears for phase transition occurring, and the
physical correlation length $\xi$ of two-point Green-functions of fields is much larger than the Planck length, 
while the inverse correlation length $\xi^{-1}$ gives the mass scale of 
low-energy excitations of the ``effective continuum theory''; 

\item
$\beta$ function $\beta(g)$, i.e., the scale dependence of the gauge coupling $g$ in the vicinity of the nontrivial ultraviolet fix points $g_c$, and renormalization-group invariant equation 
\begin{eqnarray}
\xi = {\rm constant}\cdot a \cdot\exp \int^g dg' /\beta(g'),\quad \xi\gg a,
\label{betaf}
\end{eqnarray}
in this scaling invariant region, and ``constant'' can only be obtained by nonperturbative methods. And it is a question how Eq.~(\ref{betaf}) is related to $\gamma_c$ and $\lambda_c$;

\item an effective action ${\mathcal A}^{\rm eff}_{EC}$ (\ref{par}), all relevant and renormalizable operators [one-particle irreducible (1PI) functions] with effective dimension-four to obtain an effective low-energy theory in this scaling invariant region. 
\end{enumerate}

The gauge-invariant correlation length $\xi$ can be possibly measured by the gauge-invariant two-point correlation function of the holonomy fields $X_h(v,U)$ (\ref{xs}),
\begin{eqnarray}
\langle X_h[v(x),U(x)], X^\dagger_h[v(y),U(y)]\rangle \sim ~ e^{-|x-y|/\xi},\quad |x-y|\gg \xi ,
\label{length}
\end{eqnarray}
where $|x-y|$ indicates the separation between two holonomy fields $X_h(v,U)$. Actually, Eq.~(\ref{length}) is related to the invariant curvature correlation 
function [see Eq.~(\ref{xs6})]. 
\comment{
The genuine violation of the diffeomorphism invariance at the size 
of a 4-simplex is negligible, when we consider
large scales probed by long wavelengths of fields. 
}

Although we have added the bare cosmological term (\ref{cosmo}) into the regularized action,
1PI functions ${\mathcal A}^{\rm eff}_{EC}$ (\ref{par}) 
effectively contain this dimensional 
operator (\ref{cosmo}), 
which is related to the two-point correlation function (\ref{length}). It is then a question 
what is the scaling property of this operator in terms of the low-energy scale $\xi^{-2}$ . We speculate that the 
gauge-invariant 
correlation length $\xi$, instead of the Planck length, sets the scale for the nonperturbative renormalized  cosmological constant, i.e.,
\begin{eqnarray}
\Lambda_{\rm COSM}\sim \xi^{-2},
\label{cosm1}
\end{eqnarray}
which is rather similar to the scale $\Lambda_{\rm QCD}$ calculated in the lattice QCD theory. This would possibly explain why the observed cosmological constant is much smaller than that expected in terms of the Planck scale [see Eq.~(\ref{betaf})]. We also speculate that in the pure gravity at strong gauge coupling $g\gg 1$, the scale $\xi^{-2}$ should 
measure the exponential area-decay law of holonomy fields (\ref{rpa0s},\ref{eve}) for sufficiently large loops
\begin{eqnarray}
\langle X_{\mathcal C}(v,U)\rangle \sim ~ e^{-A_{\rm min}({\mathcal C})/\xi^2},\quad A_{\rm min}({\mathcal C})\gg \xi^2,
\label{xarea}
\end{eqnarray}
where $A_{\rm min}({\mathcal C})$ is the minimal area, corresponding to the minimal number of 2-simplices $h$, that can be spanned by the loop ${\mathcal C}$ (see Ref.~\cite{hambert2007}). The scaling invariant region $g_c$, scaling law (\ref{betaf}) and correlation length $\xi$ are important to study our present Universe (see Ref.~ \cite{hambert2010}).


The effective quadralinear-fermion interactions in the continuum EC theory (\ref{eca2}) are originated by integrating over {\it static} torsion fields and the torsion-free condition is satisfied as required by the equivalence principle. In this sense, quadralinear-fermion interactions are inevitable as long as the interacting between fermion and gravitational fields is included.

The bilinear fermion action (\ref{plart}) introduces a nonvanishing torsion field (\ref{spinc1}) in the regularized EC theory. The torsion fields (\ref{spinc1}) are not exactly {\it static}, however, they are fields only surviving in short distances at the Planck scale, which is due to the quantum gravity [see for example the mean-field approximation result (\ref{ieq2}-\ref{ieq4})].  
The effective quadralinear-fermion interactions (\ref{4fl1}) is formulated by hand 
together with a torsion-free bilinear fermion action (\ref{plart}) so that they approach to the fermion action of the continuum EC theory in the continuum limit. 
In principle, it should be possible to obtain an effective action by solving the discretized Cartan structure Eq.~ (\ref{wel1}) or Eq.~(\ref{up1}) with the nonvanishing discretized torsion (\ref{spinc1}), and integrating over torsion fields at short distances, in the same way as (\ref{werelation1}-\ref{eca2}) of
the continuum EC theory. In this way, one will obtain a 
complicate effective action of fermion fields with high-order dimensional ($d>6$) operators. However, we expect that in the continuum limit the relevant operators of fermion fields should be Eq.~(\ref{4fl1}) and its continuum counterpart (\ref{4f}).    

On the other hand, due to the no-go theorem \cite{nn1981}, the bilinear fermion action (\ref{plart}) 
has the problem of either fermion doubling or chiral (parity) 
gauge symmetry breaking, which is inconsistent with the low-energy standard model for particle physics. As discussed, the effective quadralinear-fermion interactions (\ref{4fl1}) are inevitable, due to mediating very massive torsion fields in short distances at the Planck scale. We expect that in the invariant scaling region of the nontrivial ultraviolet fix points ($g_c,\,\gamma_c,\,\lambda_c$), the      
quadralinear-fermion interactions should be relevant operators, which not only give a possible resolution to the fermion doubling problem \cite{ep1986,xue1997}, but also the compelling dynamics for fermion mass generation \cite{xuemass, zubkov2010}, via the Nambu Jona-Lasinio mechanism \cite{njl}.
   
 
\section{Acknowledgment}
The anonymous referee reports played an important role in greatly improving this article. The author thanks H.~W.~Hamber and R.~M.~Williams for discussions on the  renormalization-group invariance, properties of Dirac-matrix valued tetrad fields, and the gauge invariance of the regularized Einstein-Cartan action in terms of tetrad and spin-connection fields. The 
author also thanks
H.~Kleinert and J.~ Maldacena for the discussions on invariant holonomy fields in gauge theories, and thanks to V.~Khatsymovsky and M.~A.~Zubakov for the discussions on the measure and convergences of path integrals (\ref{par}) and (\ref{eve}). The author also thanks R.~Ruffini and G.~Preparata for discussions on general relativity and Wheeler's foam \cite{wheeler1964}, which brought the author's attention to this topic.

\comment{
The detailed calculations will be presented in a lengthy article.
It is worthwhile to mention the important issue that bilinear fermion term (\ref{plart}) has the problem of either fermion doubling or chiral (parity) 
gauge symmetry breaking, due to the No-Go theorem \cite{nn1981}. Quasilinear fermionic operators can possibly be resolution to this 
problem \cite{ep1986}, and the four-fermion interactions 
(\ref{4fl1}) was studied in Ref.~\cite{xue1997}.
We have to point out that the regularization action (\ref{ecp}) is not unique, it can possibly contain non-local high-dimensional ($d>6$) operators of tetrad, link and fermion fields, permitted by diffeomorphism and {\it local}  gauge-invariances. Beside, non-local holonomies  (\ref{pa0s}) needs to be defined for a genuine non-perturbative formulation of quantum gravity. The detailed calculations will be presented in a lengthy article.   
}

\appendix

\section{}\label{con}

By using Eqs.~(\ref{link0}) and (\ref{link00}) and the identity 
$e^{\hat A}e^{\hat B}=e^{\hat A+\hat B+[\hat A,\hat B]/2}$, we calculate 
$U_{\mu\nu}(x)$ (\ref{p00})-(\ref{uu0})  
in the {\it native continuum limit}: 
$ag\omega_\mu \ll 1$. 
Expanding $U_{\mu\nu}(x)$ in powers of $ag\omega_\mu$, we have
\begin{eqnarray}
&&U_{\mu\nu}(x)=U_\mu(x)U_\nu(x+a_\mu)= 
\nonumber\\ &&=\exp\left\{iga[\omega_\mu(x)+\omega_\nu(x)]
+iga^2\partial_\mu \omega_\nu(x)
-\frac{1}{2}(ga)^2\left[\omega_\mu(x),\omega_\nu(x)\right]+{\mathcal O}(a^3)\right\}\nonumber\\
&&=\exp\left\{iga[\omega_\mu(x)+\omega_\nu(x)]
+iga^2\partial_\mu \omega_\nu(x)
-\frac{i}{2}(ga)^2[\omega^{ae}(x)\wedge\omega^b_{~e}(x)]_{\mu\nu}
\sigma_{ab}+{\mathcal O}(a^3)\right\}\nonumber\\
&&=\exp\left\{iga\sigma_{AB}G^{AB}_{\mu\nu}+{\mathcal O}(a^3)\right\},
\label{uu1}
\end{eqnarray}
where 
\begin{eqnarray}
G^{AB}_{\mu\nu}&=&[\omega^{AB}_\mu(x)+\omega^{AB}_\nu(x)]
+a\partial_\mu \omega^{AB}_\nu(x)\nonumber\\
&-&\frac{1}{2}(ga)[\omega^{Ae}(x)\wedge\omega^B_e(x)]_{\mu\nu},
\label{gmu}
\end{eqnarray} 
and ${\mathcal O}(a^3)$ indicates 
high-order powers of $ag\omega_\mu$. 
In Eq.~(\ref{uu1}), we use $\left[\sigma_{ab},\sigma_{bc}\right]=i\delta_{bb}\sigma_{ca}$ 
(no sum with index $b$), $[\gamma_5,\sigma_{ca}]=0$
and
\begin{eqnarray}
\omega_{\mu\nu}(x)\equiv
\left[\omega_\mu(x),\omega_\nu(x)\right]=[\omega^{ae}(x)\wedge\omega^{eb}(x)]_{\mu\nu}
\left[\sigma_{ae},\sigma_{eb}\right]=i[\omega^{ae}(x)\wedge\omega^b_{~e}(x)]_{\mu\nu}
\sigma_{ab}.
\label{ww}
\end{eqnarray}
For exchanging $\mu\leftrightarrow\nu$ in Eqs.~(\ref{uu1}) and (\ref{gmu})
\begin{eqnarray}
G^{AB}_{\nu\mu}&=&[\omega^{AB}_\mu(x)+\omega^{AB}_\nu(x)]
+a\partial_\nu \omega^{AB}_\mu(x)\nonumber\\
&-&\frac{1}{2}(ga)[\omega^{Ae}(x)\wedge\omega^B_e(x)]_{\nu\mu}.
\label{gmu1}
\end{eqnarray}
As a result, the curvature 
$R^{AB}_{\mu\nu}(x)$ (\ref{rcurvature})
\begin{eqnarray}
aR^{AB}_{\mu\nu}(x)&=&G^{AB}_{\mu\nu}(x)-G^{AB}_{\nu\mu}(x)\nonumber\\
&=&a[\partial_\mu \omega^{AB}_\nu(x)-\partial_\nu \omega^{AB}_\mu(x)]\nonumber\\
&-&(ga)[\omega^{Ae}(x)\wedge\omega^B_e(x)]_{\mu\nu},
\label{xs7}
\end{eqnarray}
where we use
\begin{eqnarray} [\omega^{Ae}(x)\wedge\omega^B_e(x)]_{\mu\nu}=-[\omega^{Ae}(x)\wedge\omega^B_e(x)]_{\nu\mu}.
\label{e2w}
\end{eqnarray}


\section{}\label{xcon}

The properties of the vertex fields $v_{\mu\nu}(x)$ (\ref{v2}) and (\ref{dirace}):
\begin{eqnarray}
v_{\mu\nu}&=&\gamma_5\frac{i}{2}[\gamma_a\gamma_b-\gamma_b\gamma_a]\frac{1}{2}(e^a_\mu e^b_\nu-e^a_\nu e^b_\mu)\nonumber\\
&=&\gamma_5\frac{i}{2}(e_\mu e_\nu-e_\nu e_\mu)=\frac{i}{2}\gamma_5(e\wedge e)_{\mu\nu};\label{eeq1}\\
v_{\mu\nu}^\dagger &=&\gamma_5^\dagger\sigma^\dagger_{ab}
(e^a\wedge e^b)^\dagger_{\mu\nu}
= \gamma_5\sigma_{ab}\frac{1}{2}
(e^b_\mu e^a_\nu-e^b_\nu e^a_\mu)\nonumber\\ 
&=& -\gamma_5\sigma_{ab}(e^a\wedge e^b)_{\mu\nu}=-v_{\mu\nu}=v_{\nu\mu}
\label{eeq2}
\end{eqnarray}
for the case $v_{\mu\nu}(x)=\gamma_5e_{\mu\nu}(x)$. Equations (\ref{eeq1}) and (\ref{eeq2}) are the same for the case $v_{\mu\nu}(x)=e_{\mu\nu}(x)$, because of $\gamma_5^\dagger=\gamma_5$. 
For the sake of simplifying notations in following calculations, we introduce 
\begin{eqnarray}
t_{\mu\nu}^{ab}\equiv (e^a\wedge e^b)_{\mu\nu}=\frac{1}{2}(e^a_\mu e^b_\nu-e^a_\nu e^b_\mu),\quad
[t_{\mu\nu}^{ab}]^\dagger=-t_{\mu\nu}^{ab},
\label{tmn}
\end{eqnarray}
$t_{\mu\nu}^{ab}=-t_{\nu\mu}^{ab}$, $t_{\mu\nu}^{ab}=-t_{\mu\nu}^{ba}$ and $e_{\mu\nu}=\sigma_{ab}t_{\mu\nu}^{ab}$.

We calculate the naive continuum limit of Eqs.~(\ref{xs}), (\ref{xsh1}), and (\ref{xsc+}) in powers of $ga\omega_\mu$.  First, at the order ${\mathcal O}(a^0)$, we consider 
all link fields in Eqs.~(\ref{xs}) and (\ref{xsh1}) to be identity, e.g., $U_{\mu}(x)\approx 1, U_{\rho}(x+a_\mu)\approx  1$, and $
U_{\nu}(x+a_\nu)\approx 1$.
Using Eqs.~(\ref{xsc})-(\ref{xsc+}), (\ref{eeq1}), and (\ref{eeq2}), we obtain up to order ${\mathcal O}(a^0)$
\begin{eqnarray}
X_{h}(v,U)+X^\dagger_{h}(v,U)={\rm tr}\left[v_{\nu\mu}(x)v_{\mu\rho}(x+a_\mu)v_{\rho\nu}(x+a_\nu)\right]
+{\rm h.c.}=0. 
\label{xs00}
\end{eqnarray}
Second, at the order ${\mathcal O}(a)$, we consider two link fields in Eqs.~(\ref{xs}) and (\ref{xsh1}) to be identity. The case (1): 
$U_{\nu}(x+a_\nu)\approx 1$ and $U_{\rho}(x+a_\mu)\approx 1$, we have up to order ${\mathcal O}(a)$, 
\begin{eqnarray}
X_{h}(v,U)&\approx & 
{\rm tr}\left[v_{\nu\mu}(x)U_{\mu}(x)v_{\mu\rho}(x+a_\mu)
v_{\rho\nu}(x+a_\nu)\right]\nonumber\\
&\approx &{\rm tr}\left[v_{\nu\mu}(x)v_{\mu\rho}(x+a_\mu)
v_{\rho\nu}(x+a_\nu)\right]\nonumber\\
&+& iga\omega^{AB}_\mu(x){\rm tr}[\gamma_5\sigma_{ab}\sigma_{AB}\sigma_{cd}\sigma_{ef}]
t^{ab}_{\nu\mu}(x)t^{cd}_{\mu\rho}(x+a_\mu)t^{ef}_{\rho\nu}(x+a_\nu).
\label{xs1}
\end{eqnarray}
for the case $v_{\mu\nu}(x)=\gamma_5e_{\mu\nu}(x)$.
Using Eqs.~(\ref{xs})-(\ref{xsc+}), and (\ref{xs00}), we have 
\begin{eqnarray}
X_{h}(v,U)+X^\dagger_{h}(v,U) 
&\approx & iga[\omega^{AB}_\mu(x)-\omega^{AB}_\nu(x)]\cdot\nonumber\\
&\cdot&{\rm tr}[\gamma_5\sigma_{ab}\sigma_{AB}\sigma_{cd}\sigma_{ef}]
t^{ab}_{\nu\mu}(x)t^{cd}_{\mu\rho}(x+a_\mu)t^{ef}_{\rho\nu}(x+a_\nu).
\label{xs3}
\end{eqnarray} 
The case (2): $U_{\mu}(x+a_\mu)\approx 1$ and $U_{\rho}(x+a_\mu)\approx 1$, we obtain the result with the replacement $[\omega^{AB}_\mu(x)-\omega^{AB}_\nu(x)]\rightarrow
[\omega^{AB}_\nu(x)-\omega^{AB}_\mu(x)]$ in Eq.~(\ref{xs3}). Taking into account all contributions from these cases, we obtain up to the order  ${\mathcal O}(a)$
\begin{eqnarray}
X_{h}(v,U)+X^\dagger_{h}(v,U)=0. 
\label{xs00.0}
\end{eqnarray}
These results are also valid for the case $v_{\mu\nu}(x)=e_{\mu\nu}(x)$, since the calculations of Eqs.~(\ref{xs00})-(\ref{xs3}) without $\gamma_5$ are the same.

Third, at the order ${\mathcal O}(a^2)$,
we consider one link field in Eqs.~(\ref{xs}) and (\ref{xsh1}) to be identity, e.g., 
$ U_{\rho}(x+a_\mu)\approx 1$, 
\begin{eqnarray}
X_{h} (v,U)&\approx & {\rm tr}\left[v_{\nu\mu}(x)U_{\mu}(x)v_{\mu\rho}(x+a_\mu)
v_{\rho\nu}(x+a_\nu)U_{\nu}(x+a_\nu)\right]
\nonumber\\
&\approx & 
{\rm tr}\left[v_{\nu\mu}(x)U_{\mu}(x)U_{\nu}(x)v_{\mu\rho}(x+a_\mu)
v_{\rho\nu}(x+a_\nu)\right]
,
\label{xs1.2}
\end{eqnarray}
where in the second line, 
we use Eq.~(\ref{link0}), $[\sigma_{ab},\gamma_5]=0$, $[U_\mu(x),v_{\rho\nu}]={\mathcal O}(a)$, and $U_\nu(x+a_\nu)=U_\nu(x)+{\mathcal O}(a)$. 
Using Eq.~(\ref{uu0}) or (\ref{uu1}) for $U_{\mu\nu}(x)\equiv U_\mu (x) U_\nu(x)$ and the result (\ref{xs00}), we have up to ${\mathcal O}(a^2)$
\begin{eqnarray}
X_{h} (v,U)&\approx & 
{\rm tr}\left[v_{\nu\mu}(x)U_{\mu\nu}(x)v_{\mu\rho}(x+a_\mu)
v_{\rho\nu}(x+a_\nu)\right]\nonumber\\
&=&iagG^{AB}_{\mu\nu}(x){\rm tr}[\gamma_5\sigma_{ab}\sigma_{AB}\sigma_{cd}\sigma_{ef}]
t^{ab}_{\nu\mu}(x)t^{cd}_{\mu\rho}(x+a_\mu)t^{ef}_{\rho\nu}(x+a_\nu),
\label{xs3.0}
\end{eqnarray}
for the case $v_{\mu\nu}(x)=\gamma_5\sigma_{\mu\nu}(x)$. Using the relationships $X^\dagger_{h}(v,U)=X_{h}(v,U)|_{\mu\leftrightarrow\nu}$ (\ref{xsc}) and (\ref{xsh1}) and $t_{\mu\nu}^{ab}=-t_{\nu\mu}^{ab}$ (\ref{tmn}), we have
\begin{eqnarray}
X^\dagger_{h} (v,U)&\approx & 
-iagG^{AB}_{\nu\mu}(x){\rm tr}[\gamma_5\sigma_{ab}\sigma_{AB}\sigma_{cd}\sigma_{ef}]
t^{ab}_{\nu\mu}(x)t^{cd}_{\mu\rho}(x+a_\mu)t^{ef}_{\rho\nu}(x+a_\nu).
\label{xs3.1}
\end{eqnarray}
As a result, using Eq.~(\ref{xs7}) in Appendix \ref{con}, we obtain up to ${\mathcal O}(a^2)$
\begin{eqnarray}
X_{h} (v,U)+X^\dagger_{h} (v,U)&\approx & 
ia^2gR^{AB}_{\mu\nu}(x){\rm tr}[\gamma_5\sigma_{ab}\sigma_{AB}\sigma_{cd}\sigma_{ef}]
t^{ab}_{\nu\mu}(x)t^{cd}_{\mu\rho}(x+a_\mu)t^{ef}_{\rho\nu}(x+a_\nu).
\label{xs3.2}
\end{eqnarray}
For the case $v_{\mu\nu}(x)=e_{\mu\nu}(x)$, the result is given by
Eq.~(\ref{xs3.2}) without $\gamma_5$.

In Appendix \ref{a2}, we show the calculations of 
${\rm tr}[\gamma_5\sigma_{ab}\sigma_{AB}\sigma_{cd}\sigma_{ef}]$ and ${\rm tr}[\sigma_{ab}\sigma_{AB}\sigma_{cd}\sigma_{ef}]$ in Eq.~(\ref{xs3.2}). Using these results (\ref{xs2}) and (\ref{xs2.0}), we obtain
for the case $v_{\mu\nu}(x)=\gamma_5e_{\mu\nu}(x)$,
\begin{eqnarray}
X_{h} (v,U)+X^\dagger_{h} (v,U)
&\approx & 
8a^2gR^{AB}_{\mu\nu}(x)\epsilon_{abAB}
t^{ab}_{\nu\mu}(x)t^{cd}_{\mu\rho}(x+a_\mu)t^{cd}_{\rho\nu}(x+a_\nu);
\label{xs6}
\end{eqnarray}
and for the case $v_{\mu\nu}(x)=e_{\mu\nu}(x)$,
\begin{eqnarray}
X_{h} (v,U)+X^\dagger_{h} (v,U)
&\approx & 
2i\cdot 8a^2gR^{AB}_{\mu\nu}(x)
t^{AB}_{\nu\mu}(x)t^{cd}_{\mu\rho}(x+a_\mu)t^{cd}_{\rho\nu}(x+a_\nu).
\label{xs6.1}
\end{eqnarray}

Using Eqs.~(\ref{seeq}) and (\ref{tmn}), 
we rewrite the fundamental area (\ref{areaod1}) and (\ref{areaod2}) of the 2-simplex $h(x)$ in terms of $t^{cd}_{\mu\rho}(x+a_\mu)$ and $t^{cd}_{\rho\nu}(x+a_\nu)$: 
\begin{eqnarray}
S^{\rm h}_{\mu\rho}(x+a_\mu)&=&\sigma_{cd}S^{cd}_{\mu\rho}(x+a_\mu),\quad
S^{cd}_{\mu\rho}(x+a_\mu)= -ia^2t^{cd}_{\mu\rho}(x+a_\mu),\label{xsvs1}\\
S^{\rm h}_{\rho\nu}(x+a_\nu)&=&\sigma_{cd}S^{cd}_{\rho\nu}(x+a_\nu),\quad
S^{cd}_{\rho\nu}(x+a_\nu)=-ia^2t^{cd}_{\rho\nu}(x+a_\nu),
\label{xsvs2}
\end{eqnarray}
where
$S^{\rm h}_{\mu\rho}(x+a_\mu)=-S^{\rm h}_{\rho\mu}(x+a_\mu)$ and $S^{\rm h}_{\rho\nu}(x+a_\nu)=-S^{\rm h}_{\nu\rho}(x+a_\nu)$. As discussed in Eqs.~(\ref{areaod}), (\ref{areaod1}), and (\ref{areaod2}) [see Sec.~\ref{2-simplex-area}], the area of three
area operators $S^{\rm h}_{\mu\nu}(x)$, $S^{\rm h}_{\rho\mu}(x+a_\mu)$ and $S^{\rm h}_{\nu\rho}(x+a_\nu)$ are identical. Therefore,
equivalently to Eqs.~(\ref{aread}) and (\ref{vold}), we write the volume element contributed from the 2-simplex $h(x)$ as
\begin{eqnarray}
dV_h&\equiv& S^{cd}_{\mu\rho}(x+a_\mu)S^{cd\dagger}_{\rho\nu}(x+a_\nu)= a^4t^{cd}_{\mu\rho}(x+a_\mu)t^{cd}_{\rho\nu}(x+a_\nu)\nonumber\\
&=& S^{cd}_{\mu\nu}(x)S^{cd\dagger}_{\mu\nu}(x)= a^4t^{cd}_{\mu\nu}(x)t^{cd}_{\mu\nu}(x),
\label{xsv}
\end{eqnarray}
where indexes $c,d$ are summed, while indexes $\mu,\nu$ and $\rho$ are not summed.
Using Eq.~(\ref{gs}) in Appendix \ref{a2}, we obtain
\begin{eqnarray}
dV_h(x)
&=& S^2_{\rm h}(x) = \frac{1}{8}\,{\rm tr}\left[S^{\rm h}_{\mu\nu}(x)S^{{\rm h}\dagger}_{\mu\nu}(x)\right],
\label{xsv1}
\end{eqnarray}
where $S^{\rm h}_{\mu\nu}(x)=\sigma_{ab}S^{ab}_{\mu\nu}(x)$ and $S^{ab}_{\mu\nu}(x)=-ia^2t^{ab}_{\mu\nu}(x)$.
Using Eqs.~(\ref{xs6},\ref{xs6.1}) and (\ref{xsvs1}-\ref{xsv1}), we can show the regularized Palatini action (\ref{pact}) and Host action (\ref{diract}) approach to their continuum counterparts (\ref{host}) and (\ref{host1}) in the naive continuum limit $ag\omega_\mu \ll 1$.

\section{}\label{a2}
\comment{
First we prove 
\begin{eqnarray}
{\rm tr}[\gamma_5\sigma_{ab}\sigma_{cd}\sigma_{ef}]=0. 
\label{tr0}
\end{eqnarray}
The trace ${\rm tr}[\gamma_5\sigma_{ab}\sigma_{cd}\sigma_{ef}]$ has a general form ${\rm tr}[\gamma_5\gamma_a\gamma_b\gamma_c\gamma_d\gamma_e\gamma_f]$ with $a\not= b$,
$c\not= d$, and $e\not= f$. For $\gamma_5=\gamma_0\gamma_1\gamma_2\gamma_3$, in order to have nonvanishing contribution, indexes $a,b,c,d$ must take different values, e.g., ${\rm tr}[\gamma_5\gamma_0\gamma_1\gamma_2\gamma_3\gamma_e\gamma_f]=0$, while $e\not=f$ can only take values  $0,1,2,3$. Thus it turn out ${\rm tr}[\gamma_5\gamma_a\gamma_b\gamma_c\gamma_d\gamma_e\gamma_f]\sim {\rm tr}[\gamma_5\gamma_a\gamma_b]=0$ (in six dimensions case it can be non-zero for different values of $a,b,c,d,e,f$). In other words, possible non-zero case is that two of indexes are the same, but  $a\not= b, c\not= d,e\not= f$. One can consider for instance 
$d=e$, however the index $f$ has to be one of $a,b,c$ values. Thus the result should be ${\rm tr}[\gamma_5\gamma_a\gamma_b]$ which is zero. Then we prove Eq.~(\ref{tr0}).
}

It can be shown that
$
{\rm tr}[\gamma_5\sigma_{ab}\sigma_{cd}\sigma_{ef}]=0 
$ for $\gamma_5=\gamma_0\gamma_1\gamma_2\gamma_3$ in the four-dimensional space-time.
nonvanishing contributions of the following trace 
\begin{eqnarray}
{\rm tr}[\gamma_5\sigma_{ab}\sigma_{AB}\sigma_{cd}\sigma_{ef}], 
\label{tr0}
\end{eqnarray}
come from the product of two spinor matrices $\sigma$'s in Eq.~(\ref{tr0}) being identical,  
\begin{eqnarray}
{\rm tr}[\gamma_5\sigma_{ab}\sigma_{AB}\sigma_{cd}\sigma_{ef}]\Rightarrow
{\rm tr}[\gamma_5\sigma_{ab}\sigma_{AB}].
\label{tr1}
\end{eqnarray}
In Eq.~(\ref{xs3.2}), as example, we take (i) $\sigma_{cd}\sigma_{ef}=1$ for  $c=e,d=f$ and 
(ii) $\sigma_{cd}\sigma_{ef}=-1$
$c=f,d=e$,
\begin{eqnarray}
&&\sum_{cdef}[\sigma_{cd}\sigma_{ef}]t^{cd}_{\mu\rho}(x+a_\mu)t^{ef}_{\rho\nu}(x+a_\nu)\nonumber\\
&&=\sum_{cd}[\sigma_{cd}\sigma_{cd}]t^{cd}_{\mu\rho}(x+a_\mu)t^{cd}_{\rho\nu}(x+a_\nu)+\sum_{cd}[\sigma_{cd}\sigma_{dc}]t^{cd}_{\mu\rho}(x+a_\mu)t^{dc}_{\rho\nu}(x+a_\nu),\nonumber\\
&&=\sum_{cd}t^{cd}_{\mu\rho}(x+a_\mu)t^{cd}_{\rho\nu}(x+a_\nu)-
\sum_{cd}t^{cd}_{\mu\rho}(x+a_\mu)t^{dc}_{\rho\nu}(x+a_\nu)\nonumber\\
&&=2\sum_{cd}t^{cd}_{\mu\rho}(x+a_\mu)t^{cd}_{\rho\nu}(x+a_\nu).
\label{tr2}
\end{eqnarray} 
Thus, in Eq.~(\ref{xs3.2}) we have
\begin{eqnarray}
&&{\rm tr}[\gamma_5\sigma_{ab}\sigma_{AB}\sigma_{cd}\sigma_{ef}]
t^{ab}_{\nu\mu}(x)t^{cd}_{\mu\rho}(x+a_\mu)t^{ef}_{\rho\nu}(x+a_\nu)\nonumber\\
&&=2{\rm tr}[\gamma_5\sigma_{ab}\sigma_{AB}]
t^{ab}_{\nu\mu}(x)t^{cd}_{\mu\rho}(x+a_\mu)t^{cd}_{\rho\nu}(x+a_\nu)\nonumber\\
&&= -8i\epsilon^{abAB}
t^{ab}_{\nu\mu}(x)t^{cd}_{\mu\rho}(x+a_\mu)t^{cd}_{\rho\nu}(x+a_\nu),
\label{xs2}
\end{eqnarray}
where we use the formula
\begin{eqnarray}
{\rm tr}\left(\gamma_5\sigma^{ab}\sigma^{AB}\right)&=&\frac{1}{2}{\rm tr}\left(\gamma_5\{\sigma^{ab},\sigma^{AB}\}\right)=-4i\epsilon^{abAB},
\label{g5s}
\end{eqnarray}
and Eq.~(\ref{sisi}). In the same way we calculate Eq.~(\ref{xs3.2}) for other possibilities, e.g., $\sigma_{ab}\sigma_{ef}=1$ for (i) $a=e, b=f$ and (ii) $\sigma_{ab}\sigma_{ef}=-1$ $a=f, b=e$. As a result, we obtain Eq.~(\ref{xs6}).

Analogous to the discussions for Eq.~(\ref{tr1}), nonvanishing contributions to 
${\rm tr}[\sigma_{ab}\sigma_{AB}\sigma_{cd}\sigma_{ef}]$ come from the product of two spinor matrices $\sigma$'s being identical,  
\begin{eqnarray}
{\rm tr}[\sigma_{ab}\sigma_{AB}\sigma_{cd}\sigma_{ef}]\Rightarrow
{\rm tr}[\sigma_{ab}\sigma_{AB}].
\label{tr1.0}
\end{eqnarray}
In Eq.~(\ref{xs3.2}) without $\gamma_5$, as example, we take (i) $\sigma_{cd}\sigma_{ef}=1$ for  $c=e,d=f$ and 
(ii) $\sigma_{cd}\sigma_{ef}=-1$
$c=f,d=e$, and use formula 
\begin{eqnarray}
{\rm tr}\left(\sigma^{ab}\sigma^{AB}\right)
&=&4(\delta^{aA}\delta^{bB}-\delta^{aB}\delta^{bA}).
\label{gs}
\end{eqnarray}
As a result we obtain
\begin{eqnarray}
&&{\rm tr}[\sigma_{ab}\sigma_{AB}\sigma_{cd}\sigma_{ef}]
t^{ab}_{\nu\mu}(x)t^{cd}_{\mu\rho}(x+a_\mu)t^{ef}_{\rho\nu}(x+a_\nu)\nonumber\\
&&=2{\rm tr}[\sigma_{ab}\sigma_{AB}]
t^{ab}_{\nu\mu}(x)t^{cd}_{\mu\rho}(x+a_\mu)t^{cd}_{\rho\nu}(x+a_\nu)\nonumber\\
&&= 2\cdot 8
t^{AB}_{\nu\mu}(x)t^{cd}_{\mu\rho}(x+a_\mu)t^{cd}_{\rho\nu}(x+a_\nu),
\label{xs2.0}
\end{eqnarray}
and Eq.~(\ref{xs3.2}) without $\gamma_5$ becomes Eq.~(\ref{xs6.1}).

\section{}\label{meana}

Using the properties (\ref{eeq1}) of the vertex field $v_{\mu\nu}(x)=\gamma_5e_{\mu\nu}(x)$, we have 
\begin{eqnarray}
\bar X_{h} (v,U)&=& 
\frac{i}{2}{\rm tr}\gamma_5\Big[e_\nu(x)U_{\mu}(x)U_{\rho}(x+a_\mu)U_{\nu}(x+a_\nu)e_\mu(x)\nonumber\\
&-&e_\mu(x)U_{\mu}(x)U_{\rho}(x+a_\mu)U_{\nu}(x+a_\nu)e_\nu(x)\Big]M^2_{\rm h}\nonumber\\
\bar X_{h}^\dagger(v,U)&=& 
\frac{i}{2}{\rm tr}\gamma_5\Big[e_\mu(x)U_{\nu}(x)U_{\rho}(x+a_\nu)U_{\mu}(x+a_\mu)e_\nu(x)\nonumber\\
&-&
e_\nu(x)U_{\nu}(x)U_{\rho}(x+a_\nu)U_{\mu}(x+a_\mu)e_\mu(x)\Big]M^2_{\rm h},
\label{xsg1}
\end{eqnarray}
and 
\begin{eqnarray}
&&\bar X_{h} (v,U)+\bar X_{h}^\dagger(v,U)\nonumber\\
&&=
\frac{i}{2}M^2_{\rm h}{\rm tr}\gamma_5e_\nu(x)\Big[U_{\mu}(x)U_{\rho}(x+a_\mu)U_{\nu}(x+a_\nu)
-U_{\nu}(x)U_{\rho}(x+a_\nu)U_{\mu}(x+a_\mu)\Big]e_\mu(x)\nonumber\\
&&+\frac{i}{2}M^2_{\rm h}{\rm tr}\gamma_5e_\mu(x)\Big[U_{\nu}(x)U_{\rho}(x+a_\nu)U_{\mu}(x+a_\mu)
-U_{\mu}(x)U_{\rho}(x+a_\mu)U_{\nu}(x+a_\nu)\Big]e_\nu(x)\nonumber\\
&&= 
{\rm tr}\left[e_\nu(x)\gamma_5H_{\nu\mu}(x)e_\mu(x)\right]-{\rm tr}\left[e_\mu(x)\gamma_5H_{\nu\mu}(x)e_\nu(x)\right],
\label{xsg2}
\end{eqnarray}
where 
$\gamma_5 e_\mu(x)=-e_\mu(x)\gamma_5 $ and the
tensor 
\begin{eqnarray}
H_{\nu\mu}(x)&\equiv&
\frac{i}{2}M^2_{\rm h}\left[U_{\nu}(x)U_{\rho}(x+a_\nu)U_{\mu}(x+a_\mu)-U_{\mu}(x)U_{\rho}(x+a_\mu)U_{\nu}(x+a_\nu)
\right]\nonumber\\
&=&
\frac{i}{2}M^2_{\rm h}\left[U_{\nu}(x)U_{\rho}(x+a_\nu)U_{\mu}(x+a_\mu)-U^\dagger_{\mu}(x+a_\mu)
U^\dagger_{\rho}(x+a_\nu)U^\dagger_{\nu}(x)\right]\nonumber\\
&=&
\frac{i}{2}M^2_{\rm h}\left[U_{\nu}(x)U_{\rho}(x+a_\nu)U_{\mu}(x+a_\mu)\right]+{\rm h.c.}\nonumber\\
&=&
\frac{i}{2}M^2_{\rm h}\left[U_{\nu}(x)U_{\rho}(x+a_\nu)U^\dagger_{\mu}(x)\right]+{\rm h.c.},
\label{chg2}
\end{eqnarray}
$H_{\nu\mu}=-H_{\mu\nu}$ and $H^\dagger_{\nu\mu}=H_{\nu\mu}$, 
following the relations $U_\mu(x)=U_\mu^\dagger(x+a_\mu)$, $U^\dagger_\nu(x)=U_\nu(x+a_\nu)$ and $U_{\rho}(x+a_\mu)=U^\dagger_{\rho}(x+a_\nu)$. The $H_{\mu\nu}(x)$ is a product of three edge fields $U_{\nu}(x)$, $U^\dagger_{\mu}(x)$ and $U_{\rho}(x+a_\mu)$ of the 2-simplex $h(x)$. 
For the case $v_{\mu\nu}(x)=e_{\mu\nu}(x)$, the same result can be obtained by the replacement $\gamma_5\rightarrow -1$ in Eq.~(\ref{xsg2}). The sum of two contributions gives  Eqs.~(\ref{action2d}) and (\ref{chg2t}) in the main text.   

\section{}\label{meana1}

For each 2-simplex $h\,\,(\mu\not=\nu\not=\rho)$, we have the fundamental area operator $e_\mu\wedge e_\nu\equiv e_\nu e_\mu-e_\mu e_\nu$ [see Eq.~(\ref{areaod})] and ${\rm tr }(e_\nu e_\mu-e_\mu e_\nu)=0$, we can rewrite the mean-field action (\ref{action2d}) as follows:
\begin{eqnarray}
\bar {\mathcal A}_h &=& {\rm tr }(e_\nu e_\mu-e_\mu e_\nu) + \bar {\mathcal A}_h\nonumber\\
&=& 
{\rm tr}\left[e_\nu\left(I-\Gamma^h_{\nu\mu}\right)e_\mu - e_\mu\left(I-\Gamma^h_{\nu\mu}\right)e_\nu\right]\nonumber\\
&=& {\rm tr}\left\{\left(\matrix{e_\nu & e_\mu}\right)\left[\matrix{0 & (I-\Gamma^h_{\nu\mu})\cr
-(I-\Gamma^h_{\nu\mu}) & 0}\right]\left(\matrix{e_\nu\cr e_\mu}\right)\right\}.
\label{action2d1}
\end{eqnarray}
where $I$ is the identity matrix. For  each single 2-simplex $h$, we have the integrations 
\begin{eqnarray}
\int_h d e_\mu de_\nu \exp -\bar{\mathcal A}_h
&=&{\rm det}^{-1}[I-\Gamma^h],
\label{det1}\\
\int_h d e_\mu de_\nu (e_\mu e_\nu) \exp -\bar{\mathcal A}_h
&=&\frac{1}{2}[I-\Gamma^h]_{\mu\nu}^{-1}{\rm det}^{-1}[I-\Gamma^h],
\label{det2}\\
\int_h d e_\mu de_\nu \,e_{\mu\nu}\, \exp -\bar{\mathcal A}_h
&=&\frac{i}{4}\Big\{[I-\Gamma^h]_{\mu\nu}^{-1}-[I-\Gamma^h]_{\nu\mu}^{-1}\Big\}{\rm det}^{-1}[I-\Gamma^h].
\label{det3}
\end{eqnarray}  

Using Eqs.~(\ref{det0}) and (\ref{det}), we calculate the mean-field partition function (\ref{meanpar})
\begin{eqnarray}
\bar Z_{EC}&=&\prod_{h\in {\mathcal M}} \int_h dU_\mu dU_\nu dU_\rho{\rm det}^{-1}[I-\Gamma^h]\nonumber\\
&=&\prod_{h\in {\mathcal M}} \int_h dU_\mu dU_\nu dU_\rho\Big[ 1+\sum_a\Gamma^h_{aa}+\frac{1}{2}\sum_{a,b}(\Gamma^h_{aa}\Gamma^h_{bb}+\Gamma^h_{ab}\Gamma^h_{ba})+\cdot\cdot\cdot\Big].
\label{udet}
\end{eqnarray}
In Eq.~(\ref{udet}), the first term  is one due to the formula (\ref{mea20}), the second term vanishes due to the formula (\ref{mea21}), and 
nonvanishing contribution, due to Eqs.~(\ref{mea21}) and (\ref{mea2}), 
comes from the term $\Gamma^h_{ab}\Gamma^h_{ba}$ in the third term. 
Using Eqs.~(\ref{chg2t}), (\ref{mea21}), and (\ref{mea2}), we have
\begin{eqnarray}
 \int_h dU_\mu dU_\nu dU_\rho \frac{1}{2}\sum_{a,b}\Gamma^h_{ab}\Gamma^h_{ba} &=&
\frac{1}{2} \left(\frac{1}{8g^2}\right)^2M_{\rm h}^4 \left(\frac{i}{2}\right)\left(\frac{-i}{2}\right)\int_h dU_\mu dU_\nu dU_\rho\cdot\nonumber\\  &\cdot&    2\Big[\Big(\gamma_5-\frac{1}{\gamma}\Big)_{aj}[U_\nu]_{jl}[U_\rho]_{ln}[U^\dagger_\mu]_{nb} \Big(\gamma_5-\frac{1}{\gamma}\Big)_{bm}[U_\mu]_{mk}[U^\dagger_\rho]_{ki}[U^\dagger_\nu]_{ia}\Big]\nonumber\\
&=&
\frac{1}{2} \left(\frac{1}{8g^2}\right)^2M_{\rm h}^4 \left(\frac{1}{4}\right)    
\frac{2}{d_j^3}{\rm tr}\Big[\Big(\gamma_5-\frac{1}{\gamma}\Big)^2\Big]\nonumber\\
&=&\left(\frac{1}{8g^2}\right)^2M_{\rm h}^4    
\frac{1}{d_j^3}\Big(1+\frac{1}{\gamma^2}\Big).
\label{udetf}
\end{eqnarray} 
As a result, we obtain the mean-field partition function (\ref{par2}) in the main text.

Using Eq.~(\ref{det3}), we calculate the mean-field value of the mean-field action $\bar{\mathcal A}_h$ (\ref{action2d}) of  the single 2-simplex $h$,
\begin{eqnarray}
\langle\bar{\mathcal A}_h\rangle_\circ &=&\langle{\rm tr}\left[e_\nu\Gamma^h_{\nu\mu}e_\mu-e_\mu\Gamma^h_{\nu\mu}e_\nu\right]\rangle_\circ
\nonumber\\
&=&\frac{1}{2\bar Z_h}\int_h {\mathcal D}U\, {\rm tr} \Big\{\frac{\Gamma^h_{\nu\mu}}{I-\Gamma^h_{\nu\mu}}-\frac{\Gamma^h_{\nu\mu}}{I-\Gamma^h_{\mu\nu}}\Big\} 
{\rm det}^{-1}[I-\Gamma^h]\nonumber\\
&=&\frac{1}{2\bar Z_h}\int_h {\mathcal D}U\,{\rm tr} \Big\{2\Gamma^h_{\nu\mu}\Gamma^h_{\nu\mu} + \cdot\cdot\cdot\,\,\Big\} 
{\rm det}^{-1}[I-\Gamma^h]\nonumber\\
&=&\frac{1}{\bar Z_h}\left(\frac{1}{8g^2}\right)^2M_{\rm h}^4 \left(\frac{1}{4}\right)    
\frac{2}{d_j^3}{\rm tr}\Big[\Big(\gamma_5-\frac{1}{\gamma}\Big)^2\Big]\nonumber\\
&=&\frac{1}{\bar Z_h}\left(\frac{1}{8g^2}\right)^2M_{\rm h}^4     
\frac{2}{d_j^3}\Big(\frac{\gamma^2+1}{\gamma^2}\Big),
\label{va1}
\end{eqnarray}
which gives Eq.~(\ref{aqj}) in the main text.

Using Eqs.~(\ref{det1}), (\ref{det2}), and (\ref{udet}) and $(\Gamma^h)_{\mu\rho}=-(\Gamma^h)_{\rho\mu}$ [see Eqs.~(\ref{chg2t}) and (\ref{chg2})], 
we have 
\begin{eqnarray}
\langle [e_{\mu\rho}]\rangle^h_\circ &=& \frac{i}{4}\frac{1}{\bar Z_h}\int_h {\mathcal D}U \left\{[I-\Gamma^h]^{-1}_{\mu\rho}-[I-\Gamma^h]^{-1}_{\rho\mu}\right\}{\rm det}^{-1}[I-\Gamma^h]\nonumber\\
&=&\frac{i}{4}\frac{1}{\bar Z_h}\int_h {\mathcal D}U\left[\,2\Gamma^h_{\mu\rho}
+\cdot\cdot\cdot\,
\right]
{\rm det}^{-1}[I-\Gamma^h]\nonumber\\
&=&\frac{i}{4}\frac{2}{\bar Z_h}\left(\frac{1}{8g^2}\right)^2M_{\rm h}^4 \left(\frac{1}{4}\right)    
\frac{2}{d_j^3}\Big[\Big(\gamma_5-\frac{1}{\gamma}\Big)^2\Big],
\label{va0}
\end{eqnarray}
and $\langle [e_{\rho\mu}]\rangle^h_\circ=-\langle [e_{\mu\rho}]\rangle^h_\circ$. 
As a result, Eq.~(\ref{ao30}) becomes 
\begin{eqnarray}
\langle {\mathcal A}_{EC}\rangle_\circ &\approx & \sum_{h\in {\mathcal M}} \frac{(\bar Z_h)^2}{4M^2_{\rm h}}\left\{\langle{\rm tr}\left[e_\nu\Gamma^h_{\nu\mu}e_\mu-e_\mu\Gamma^h_{\nu\mu}e_\nu\right]\rangle_\circ{\rm tr}\Big[\langle [e_{\mu\rho}]\rangle^h_\circ\,\, \langle [e_{\rho\nu}]\rangle^h_\circ\Big]\right\}\nonumber\\
&= & \sum_{h\in {\mathcal M}} \frac{1}{M^2_{\rm h}}
\left(\frac{1}{\bar Z_h}\right)\left(\frac{1}{8g^2}\right)^6(M_{\rm h}^4)^3 \left(\frac{1}{4}\right)\left(    
\frac{2}{d_j^3}\right)^3\Big(\frac{\gamma^2+1}{\gamma^2}\Big)\Big[\Big(\frac{\gamma^2+1}{\gamma^2}\Big)^2+\frac{4}{\gamma^2}\Big],
\label{fao30}
\end{eqnarray}
and we obtain Eq.~(\ref{fao}) in the main text.

\comment{
Let us see Eq.~(\ref{ao31}). Using Eqs.~(\ref{action2d},\ref{chg2t}), we can approximately have 
\begin{eqnarray}
\langle {\rm tr}(v_{\mu\rho})\rangle^h_\circ \langle {\rm tr}(v_{\rho\nu})\rangle^h_\circ &\approx & \frac{1}{W}
\langle \bar{\mathcal A}_h\rangle^h_\circ \langle \bar{\mathcal A}_h\rangle^h_\circ,
\label{aao310}
\end{eqnarray}
where the normalization factor
\begin{eqnarray}
W&= & \int_h dU_\mu dU_\nu dU_\rho \Gamma^h_{\mu\nu\rho}\Gamma^h_{\mu\nu\rho}
=2\left(\frac{M^2_{\rm h}}{8g^2}\right)^2\frac{\gamma^2+1}{\gamma^2 d_j^3},
\label{aao31}
\end{eqnarray}
see Eq.~(\ref{udetf}) in above.
}

\end{document}